\documentclass[a4paper,10pt,titlepage,twoside,openright]{book}
\usepackage{amssymb,amsmath,illcmolthesis,stmaryrd, amsthm,url,color,fancyhdr}
\usepackage[Bjornstrup]{fncychap}
\usepackage[all]{xy}

\newcommand{\leftand}{~
     \mathbin{\setlength{\unitlength}{1ex}
     \begin{picture}(1.4,1.8)(-.3,0)
     \put(-.6,0){$\wedge$}
     \put(-.54,-0.2){\textcolor{white}{\circle*{0.6}}}
     \put(-.54,-0.2){\circle{0.6}}
     \end{picture}
     }}
         
\newcommand{\leftor}{~
     \mathbin{\setlength{\unitlength}{1ex}
     \begin{picture}(1.4,1.8)(-.3,0)
     \put(-.6,0){$\vee$}
     \put(-.54,1.54){\textcolor{white}{\circle*{0.6}}}
     \put(-.54,1.54){\circle{0.6}}
     \end{picture}
     }}
     
\newcommand{\leftandpf}{~
     \setlength{\unitlength}{1ex}
     \begin{picture}(1.4,1.8)(-.3,0)
     \put(-.6,0){$\wedge$}
     \put(-.54,-0.2){\textcolor{white}{\circle*{0.6}}}
     \put(-.54,-0.2){\circle{0.6}}
     \end{picture}
     }

\newcommand{\monus}{~
		 \mathbin{\setlength{\unitlength}{1ex}
		 \begin{picture}(1.4,1.8)(0,0)
		 \put(-.6,0){$-$}
		 \put(0,0.5){$\cdot$}
		 \end{picture}
		 }}

\newcommand{\lland}{\mathbin{\&}
}
\newcommand{\llor}{\mathbin{|}
}
\newcommand{\proj}{\text{pgaul2pgau}}
\newcommand{\iproj}{\text{\emph{pgaul2pgau}}}
\newcommand{\dlaf}{DLA$_{\text{f}}$}
\newcommand{\dlcaf}{DLCA$_{\text{f}}$}
\newcommand{\dltaf}{DLTA$_{\text{f}}$}
\newcommand{\?}{\text{?}}
\newcommand{\iq}{\text{\emph{?}}}
\newcommand{\e}{\text{!}}

\newcommand{\pgaul}{PGA$_{\text{ul}}$}
\newcommand{\pgaulf}{PGA$_{\text{ul}}^{\text{fin}}$}
\newcommand{\pgau}{PGA$_{\text{u}}$}
\newcommand{\pgauf}{PGA$_{\text{u}}^{\text{fin}}$}

\newcommand{\un}{\ensuremath{\textup{\textbf{u}}}}

\newcommand{\lef}{\ensuremath{\triangleleft}}
\newcommand{\rig}{\ensuremath{\triangleright}}

\newcommand{\Eq}{\axname{Eq}}

\newcommand{\MSCL}{\axname{MSCL}}
\newcommand{\SCLe}{\axname{EqFSCL}}
\newcommand{\SCL}{\axname{SCL}}
\newcommand{\RPSCL}{\axname{RPSCL}}

\newcommand{\SSCL}{\axname{SSCL}}

\newcommand{\FreeSCL}{\axname{FSCL}}
\newcommand{\export}{~\Box~}

\newcommand{\CP}{\axname{CP}}

\newcommand{\axname}[1]{\textup{\ensuremath{\textrm{#1}}}}

\newtheorem{theorem}{Theorem}[section]
\newtheorem{lemma}[theorem]{Lemma}
\newtheorem{proposition}{Proposition}

\newtheorem*{notation}{Notation}

\newtheorem{definition}{Definition}  
\theoremstyle{definition}

\newcommand{\ownqed}{\nobreak \ifvmode \relax \else
      \ifdim\lastskip<1.5em \hskip-\lastskip
      \hskip1.5em plus0em minus0.5em \fi \nobreak
      \vrule height0.5em width0.5em depth0em\fi}
      
\author{L.L.~Wortel}
\title{Side Effects in Steering Fragments}
\birthdate{12 October 1984}
\birthplace{Heemskerk}
\defensedate{September 5th, 2011}
\supervisor{Alban Ponse}
\supervisor{Paul Dekker}
\committeemember{Dr Alban Ponse}
\committeemember{Dr Paul Dekker}
\committeemember{Prof Dr Jan van Eijck}
\committeemember{Dr Sara Uckelman}
\committeemember{Prof Dr Benedikt L\"owe}
\begin{document}
\maketitle
\pagestyle{empty}
\cleardoublepage
\addtolength{\oddsidemargin}{.35in}
\emph{I'm dedicating this thesis to my parents, without whom I would never have gotten to this point. I have tested their patience by taking my time to graduate, but they kept supporting every silly little thing I have ever done. Thanks guys, you're the best.}

\chapter*{Abstract}
\thispagestyle{empty}
In this thesis I will give a formal definition of side effects. I will do so by modifying a system for modelling program instructions and program states, Quantified Dynamic Logic, to a system called \dlaf\ (for Dynamic Logic with Assignments as Formulas), which in contrast to QDL allows assignments in formulas and makes use of short-circuit evaluation. I will show the underlying logic in those formulas to be a variant of short-circuit logic called repetition-proof short-circuit logic.

Using \dlaf\ I will define the actual and the expected evaluation of a single instruction. The side effects are then defined to be the difference between the two. I will give rules for composing those side effects in single instructions, thus scaling up our definition of side effects to a definition of side effects in deterministic \dlaf-programs. Using this definition I will give a classification of side effects, introducing as most important class that of marginal side effects. Finally, I will show how to use our system for calculating the side effects in a real system such as Program Algebra (PGA).

\chapter*{Acknowledgements}
\thispagestyle{empty}
I would first and foremost like to thank my supervisor Alban Ponse for the big amounts of time and energy he put into guiding me through this project. His advice has been invaluable to me and his enthusiasm has been a huge motivation for me throughout.

A thank you also goes out to Jan van Eijck for pointing me in the right direction halfway through the project.

Finally I would like to thank my entire thesis committee, consisting of Alban Ponse, Paul Dekker, Jan van Eijck, Sara Uckelman and Benedikt L\"owe, for taking the time to read and grade my thesis.

\hfill --- Lars Wortel, August 2011
\pagenumbering{Roman}
\newpage
\setcounter{page}{0}
\newpage
\thispagestyle{empty}
\pagestyle{fancy}
\tableofcontents
\thispagestyle{plain}

\chapter{Introduction}
\label{ch:introduction}
\pagenumbering{arabic}

\section{What are side effects?}
In programming practice, \emph{side effects} are a well-known phenomenon, even though nobody seems to have an exact definition of what they are. To get a basic idea, here are some examples from natural language and programming that should explain the intuition behind side effects.

Suppose you and your wife have come to an agreement regarding grocery shopping. Upon leaving for work, she told you that ``if I don't call, you do not have to do the shopping''. Later that day, she calls you to tell you something completely different, for instance that she is pregnant. This call now has as side effect that you no longer know whether you have to do grocery shopping or not, even though the meaning of the call itself was something completely different.

Another example is taken from \cite{Dekker}. Suppose someone tells you that ``Phoebe is waiting in front of your door, and you don't know it!'' This is a perfectly fine thing to say, but you cannot say it twice because then it will no longer be true that you don't know that Phoebe is waiting (after all, you were just told). Here, the side effect is that your knowledge gets updated by the sentence, which makes the latter part of that sentence, which is a statement about your knowledge, false.

As said, in programming practice, side effects are a well-known phenomenon. Logically, they are interesting because the possible presence of side effects in a program instruction sequence invalidates principles of propositional logic such as commutativity ($\phi \land \psi \leftrightarrow \psi \land \phi$) and idempotency ($\phi \land \phi \leftrightarrow \phi$). The textbook example is the following program:
\begin{verbatim}
x:=1
if (x:=x+1 and x=2) then y
\end{verbatim}
Here the operator $:=$ stands for assignment and $=$ for an equality test. Assuming an assignment instruction always succeeds (that is, yields the reply \verb$true$), in the above example the test $\phi \land \psi$, where $\phi$ is the instruction \verb$x:=x+1$ and $\psi$ the instruction \verb$x=2$, will succeed and therefore, $y$ will be executed. However, should the order of those instructions be reversed ($\psi \land \phi$), this no longer will be the case. The reason is that the instruction $\phi$ has a side effect: apart from returning \verb$true$, it also increments the variable $x$ with $1$, thus making it $2$. If $\phi$ is executed before $\psi$, the test in $\psi$ (\verb$x=2$) will yield \verb$true$. Otherwise, it will yield \verb$false$.

It is easy to see that should $\phi \land \psi$ be executed twice, the end result will also be \verb$false$. Therefore, for $\chi = \phi \land \psi$, we have that $\chi \land \chi \not\leftrightarrow \chi$.

\section{What are steering fragments?}
Now that I have given a rough idea of what side effects are, the reader is probably wondering about the second part of my thesis title: that of steering fragments. A \emph{steering fragment} or \emph{test} is a program fragment which is concerned with the control flow of the execution of that program. To be exact, a steering fragment will use the evaluation result of a formula (which is a Boolean) and depending on the outcome, will steer further execution of the program. Thus, a steering fragment consists of two parts: a formula and a control part which decides what to do with the evaluation result of that formula. Throughout this thesis, I will be using the terms steering fragment and test interchangeably.

The formula in a steering fragment can either be a primitive or a compound formula. The components of a compound formula are usually connected via logical connectives such as $\land$ and $\lor$, or involve negation. If the formula of a steering fragment is compound, we say that the steering fragment is a \emph{complex steering fragment}.

We have already seen a classical example of a (complex) steering fragment in the previous section: the \emph{if $\ldots$ then} instruction. In the example above, the formula is a compound formula with $x:=x+1$ and $x=2$ as its components, connected via the logical connective $\land$. The control part of this steering fragment consists of \emph{if} and \emph{then} and the prescription to execute $y$ if evaluation of \verb$x:=x+1 and x=2$ yields true.

\section{Related work}
\label{sec:relatedwork}
The main contribution of this thesis is to construct a formal model of side effects in dynamic logic. Because of that, I only had limited time and space to properly research related work done in this area. Despite that, I will briefly describe some references I have come across throughout this project.

Currently, a formal definition of side effects appears to be missing in literature. That is not to say that side effects have been completely ignored. Attempts have been made to create a logic which admits the possibility of side effects by Bergstra and Ponse \cite{SCL}. Furthermore, an initial, informal classification of side effects has been presented by Bergstra in \cite{SF}. I will return to those references later in this thesis.

Black and Windley have made an attempt to reason in a setting with side effects in \cite{BlackWindley1, BlackWindley2}. In their goal to verify a secure application written in C using Hoare axiomatic semantics to express the correctness of program statements, they encountered the problem of side effects occurring in the evaluation of some C-expressions. They solved the problem by creating extra inference rules which essentially separate the evaluation of the side effect from the evaluation of the main expression.

Also working with C is Norrish in \cite{Norrish}. He presents a formal semantics for C and he, too, runs into side effects in the process. Norrish claims that a semantics gives a program meaning by describing the way in which it changes a program state. Such a program state would both include the computer's memory as well as what is commonly known as the environment (types of variables, mapping of variable names to addresses in memory etc.). Norrish claims that in C, changes to the former come about through the actions of side effects, which are created by evaluating certain expression forms such as assignments. Norrish' formal semantics for C is able to handle these side effects.

B\"ohm presents a different style of axiomatic definitions for programming languages \cite{Boehm}. Whereas other authors such as Black and Windley above use Hoare axiomatic semantics which bases the logic on the notion of pre- or postcondition, B\"ohm uses the value of a programming language expression as the underlying primitive. He relies on the fact that the underlying programming language is an expression language such as Algol 68 \cite{Wijngaarden}. Expressions are allowed to have arbitrary side effects and the notions of statement and expression coincide. B\"ohm claims that his formalism is just as intuitive as Hoare-style logic and that the notion of `easy axiomatizability' --- which is a major measurement of the quality of a programming language --- is a matter of a choice of formalism, which in turn is arbitrary.

In this thesis I will develop a variant of Dynamic Logic to model side effects. Dynamic Logic is used for a wide range of applications, ranging from modelling key constructs of imperative programming to developing dynamic semantic theories for natural language. An early overview of dynamic logic is given by Harel in \cite{Harel2}. More recently, Van Eijck and Stokhof have given an extensive overview of various systems of dynamic logic in \cite{EijckStokhof}.

\section{Overview of this thesis}
Intuitively, a side effect of a propositional statement is a change in state of a program or model other than the effect (or change in state) it was initially executed for. In this thesis I will present a system that makes this intuition explicit.

First, in Chapter \ref{ch:qdl} I will present the preliminaries on which my system, that can model program instructions and their effect on program states, is based. This system, which I present in Chapter \ref{ch:modifying}, will be a modified version of Quantified Dynamic Logic, overviews of which can be found in \cite{Harel2,EijckStokhof}.

After introducing some terminology and exploring the logic behind this system in Chapters \ref{ch:terminology} and \ref{ch:logicalstructure}, I can formally define side effects, which I will do in Chapter \ref{ch:treatment}. In Chapter \ref{ch:classification} I will proceed to giving a classification of side effects, introducing marginal side effects as the most important class.

In Chapter \ref{ch:pga} I will present a case study to see this definition of side effects in action. For this I will use an --- again slightly modified --- version of Program Algebra \cite{PGA}. I will end this thesis with some conclusions and some pointers for future work.

\chapter{Preliminaries}
\label{ch:qdl}

\section{Introduction}
\label{sec:introsideeffects}
In order to say something useful about side effects, we need a formal definition. Such a definition can be found using dynamic logics. The basic idea here is that the update of a program instruction is the change in program state it causes. This allows us to introduce an expected and an actual evaluation of a program instruction. The expected evaluation of a program instruction is the change you would expect a program instruction to make to the program state upon evaluation. This may differ, however, from the actual evaluation, namely when a side effect occurs when actually evaluating the program instruction. The side effect of a program instruction then is defined as the difference in expected and actual evaluation of a program instruction.

To flesh this out in a formal definition, we first need a system that is able to model program states and program instructions. \emph{Quantified Dynamic Logic} (QDL) is such a system. QDL was developed by Harel \cite{Harel} and Goldblatt \cite{Goldblatt}. It can be seen as a first order version of \emph{Propositional Dynamic Logic} (PDL), which was developed by Pratt in \cite{Pratt76, Pratt80}. Much of the overview of both PDL and QDL I will give below is taken from the overview of dynamic logic by Van Eijck and Stokhof \cite{EijckStokhof}.

Dynamic logic can be viewed as dealing with the logic of action and the result of action \cite{EijckStokhof}. Although various kinds of actions can be modelled with it, one is of particular interest for us: the actions performed on computers, i.e. computations. In essence, these are actions that change the memory state of a machine, or on a somewhat higher level the program state of a computer program.

Regardless of what kinds of actions are modelled, the core of dynamic logic can in many cases be characterized in a similar way via the logic of `labelled transistion systems'. A labelled transition system or LTS over a signature $\langle P,A \rangle$, with $P$ a set of propositions and $A$ a set of actions, is a triple $\langle S,V,R \rangle$ where $S$ is a set of states, $V : S \rightarrow \mathcal{P}(P)$ is a valuation function and $R = \{ \overset{a}{\rightarrow} \subseteq S \times S \mid a \in A\}$ is a set of labelled transitions (one binary relation on $S$ for each label $a$).

There are various versions of dynamic logic. Before I will introduce two of these, I will first describe the setting I will be using in my examples. This setting consists of a toy programming language that is expressive enough to model the working examples I need to discuss side effects.

\section{Toy language}
\label{sec:toy}
My toy language should be able to handle assignments and steering fragments. The steering fragment can possibly be complex, so our toy language should be able to handle compound formulas: multiple formulas (such as equality tests) connected via logical connectives. In particular, I will be using short-circuit left and ($\leftand$) and short-circuit left or ($\leftor$) as connectives. Finally, assignments should be allowed in tests as well: they are, in line with what one would expect, defined to always return \verb$true$.

As toy language I will first present the WHILE language defined by Van Eijck in \cite{EijckStokhof}. We will see soon enough that we will actually need more functionality than it offers, but it will serve us well in the introduction of PDL, QDL and the illustration of the problems we will run into.

The WHILE language works on natural numbers and defines arithmetic expressions, Boolean expressions and programming commands. Arithmetic expressions $a$ with $n$ ranging over numerals and $v$ over variables from a set $\mathcal{V}$ are defined as follows:
\begin{equation*}
a ::= n \mid v \mid a_{1}+a_{2} \mid a_{1}*a_{2} \mid a_{1}\monus a_{2}
\end{equation*}
Boolean expressions are defined as:
\begin{equation*}
B ::= \top \mid a_{1}=a_{2} \mid a_{1} \leq a_{2} \mid \lnot B \mid B_{1} \lor B_{2}
\end{equation*}
Finally, we define the following programming commands:
\begin{equation*}
C ::= \text{SKIP} \mid \text{ABORT} \mid v:=a \mid C_{1};C_{2} \mid \text{IF } B \text{ THEN } C_{1} \text{ ELSE } C_{2}
\end{equation*}
For the sake of simplicity, we will postpone the introduction of the WHILE command until after we have presented our modified system in Chapter \ref{ch:modifying}.

The semantics of the arithmetic expressions are fairly self-explanatory. We assume that every numeral $n$ in $N$ has an interpretation $I(n) \in \mathbb{N}$ and let $g$ be a mapping from $\mathcal{V}$ to $\mathbb{N}$. We then have the following interpretations of the arithmetic expressions, relative to initial valuation or initial program state $g$:
\begin{align*}
\llbracket n \rrbracket_{g} &:= I(n)\\
\llbracket v \rrbracket_{g} &:= g(v)\\
\llbracket a_{1}+a_{2} \rrbracket_{g} &:= \llbracket a_{1} \rrbracket_{g} + \llbracket a_{2}\rrbracket_{g}\\
\llbracket a_{1}*a_{2} \rrbracket_{g} &:= \llbracket a_{1} \rrbracket_{g} * \llbracket a_{2}\rrbracket_{g}\\
\llbracket a_{1}\monus a_{2} \rrbracket_{g} &:= \llbracket a_{1} \rrbracket_{g} \monus \llbracket a_{2}\rrbracket_{g}
\end{align*}
The semantics of the Boolean expressions are standard as well, writing $T$ for true and $F$ for false:
\begin{align*}
\llbracket \top \rrbracket_{g} &:=\ T\\
\llbracket a_{1}=a_{2} \rrbracket_{g} &:=
\begin{cases}
T \text{ if } \llbracket a_{1}\rrbracket_{g} = \llbracket a_{2}\rrbracket_{g}\\
F \text{ otherwise}
\end{cases}\\
\llbracket a_{1}\leq a_{2} \rrbracket_{g} &:=
\begin{cases}
T \text{ if } \llbracket a_{1}\rrbracket_{g} \leq \llbracket a_{2}\rrbracket_{g}\\
F \text{ otherwise}
\end{cases}\\
\llbracket \lnot B \rrbracket_{g} &:=
\begin{cases}
T \text{ if } \llbracket B\rrbracket_{g} = F\\
F \text{ otherwise}
\end{cases}\\
\llbracket B_{1}\lor B_{2} \rrbracket_{g} &:=
\begin{cases}
T \text{ if } \llbracket B_{1}\rrbracket_{g} = T \text{ or } \llbracket B_{2}\rrbracket_{g} = T\\
F \text{ otherwise}
\end{cases}
\end{align*}
The semantics of the commands of the toy language can be given in various styles. Here I take a look at a variant called structural operational semantics \cite{EijckStokhof}. It is specified using a transition system from pairs of a state and a command, to either a state or again a state and a (new) command.

First I will give the transitions for the assignment command. It looks like this, where we write $g[v \mapsto t]$ for the valuation which is like valuation $g$ except for the variable $v$, which has been mapped to $t$:
\begin{equation*}
(g, v:=t) \Longrightarrow g[v \mapsto \llbracket t \rrbracket_{g}]
\end{equation*}
Here we have the pair of state $g$ and the assignment command $v:=a$ at the start of the transition. After the transition, we only have a new state left, since the execution of this command has finished in a single step.

The SKIP command does nothing: it does not change the state and it finishes in a single step.
\begin{equation*}
(g,\text{SKIP}) \Longrightarrow g
\end{equation*}

In structural operational semantics, there are two rules for sequential composition, one for when program $C_{1}$ finishes in a single step and one for which it does not.
\begin{equation*}
\frac{\displaystyle (g, C_{1}) \Longrightarrow g'}{\displaystyle (g,C_{1};C_{2}) \Longrightarrow (g',C_{2})}
\end{equation*}
\begin{equation*}
\frac{\displaystyle (g, C_{1}) \Longrightarrow (g',C_{1}')}{\displaystyle (g,C_{1};C_{2}) \Longrightarrow (g',C_{1}';C_{2})}
\end{equation*}

Finally, we have the rules for conditional action. There are two (similar) rules, depending on the outcome of the test:
\begin{equation*}
\frac{}{\displaystyle (g, \text{ IF } B \text{ THEN } C_{1} \text{ ELSE } C_{2}) \Longrightarrow (g,C_{1})} \llbracket B \rrbracket_{g} = T
\end{equation*}
\begin{equation*}
\frac{}{\displaystyle (g, \text{ IF } B \text{ THEN } C_{1} \text{ ELSE } C_{2}) \Longrightarrow (g,C_{2})} \llbracket B \rrbracket_{g} = F
\end{equation*}

\section{Propositional Dynamic Logic}
\label{sec:pdl}
Now that I have introduced the toy language, it is time to take a look at the first version of dynamic logic we are interested in: Propositional Dynamic Logic (PDL in short). The language of PDL consists of formulas $\phi$ (based on basic propositions $p \in P$) and programs $\alpha$ (based on basic actions $a \in A$):
\begin{align*}
\phi &::= \top \mid p \mid \lnot\phi \mid \phi_{1} \lor \phi_{2} \mid \langle\alpha\rangle\phi\\
\alpha &::= a \mid \?\phi \mid \alpha_{1};\alpha_{2} \mid \alpha_{1} \cup \alpha_{2} \mid \alpha^{*}
\end{align*}
As the name suggests, PDL is based on propositional logic. This means that the usual properties such as associativity and duality are valid and will be used throughout. Furthermore, we can use the following abbreviations:
\begin{align*}
\bot &= \lnot\top\\
\phi_{1} \land \phi_{2} &= \lnot(\lnot \phi_{1} \lor \lnot \phi_{2})\\
\phi_{1} \rightarrow \phi_{2} &= \lnot\phi_{1} \lor \phi_{2}\\
\phi_{1} \leftrightarrow \phi_{2} &= (\phi_{1} \rightarrow \phi_{2}) \land (\phi_{2} \rightarrow \phi_{1})\\
[\alpha]\phi &= \lnot\langle\alpha\rangle\lnot\phi
\end{align*}

The relational composition $R_{1} \circ R_{2}$ of binary relations $R_{1}, R_{2}$ on state set $S$ is given by:
\begin{equation*}
R_{1} \circ R_{2} = \{(t_{1},t_{2}) \in S \times S \mid \exists t_{3}((t_{1},t_{3})\in R_{1} \land (t_{3},t_{2}\in R_{2}))\}
\end{equation*}
The $n$-fold composition $R^{n}$ of a binary relation $R$ on $S$ with itself is recursively defined as follows, with $I$ the identity relation on $S$:
\begin{align*}
R^{0} &= I\\
R^{n} &= R \circ R^{n-1}
\end{align*}
Finally, the reflexive transitive closure of $R$ is given by:
\begin{equation*}
R^{*} = \bigcup_{n\in\mathbb{N}} R^{n}
\end{equation*}
To define the semantics of PDL over basic propositions $P$ and basic actions $A$, we need the labelled transistion system $T = \langle S_{T},V_{T},R_{T}\rangle$ for signature $\langle P,A\rangle$. The formulas of PDL are interpreted as subsets of $S_{T}$, the actions as binary relations on $S_{T}$. This leads to the following interpretations:
\begin{align*}
\llbracket \top \rrbracket^{T} &= S_{T}\\
\llbracket p \rrbracket^{T} &= \{s \in S_{T} \mid p \in V_{T}(s)\}\\
\llbracket \lnot \phi \rrbracket^{T} &= S_{T} - \llbracket \phi \rrbracket^{T}\\
\llbracket \phi_{1} \lor \phi_{2} \rrbracket^{T} &= \llbracket \phi_{1} \rrbracket^{T} \cup \llbracket \phi_{2} \rrbracket^{T}\\
\llbracket \langle\alpha\rangle\phi \rrbracket^{T} &= \{s \in S_{T} \mid \exists t (s,t) \in \llbracket \alpha \rrbracket^{T} \text{ and } t \in \llbracket \phi \rrbracket^{T}\}\\
\\
\llbracket a \rrbracket^{T} &=\ \xrightarrow{a}_{T}\\
\llbracket \?\phi \rrbracket^{T} &= \{(s,s)\in S_{T} \times S_{T} \mid s \in \llbracket \phi \rrbracket^{T}\}\\
\llbracket \alpha_{1};\alpha_{2} \rrbracket^{T} &= \llbracket \alpha_{1} \rrbracket^{T} \circ \llbracket \alpha_{2} \rrbracket^{T}\\
\llbracket \alpha_{1} \cup \alpha_{2} \rrbracket^{T} &= \llbracket \alpha_{1} \rrbracket^{T} \cup \llbracket \alpha_{2} \rrbracket^{T}\\
\llbracket \alpha^{*} \rrbracket^{T} &= (\llbracket \alpha \rrbracket^{T})^{*}
\end{align*}
The programming constructs in our toy language are expressed in PDL as follows:
\begin{align*}
\text{SKIP} &:= \?\top\\
\text{ABORT} &:= \?\bot\\
\text{IF } \phi \text{ THEN } \alpha_{1} \text{ ELSE } \alpha_{2} &:= (\?\phi;\alpha_{1}) \cup (\?\lnot\phi; \alpha_{2})
\end{align*}

Although PDL is a powerful logic, it is not enough yet to properly model the toy language we need. The reason for that is the need for assignments. Since assignments change relational structures, the appropriate assertion language is first order predicate logic, and not propositional logic \cite{EijckStokhof}. So instead of PDL, which as the name suggests uses propositional logic, we need a version of dynamic logic that uses first order predicate logic. This is where Quantified Dynamic Logic (QDL in short) comes in.

\section{Quantified Dynamic Logic}
\label{sec:qdl}
The language of QDL consists of terms $t$, formulas $\phi$ and programs $\pi$. For functions $f$ and relational symbols $R$ we have:
\begin{gather*}
t ::= v \mid ft_{1}\ldots t_{n}\\
\phi ::= \top \mid Rt_{1}\ldots t_{n} \mid t_{1}=t_{2} \mid \lnot \phi \mid \phi_{1} \lor \phi_{2} \mid \exists v\phi \mid  \langle \pi \rangle \phi\\
\pi ::= v:=\? \mid v:=t \mid \?\phi \mid \pi_{1};\pi_{2} \mid \pi_{1} \cup \pi_{2} \mid \pi^{*}
\end{gather*}
In the case of natural numbers, examples of $f$ are $+,*$ etc. and examples of $R$ are $\leq$ and $\geq$. The same abbreviations as in PDL are used, most notably $\bot = \lnot \top$ and $[\pi]\phi = \lnot\langle\pi\rangle\lnot\phi$.

The random assignment ($v:=\?$) does not increase the expressive power of QDL \cite{EijckStokhof}. It can, however, be nicely used to express the universal and existential quantifier:
\begin{align*}
\exists v\phi &\leftrightarrow \langle v:=\?\rangle\phi\\
\forall v\phi &\leftrightarrow [v:=\?]\phi
\end{align*}

The pair $(f,R)$ is called a first order signature. A model for such a signature is a structure of the form
\begin{equation*}
M = (E^{M},f^{M},R^{M})
\end{equation*}
where $E$ is a non-empty set, the $f^{M}$ are interpretations in $E$ for the members of $f$ and the $R^{M}$ similarly are the interpretations in $E$ for the members of $R$. Now let $\mathcal{V}$ be the set of variables of the language. Interpretation of terms in $M$ is defined relative to an initial valuation $g:\mathcal{V}\rightarrow E^{M}$:
\begin{gather*}
\llbracket v \rrbracket_{g}^{M} = g(v)\tag{QDL1}\\
\llbracket ft_{1}\ldots t_{n} \rrbracket_{g}^{M} = f^{M}(\llbracket t_{1}\rrbracket_{g}^{M},\ldots,\llbracket t_{n}\rrbracket_{g}^{M})\tag{QDL2}
\end{gather*}
Truth in $M$ for formulas is defined by simultaneous recursion, where $g \sim_{v} h$ then means that $h$ differs at most from $g$ on the assignment it gives to variable $v$:
\begin{align*}
M \models_g \top & \  \text{always}\tag{QDL3}\\
M \models_g Rt_{1}\ldots t_{n} &\text{ iff } (\llbracket t_{1}\rrbracket_{g}^{M},\ldots,\llbracket t_{n}\rrbracket_{g}^{M})\in R^{M}\tag{QDL4}\\
M \models_g t_{1}=t_{2} &\text{ iff } \llbracket t_{1}\rrbracket_{g}^{M} = \llbracket t_{2}\rrbracket_{g}^{M}\tag{QDL5}\\
M \models_g \lnot \phi &\text{ iff } M \not\models_{g} \phi\tag{QDL6}\\
M \models_g \phi_{1} \lor \phi_{2} &\text{ iff } M \models_{g} \phi_{1} \text{ or } M \models_{g} \phi_{2}\tag{QDL7}\\
M \models_g \exists v\phi &\text{ iff } \text{for some } h \text{ with } g \sim_v h, M \models_h \phi\tag{QDL8}\\
M \models_g \langle \pi \rangle \phi &\text{ iff } \text{for some }h \text{ with }_{g}\llbracket \pi \rrbracket_{h}^{M},M\models_{h} \phi\tag{QDL9}
\end{align*}
The same goes for the relational meaning in $M$ for programs:
\begin{align*}
_{g}\llbracket v:=t \rrbracket_{h}^{M} \text{ iff } & h = g[v \mapsto \llbracket t \rrbracket_{g}^{M}]\tag{QDL10}\\
_{g}\llbracket \?\phi \rrbracket_{h}^{M} \text{ iff } & g=h \text{ and } M \models_{g} \phi\tag{QDL11}\\
_{g}\llbracket \pi_{1};\pi_{2} \rrbracket_{h}^{M} \text{ iff } & \exists f\text{ with } _{g}\llbracket \pi_{1}\rrbracket_{f}^{M} \text{ and } _{f}\llbracket \pi_{2}\rrbracket_{h}^{M} \tag{QDL12}\\
_{g}\llbracket \pi_{1} \cup \pi_{2} \rrbracket_{h}^{M} \text{ iff } & _{g}\llbracket \pi_{1}\rrbracket_{h}^{M} \text{ or } _{g}\llbracket \pi_{2}\rrbracket_{h}^{M}\tag{QDL13}\\
_{g}\llbracket \pi^{*} \rrbracket_{h}^{M} \text{ iff } & g = h \text{ or } _{g}\llbracket \pi;\pi^{*} \rrbracket_{h}^{M}\tag{QDL14}
\end{align*}
The above definition makes concatenation ($;$) an associative operator:
\begin{equation*}
(\pi_{1};\pi_{2});\pi_{3} = \pi_{1};(\pi_{2};\pi_{3})
\end{equation*}
As a convention, we omit the brackets wherever possible.

Although QDL goes a long way to modelling our toy language and program states, we are not quite there yet. The modifications we have to make come to light when we examine the expressive power of QDL. QDL currently has more expressive power than it has semantics defined for. This problem surfaces when the modality operator is nested within a test, like this:
\begin{equation*}
\?(\langle v:=t \rangle \top)
\end{equation*}
This is the program $\?\phi$, with $\phi = \langle \pi \rangle \psi$, $\pi = v:=t$ and $\psi=\top$. As the semantics of QDL are currently defined, the program $\pi$ will make a change to an initial valuation $g$ if it is interpreted in it, returning valuation $h$ where the assignment $g$ had for variable $v$ will be expressed by $t$. This is expressed by QDL10. However, the current semantics only assign relational meaning to a test instruction $\?\phi$ as long as $g = h$, as expressed by QDL11.

Another similar example is the following:
\begin{equation*}
\?(\langle v:=v+1;v:=v\monus 1 \rangle \top)
\end{equation*}
Although this situation should be similar as above, it is not: because the program state gets changed twice, QDL now \emph{is} able to assign semantics to this program since the program state gets returned to the original state by the second program instruction (and we therefore have $g=h$).

So, not only can we devise even a very simple correct QDL-program for which there are no semantics defined, we can also give a very similar example for which QDL does define semantics. Not only does that somewhat erratic behavior seem undesirable, but the nature of the examples here present us with a problem when we are considering side effects. Exactly for the situations in which side effects occur, namely when an instruction in a test causes a change in the program state, there are no semantics defined in QDL. Therefore, I am going to have to modify QDL so that it does define semantics in those situations.

\chapter{Modifying QDL to \dlaf}
\label{ch:modifying}

\section{Introducing \dlaf}
In this chapter I will present \emph{Dynamic Logic with Assignments as Formulas}, or \dlaf\ in short, the resulting dynamic logic after making two major modifications to QDL. The modifications I will make are such that \dlaf\ can model the specific kinds of constructions that we are interested in. This means that, like the name suggests, we have to introduce semantics for assignments in formulas. Furthermore, we will drop or modify some other QDL-instructions that we do not need. Because of that \dlaf\ evades the problem of QDL mentioned in Section \ref{sec:qdl} of the previous chapter and one other problem I will get back to in Section \ref{sec:while}. Before I introduce \dlaf, however, I will show the modifications that need to be done to Van Eijck's WHILE language so that it can model the instructions we need.

In the WHILE language, Boolean expressions are assumed to cause no state change upon evaluation. However, for our purpose this is inadequate. We want to allow assignments in tests as well and they cause a state change. This warrants the first modification to the WHILE language and its semantics: assignments are allowed in Boolean expressions. The second modification is that the Boolean OR function will be replaced by a short-circuit version:
\begin{equation*}
B ::= \top \mid a_{1}=a_{2} \mid a_{1} \leq a_{2} \mid \lnot B \mid B_{1} \leftor B_{2} \mid v:=a
\end{equation*}
The new semantics for Boolean expressions are like the semantics defined by Van Eijck, with as major difference that there are now semantics defined for assignments:
\begin{equation*}
\llbracket v:=a \rrbracket_{g} := T
\end{equation*}
Furthermore, Boolean expressions now might introduce a state change, so every command containing a Boolean expression (which for now only is the IF THEN ELSE command) should account for that. In structural operational semantics, we take a look at how the Boolean expression changes the state and perform the remaining actions in that new state:
\begin{equation*}
\frac{\displaystyle (g,B) \Longrightarrow g'}{\displaystyle (g,\text{IF }B\text{ THEN }C_{1}\text{ ELSE }C_{2})\Longrightarrow (g',C_{1})}\llbracket B \rrbracket_{g} = T
\end{equation*}
And similar for the case that $\llbracket B \rrbracket_{g} = F$.

As said, there is one more thing that needs to be modified in the language above. In order to be properly able to reason about side effects, the order in which the tests get executed is important. Because of that, the OR construct in Boolean expressions needs to be replaced by a short-circuit directed version:
\begin{equation*}
\llbracket B_{1} \leftor B_{2} \rrbracket_{g} :=
\begin{cases}
T \text{ if } \llbracket B_{1} \rrbracket_{g} = T\\
T \text{ if } \llbracket B_{1} \rrbracket_{g} = F \text{ and for } (g,B) \Longrightarrow g', \llbracket B_{2} \rrbracket_{g'} = T\\
F \text{ otherwise}
\end{cases}
\end{equation*}
We will make use of its dual, the short-circuit left and ($\leftand$) too. It is defined similarly as above. As a convention, from here on $\leftor$ and $\leftand$ can be used interchangeably in definitions, unless explicitly stated otherwise. Both $\leftor$ as well as $\leftand$ are associative. We again omit brackets wherever possible.

All we have left to define now is the state change a Boolean can cause. This is defined as follows:
\begin{align*}
(g,B) \Longrightarrow
\begin{cases}
g[v \mapsto \llbracket t\rrbracket_g] &\text{if } B = (v:=t)\\
g &\text{o.w.}
\end{cases}
\end{align*}

Missing in the above WHILE language are the random assignment and the existential quantifier. This is because I have decided to drop them. The reason for that is that they can cause non-deterministic behavior and in this thesis, we are not interested in the (side effects of) non-deterministic programs. In fact it is questionable whether we can say anything about side effects in non-deterministic programs, but I will return to that in my possibilities for future work in Chapter \ref{ch:conclusions}. Aside from that, in our context of (imperative) programs, the random assignment is an unusual concept at best. The same goes for the formula $\exists v \phi$.

With those modifications to the toy language in mind, we can take a look at the similar modifications that need to be made to QDL. In the resulting dynamic logic \dlaf, we keep the same terms:
\begin{equation*}
t ::= v \mid ft_{1}\ldots t_{n}
\end{equation*}

In \dlaf\ we of course drop the random assignment and existential quantifier, too. By dropping them, we lose the quantified character of QDL. Because of that, the resulting logic is no longer called a quantified dynamic logic. The first major change to QDL, besides the absence of the random assignment and the existential quantifier, is that I replace the $\langle \pi \rangle\phi$ command with the weaker $[ v:=t ] \top$:
\begin{equation*}
\phi ::= \top \mid Rt_{1}\ldots t_{n} \mid t_{1}=t_{2} \mid \lnot \phi \mid \phi_{1} \leftor \phi_{2} \mid \phi_1 \leftand \phi_2 \mid [ v:=t ] \top
\end{equation*}
This modification explicitly expresses the possibility of assignments in formulas. All other programs, however, are no longer allowed in formulas. Because of this modification we will avoid a number of problems that QDL has, while keeping the desired functionality that there should be room for assignments in formulas. I will address these problems in detail in Section \ref{sec:while}.

We have also replaced the $\lor$ connective with its short-circuit variant ($\leftor$) and for convenience, have explicitly introduced its dual ($\leftand$). We will return to the motivation for this change at the end of this chapter.

We also need to replace the QDL-formula associated with this command (QDL9). The truth in $M$ for the new command is defined as follows:
\begin{equation*}
M \models_{g} [ v:=t ] \top \text{ always} \tag{DLA9}
\end{equation*}
It should come as no surprise that this always succeeds, since assignments always succeed and yield \verb$true$. Since this formula always succeeds, we replaced the possibility modality ($\langle v:=t \rangle\top$) for the necessity modality ($[v:=t]\top$). The reason we keep this formula in the form of a modality at all (and not just $v:=t$), is because formulas of this form can change the initial valuation. This is in sharp contrast to the basic formulas $t_{1} = t_{2}$ and $Rt_{1}\ldots t_{2}$, which do not change the initial valuation and are typically not modalities. Because of that, it is unintuitive to write the assignment formula as $v:=t$.

On a side note: in our toy language we \emph{do} simply write $v:=t$ for the assignment, regardless of where it occurs. This is because in the world of (imperative) programming, assignments are allowed in steering fragments.

We will see below that we are going to accept possible state changes in formulas, in contrast to the original QDL versions. For this we will use a mechanism to determine when a state change happens, that is, a function that returns the program(s) that are encountered when evaluating a formula $\phi$. This function is defined as follows:
\begin{definition}
The \textbf{program extraction function} $\Pi_{g}^{M} : \phi \rightarrow \pi$ returns for formula $\phi$ the program(s) that are encountered when evaluating the formula given modal $M$ and initial valuation $g$. It is defined recursively as follows:
\begin{align*}
\Pi_{g}^{M}(\top) &= \iq\top\\
\Pi_{g}^{M}(Rt_{1}\ldots t_{n}) &= \iq\top\\
\Pi_{g}^{M}(t_{1}=t_{2}) &= \iq\top\\
\Pi_{g}^{M}(\lnot \phi) &= \Pi_{g}^{M}(\phi)\\
\Pi_{g}^{M}(\phi_{1} \leftor \phi_{2}) &= 
\begin{cases}
\Pi_{g}^{M}(\phi_{1}) & \text{if } M \models_{g} \phi_{1}\\
\Pi_{g}^{M}(\phi_{1});\Pi_{h}^{M}(\phi_2) & \text{if } M \not\models_g \phi_1 \text{ and }_{g}\llbracket \Pi_g^M(\phi_{1}) \rrbracket_{h}^{M}
\end{cases}\\
\Pi_{g}^{M}(\phi_{1} \leftand \phi_{2}) &= 
\begin{cases}
\Pi_{g}^{M}(\phi_{1}) & \text{if } M \not\models_{g} \phi_{1}\\
\Pi_{g}^{M}(\phi_{1});\Pi_{h}^{M}(\phi_2) & \text{if } M \models_g \phi_1 \text{ and }_{g}\llbracket \Pi_g^M(\phi_{1}) \rrbracket_{h}^{M}
\end{cases}\\
\Pi_{g}^{M}([ v:=t ] \top) &= (v:=t)
\end{align*}
\end{definition}
In the first three cases, no programs are encountered. Therefore, the program extraction function returns the empty program ($\?\top$). The formula $\lnot\phi$ is transparent, that is, it returns any program encountered in its subformula $\phi$. Because of the short-circuit character of $\leftor$ and $\leftand$, a case distinction is made here: in case of $\leftor$, $\phi_{2}$ will not be evaluated if $\phi_{1}$ yields true, therefore only the program(s) encountered in $\phi_{1}$ will be returned. Otherwise, the result is a concatenation of the program(s) encountered in $\phi_{1}$ and $\phi_{2}$. Obviously, for $\leftand$ the opposite is the case and this clause is derivable from the previous one using duality. Finally, if the formula is an assignment, the program equivalent of that assignment is returned.

Because the evaluation of a formula now can cause a state change, the original definition for the truth in $M$ of $\leftor$ (QDL7) is no longer valid. In case $\phi_{1}$ contains an assignment, $\phi_{2}$ must be evaluated in a different valuation, namely the one resulting after evaluating $\phi_{1}$ in the initial valuation:
\begin{equation*}
M \models_{g} \phi_{1} \leftor \phi_{2} \text{ iff for }_{g}\llbracket \Pi_{g}^{M}(\phi_{1}) \rrbracket_{h}^{M}, M \models_{g} \phi_{1} \text{ or } M \models_{h} \phi_{2} \tag{DLA7a}
\end{equation*}
Since we have added $\leftand$ to formulas as well, we also explicitly have to define the truth in $M$ for $\leftand$, which is similar to the updated definition of $\leftor$:
\begin{equation*}
M \models_{g} \phi_{1} \leftand \phi_{2} \text{ iff for } _{g}\llbracket \Pi_{g}^{M}(\phi_{1}) \rrbracket_{h}^{M}, M \models_{g} \phi_{1} \text{ and } M \models_{h} \phi_{2} \tag{DLA7b}
\end{equation*}
Although $\leftor$ and $\leftand$ use short-circuit evaluation, we do not explicitly have to define them as such above because we will make sure, via the program extraction function and an updated version of QDL11 (see below), that the valuation does not change as a result of $\phi_{2}$ when $M \models_{g} \phi_{1}$ is true (in case of $\leftor$) or false (in case of $\leftand$).

We can now turn our attention to programs in \dlaf. Besides the absence of the random assignment, what a program $\pi$ can be does not change:
\begin{equation*}
\pi ::= v:=t \mid \?\phi \mid \pi_{1};\pi_{2} \mid \pi_{1} \cup \pi_{2} \mid \pi^{*}
\end{equation*}
To remedy the problem that more things can be expressed in QDL than there are semantics for, we need, as mentioned earlier, to accept that a state change can occur when evaluating a program containing formulas. In the case of QDL, that only is the test instruction, given semantics earlier in QDL11. So, as second major change we need to replace QDL11 by:
\begin{equation*}
_{g}\llbracket \?\phi \rrbracket_{h}^{M} \text{ iff }
\begin{cases}
M \models_{g} \phi \text{ and } g = h & \text{if }\Pi_{g}^{M}(\phi)=\?\top\\
M \models_{g} \phi \text{ and } _{g}\llbracket\Pi_{g}^{M}(\phi)\rrbracket_{h} & \text{otherwise}
\end{cases}\tag{DLA11}
\end{equation*}
The choice here is in place to avoid looping behavior when evaluating $_g\llbracket \?\top \rrbracket_h$.

The definitions above make extensive use of the empty program ($\?\top$). In what follows, it will be handy to know that the empty program is truly empty. In particular, we would like to have $\pi;\?\top = \pi$ and $\?\top;\pi = \pi$. I will prove that below.

\begin{lemma}
\label{emptyprogram}
For any program $\pi$, initial valuation $g$, output valuation $h$ and model $M$
\begin{equation*}
_{g}\llbracket\pi;\iq\top\rrbracket_{h}^{M} \text{ iff } _{g}\llbracket\pi\rrbracket_{h}^{M}
\end{equation*}
\end{lemma}
\begin{proof}
The proof follows from the above defined QDL-axioms:
\begin{equation*}
_{g}\llbracket\pi;\?\top\rrbracket_{h}^{M} \text{ iff } \exists f \  _{g}\llbracket\pi\rrbracket_{f}^{M} \text{ and } _{f}\llbracket \?\top\rrbracket_{h}
\end{equation*}
Since we have $_{f}\llbracket \?\top\rrbracket_{h}$ iff $f=h$ and $M \models_{f} \top$, and since the latter is always true, we have
\begin{equation*}
_{g}\llbracket\pi;\?\top\rrbracket_{h}^{M} \text{ iff } _{g}\llbracket\pi\rrbracket_{h}^{M}
\end{equation*}
\end{proof}
\begin{lemma}
For any program $\pi$, initial valuation $g$, output valuation $h$ and model $M$
\begin{equation*}
_{g}\llbracket\iq\top;\pi\rrbracket_{h}^{M} \text{ iff } _{g}\llbracket\pi\rrbracket_{h}^{M}
\end{equation*}
\end{lemma}
\begin{proof}
Similar as for Lemma \ref{emptyprogram}.
\end{proof}

The change to QDL11 has remedied the problem that there are expressions in QDL for which there are no semantics defined. Of course I made a second major change --- namely replacing $\langle \pi \rangle \phi$ by $[ v:=t ] \top$. The reason for that will come to light as soon as I will reintroduce the WHILE command in Section \ref{sec:while}. Before I will do that, however, I will first discuss a working example to provide some more insight into the inner workings of \dlaf.

\section{A working example}
\label{sec:example}

In this section I will present a working example to illustrate how \dlaf\ works. I will use the following program, presented here in our toy language:
\begin{flalign*}
&x := 1;&\\
&\text{IF }(x:=x+1 \leftand x = 2)&\\
&\text{THEN }y:=1&\\
&\text{ELSE } y:=2&
\end{flalign*}
In \dlaf, this translates to:
\begin{flalign*}
&x:=1;&\\
&(\?([ x:=x+1]\top \leftand x=2);y:=1)&\\
&\cup&\\
&(\?\lnot([ x:=x+1]\top \leftand x=2);y:=2)&
\end{flalign*}
The valuations $g,h,\ldots$ are defined for all variables $v \in \mathcal{V}$, i.e. they are total functions. Usually we are only interested in a small number of variables, e.g. $x$ and $y$, in which case we talk about a valuation $g$ such that $g(x) = \llbracket t\rrbracket_{g}^{M},g(y)=\llbracket t'\rrbracket_{g}^{M}$, or if valuation $h$ is an update of valuation $g$, $h = g[x \mapsto \llbracket t\rrbracket_{g}^{M}, y \mapsto \llbracket t'\rrbracket_{g}^{M}]$ (which is a shorthand for $g[x \mapsto \llbracket t\rrbracket_{g}^{M}][y \mapsto \llbracket t'\rrbracket_{g}^{M}]$). In all examples we discuss we take for $M$ the model of the natural numbers and we use numerals to denote its elements.

Since we are working on natural numbers, as constants we have $n$ ranging over numerals, as functions we have $+, *$ and $\monus$, and as extra relation we have $\leq$. Our model $M$ contains those constants, functions and relations. Assume we have an initial valuation $g$ that sets $x$ and $y$ to $0$: $g(x) = g(y) = 0$. We will now first show how the program in our toy language gets evaluated using the structural operational semantics we provided in Chapter \ref{ch:qdl}:
\begin{align*}
&(g,\big(x := 1;\text{IF }(x:=x+1 \leftand x = 2)\text{ THEN }y:=1\text{ ELSE } y:=2\big)) \Longrightarrow\\
&(g[x \mapsto 1],\big(\text{IF }(x:=x+1 \leftand x = 2)\text{ THEN }y:=1\text{ ELSE } y:=2\big))
\end{align*}
We now need to know if $\llbracket (x:=x+1 \leftand x=2)\rrbracket_{g[x \mapsto 1]} = T$. We can easily see that it is and furthermore updates the valuation again by incrementing $x$ by $1$. Thus we get as valuation $g[x \mapsto 2]$ and we can finish our evaluation as follows:
\begin{align*}
(g[x \mapsto 2],\big(y:=1\big)) \Longrightarrow g[x \mapsto 2, y \mapsto 1]
\end{align*}
Having seen how our example program evaluates using the semantics for our toy language, we can turn our attention to the evaluation using \dlaf. We need to ask ourselves if $_{g}\llbracket \pi \rrbracket_{h}^{M}$ exists (with $\pi$ the program above), that is, if there is a valuation $h$ that models the state of the program after being executed on initial valuation $g$.

Schematically, $\pi$ can be broken down as follows:
\begin{flalign*}
\pi &::= \pi_{0};\pi_{1}&\\
\pi_{0} &::= x:=1&\\
\pi_{1} &::= (\?\phi_{0}; \pi_{2}) \cup (\?\lnot \phi_{0}; \pi_{3})&\\
\pi_{2} &::= y:=1&\\
\pi_{3} &::= y:=2&\\
\phi_{0} &::= \phi_{1} \leftand \phi_{2}&\\
\phi_{1} &::= [ x:=x+1] \top&\\
\phi_{2} &::= x=2&
\end{flalign*}

The break-down above paves the way to evaluate $_{g}\llbracket \pi \rrbracket_{h}^{M}$ using the \dlaf-axioms given in the previous sections. We start by applying QDL12:
\begin{align*}
_{g}\llbracket\pi\rrbracket_{h}^{M} = \ & _{g}\llbracket\pi_{0};\pi_{1}\rrbracket_{h}^{M}\\
\text{iff }& \exists f \text{ s.th. }_{g}\llbracket \pi_{0}\rrbracket_{f}^{M} \text{ and } _{f}\llbracket \pi_{1}\rrbracket_{h}^{M}
\end{align*}
We find $f$ by evaluating $_{g}\llbracket x:=1\rrbracket_{f}^{M}$ using QDL10 and QDL1:
\begin{align*}
_{g}\llbracket x:=1\rrbracket_{f}^{M} \text{ iff } f &= g[x \mapsto \llbracket 1 \rrbracket_{g}^{M}]\\
&= g[x \mapsto 1]
\end{align*}
Now we need to evaluate $_{f}\llbracket(\?\phi_{0}; \pi_{2}) \cup (\?\lnot \phi_{0}; \pi_{3})\rrbracket_{h}^{M}$. Using QDL13, we get:
\begin{equation*}
_{f}\llbracket(\?\phi_{0}; \pi_{2}) \cup (\?\lnot \phi_{0}; \pi_{3})\rrbracket_{h}^{M} \text{ iff } _{f}\llbracket\?\phi_{0}; \pi_{2}\rrbracket_{h}^{M} \text{ or } _{f}\llbracket\?\lnot\phi_{0}; \pi_{3}\rrbracket_{h}^{M}
\end{equation*}
First we turn our attention to $_{f}\llbracket\?\phi_{0}; \pi_{2}\rrbracket_{h}^{M}$. Using QDL12 again we get $\exists d$ such that $_{f}\llbracket \?\phi_{0}\rrbracket_{d}^{M}$ and $_{d}\llbracket \pi_{2}\rrbracket_{h}^{M}$. To evaluate the former, we need to use our own rule DLA11. Here we need the program extraction function $\Pi$ for the first time:
\begin{align*}
_{f}\llbracket \?\phi_{0}\rrbracket_{d}^{M} &= \ _{f}\llbracket \?([ x:=x+1]\top \leftand (x=2))\rrbracket_{d}^{M}\\
&\text{iff } M \models_{f} [ x:=x+1]\top \leftand (x=2)\\
&\text{and } _{f}\llbracket \Pi_{f}^{M}([ x:=x+1]\top \leftand (x=2)) \rrbracket_{d}^{M}
\end{align*}
We will first have a look at the program extraction function $\Pi$. Below we will see how it calculates the programs that are encountered while evaluating the formula $(x:=x+1) \leftand (x=2)$:
\begin{align*}
\Pi_{f}^{M}([ x:=x+1]\top \leftand (x=2)) &= \Pi_{f}^{M}([ x:=x+1]\top);\Pi_{f}^{M}(x=2)\\
&= (x:=x+1);\?\top
\end{align*}
Therefore, we have:
\begin{align*}
_{f}\llbracket \?\phi_{0}\rrbracket_{d}^{M} &= \ _{f}\llbracket \?([ x:=x+1]\top \leftand (x=2))\rrbracket_{d}^{M}\\
&\text{iff } M \models_{f} [ x:=x+1]\top \leftand (x=2)\\
&\text{and } _{f}\llbracket x:=x+1;\?\top \rrbracket_{d}^{M} \text{ iff }  _{f}\llbracket x:=x+1 \rrbracket_{d}^{M}
\end{align*}
The first of these two, $M \models_{f} [ x:=x+1]\top \leftand (x=2)$, nicely shows why we need an updated version of $\leftand$ and $\leftor$. As we already noticed the test $\phi_{0}$ contains a program (the assignment $x:=x+1$) and therefore the state (valuation) changes. As we will see, this will change the outcome of the second part of the test. We need DLA7b and our program extraction function $\Pi$ here:
\begin{align*}
M \models_{f} (x:=x+1) \leftand (x=2) \text{ iff for } _{f}\llbracket x:=x+1 \rrbracket_{c}^{M}, &M \models_{f} (x:=x+1) \text{ and }\\
&M \models_{c} (x=2)
\end{align*}
$M \models_{f} (x:=x+1)$ is defined by DLA8 to be always true. Applying QDL10 on $_{f}\llbracket x:=x+1 \rrbracket_{c}^{M}$ will give us $c = f[x \mapsto 2]$. We can then apply QDL5 on $M \models_{c} (x=2)$:
\begin{equation*}
M \models_{c} (x=2) \text{ iff } \llbracket x \rrbracket_{c}^{M} = \llbracket 2 \rrbracket_{c}^{M}
\end{equation*}
We can easily see (using QDL1) that $\llbracket x \rrbracket_{c}^{M} = c(x) = 2 = \llbracket 2 \rrbracket_{c}^{M}$. Therefore, we have $M \models_{c} (x=2)$ and thus $M \models_{f} [ x:=x+1]\top \leftand (x=2)$.

We now need to finish the evaluation of DLA11 by evaluating $_{f}\llbracket x:=x+1 \rrbracket_{d}^{M}$. This can again be done using QDL10 and gives us $d = f[x \mapsto 2]$. Because the test $\phi_{0}$ has now succeeded, we can continue to the evaluation of $_{d}\llbracket \pi_{2}\rrbracket_{h}^{M} = \ _{d}\llbracket y:=1\rrbracket_{h}^{M}$. This will give us $h = d[y \mapsto 1]$. Having already established that $\?\phi_{0}$ succeeds, we also know that $\?\lnot\phi$ will not succeed. Therefore, we are done with the evaluation of this program $\pi$, getting that $_{g}\llbracket \pi \rrbracket_{h}^{M}$ with $g(x)=g(y)=0$ is indeed possible with $h = g[x \mapsto 2, y\mapsto 1]$.

\section{Re-introducing WHILE}
\label{sec:while}
In Section \ref{sec:toy} I introduced our toy language, which was like Van Eijck's WHILE language, but without a WHILE (or: guarded iteration) programming command. Now that we have seen \dlaf\ in action in our simplified toy language, it is time to re-introduce the WHILE command. After doing that, we will see that the re-introduction of WHILE raises some more issues that warrant the second modification I made to QDL, namely replacing the formula $\langle \pi \rangle \phi$ with $[v:=t]\top$.

\subsection{The WHILE command}
The WHILE command takes the form WHILE $B$ DO $C$. The complete list of programming commands in our toy language then is:
\begin{align*}
C ::= &\text{ SKIP} \mid \text{ABORT} \mid v:=a \mid C_{1};C_{2} \mid \text{IF } B \text{ THEN } C_{1} \text{ ELSE } C_{2} \mid\\
&\text{ WHILE } B \text{ DO } C
\end{align*}
In structural operational semantics, the semantics for the guarded iteration are as follows. There are two options: if the guard ($B$) is not satisfied, command $C$ is not executed. Instead, the command finishes, with as only (possible) change the change that the evaluation of guard $B$ has made to the state:
\begin{equation*}
\frac{\displaystyle (g,B) \Longrightarrow g'}{\displaystyle (g,\text{WHILE }B\text{ DO }C)\Longrightarrow g'}\llbracket B \rrbracket_{g} = F
\end{equation*}
If the guard \emph{is} satisfied, the rule becomes a little more complicated because command $C$ gets executed in a state which is possibly changed by guard $B$. Like before, we have two cases: one for which $C$ finishes in a single step and one for which it does not.
\begin{equation*}
\frac{\displaystyle (g,B) \Longrightarrow g' \ \ \ \ \ (g',C) \Longrightarrow g''}{\displaystyle (g,\text{WHILE }B\text{ DO }C)\Longrightarrow (g'',\text{WHILE }B \text{ DO }C)}\llbracket B \rrbracket_{g} = T
\end{equation*}
\begin{equation*}
\frac{\displaystyle (g,B) \Longrightarrow g' \ \ \ \ \ (g',C) \Longrightarrow (g'',C')}{\displaystyle (g,\text{WHILE }B\text{ DO }C)\Longrightarrow (g'',C';\text{WHILE }B \text{ DO }C)}\llbracket B \rrbracket_{g} = T
\end{equation*}



\subsection{WHILE in \dlaf}
\label{sec:whileinqdla}

In PDL, and therefore QDL and \dlaf, WHILE is expressed as follows:
\begin{equation*}
\text{WHILE } \phi \text{ DO } \alpha := (\?\phi;\alpha)^{*};\?\lnot\phi
\end{equation*}
Thanks to the updated rule for $\?\phi$ (DLA11), \dlaf\ is able to handle programs with WHILE perfectly. To see how this works, consider the following example:
\begin{flalign*}
&x := 0;&\\
&y := 0;&\\
&\text{WHILE }(x:=x+1 \leftand x \leq 2)&\\
&\text{DO }y:= y+ 1&
\end{flalign*}
In \dlaf, this translates to:
\begin{flalign*}
&x:=0;&\\
&y:=0;&\\
&(\?([ x:=x+1]\top \leftand x\leq 2);y:=y+1)^{*};&\\
&\?\lnot([ x:=x+1]\top \leftand x\leq 2)&
\end{flalign*}
After the first two commands, we have $g(x) = g(y)=0$. We now need to look at how the $^{*}$ operator is evaluated. QDL14 states that $_{g}\llbracket\pi^{*}\rrbracket_{h}^{M}$ iff $g=h$ or $_{g}\llbracket\pi;\pi^{*}\rrbracket_{h}^{M})$. This means that $\pi$ is either executed not at all (in which case $g=h$) or at least once. In our case, $\pi = \?([ x:=x+1]\top \leftand x\leq 2);y:=y+1$. 

The first option is that $\pi$ is executed not at all, in which case $g=h$. However, under this valuation $h$ there is no possible valuation $h'$ after evaluation of the next program command ($\?\lnot([ x:=x+1]\top \leftand x\leq 2)$). In other words, $_{h}\llbracket\?\lnot([ x:=x+1]\top \leftand x\leq 2)\rrbracket_{h'}^{M}$ is false. Therefore, we have to turn our attention to the other option given by the $^{*}$ command, which is $_{g}\llbracket\pi;\pi^{*}\rrbracket_{h}^{M}$. For the evaluation of this we first need QDL12, which tells us that there has to be an $f$ such that $_{g}\llbracket \pi \rrbracket_{f}^{M}$ and $_{f}\llbracket\pi^{*}\rrbracket_{h}^{M}$. In Section \ref{sec:example} we have already seen how $_{g}\llbracket\pi\rrbracket_{f}^{M}$ evaluates; it will succeed and result in a new valuation $f = g[x \mapsto 1, y \mapsto 1]$.

Now we need to evaluate $\pi^{*}$ again, but this time with a different initial valuation (namely $f$). This loop continues until we arrive at a valuation $f'$ for which the final program command (the test $\?\lnot([ x:=x+1]\top \leftand x\leq 2)$) \emph{will} succeed. In our example, this happens in the second iteration, when we have $f' = g[x \mapsto 2, y \mapsto 2]$, giving us a resulting valuation $h = g[x \mapsto 3, y \mapsto 2]$, which is exactly what we would expect given this WHILE loop.

\subsection{Looping behavior and abnormal termination}
An interesting problem regarding the WHILE language and QDL is that WHILE $T$ DO SKIP (looping behavior) and ABORT (abnormal termination) are indistinguishable. In some semantics, such as natural semantics, this is also the case \cite{EijckStokhof}. In structural operational semantics, however, there is an (infinite) derivation sequence for WHILE $T$ DO SKIP, whereas there is no derivation sequence for ABORT.

Using the standard lemma that $\langle \pi_{1};\pi_{2}\rangle\phi \leftrightarrow \langle \pi_{1}\rangle \langle \pi_{2} \rangle \phi$ (cf. \cite{Harel2,EijckStokhof}) we can prove the equivalence of WHILE $T$ DO SKIP and ABORT in QDL. To do so, we need to ask if $\langle(\?\top;\?\top)^{*};\?\bot\rangle\phi \leftrightarrow \langle \?\bot \rangle\phi$.
\begin{theorem}
In QDL, looping behavior and abnormal termination are equivalent: for any $\phi$
\begin{equation*}
\langle(\iq\top;\iq\top)^{*};\iq\bot\rangle\phi \leftrightarrow \langle \iq\bot \rangle\phi
\end{equation*}
\end{theorem}
\begin{proof}
We will work out the left part first:
\begin{equation*}
\langle(\?\top;\?\top)^{*};\?\bot\rangle\phi \leftrightarrow \langle(\?\top;\?\top)^{*}\rangle\langle\?\bot\rangle\phi
\end{equation*}
So we have $\langle(\?\top;\?\top)^{*}\rangle\psi$ with $\psi = \langle\?\bot\rangle\phi$. Truth of the former in a random model $M$ and for an initial valuation $g$ is defined as follows:
\begin{equation*}
M \models_{g} \langle (\?\top;\?\top)^{*} \rangle\psi \text{ iff for some }h\text{ with } _{g}\llbracket (\?\top;\?\top)^{*} \rrbracket_{h}^{M}, M \models_{h} \psi
\end{equation*}
Furthermore we have
\begin{equation*}
_{g}\llbracket (\?\top;\?\top)^{*} \rrbracket_{h}^{M} \text{ iff } g=h \text{ or } _{g}\llbracket (\?\top;\?\top);(\?\top;\?\top)^{*}\rrbracket_{h}^{M}
\end{equation*}
We have seen in the previous section how such a formula evaluates; after one iteration we will have $_{g}\llbracket \?\top;\?\top \rrbracket_{f}^{M}$, with $f=h$, as one of the options the $^{*}$ command gives us. Finally we have
\begin{align*}
_{g}\llbracket \?\top;\?\top \rrbracket_{h}^{M} =& \ _{g}\llbracket \?\top \rrbracket_{h}^{M}\\
\text{iff }& g = h \text{ and } M \models_{g} \top
\end{align*}
This is always the case, so indeed there is an $h$ such that $_{g}\llbracket (\?\top;\?\top)^{*} \rrbracket_{h}^{M}$ (namely $h=g$). Therefore, determining the truth of $M \models_{g} \langle (\?\top;\?\top)^{*} \rangle\psi$ comes down to determining the truth of $M \models_{g} \psi$, which is $M\models_{g} \langle \?\bot \rangle\phi$.

Since that is exactly the right hand side of the equation we started out with, we indeed have that
\begin{equation*}
\langle(\?\top;\?\top)^{*};\?\bot\rangle\phi \leftrightarrow \langle \?\bot \rangle\phi
\end{equation*}
\end{proof}
Not being able to distinguish between looping behavior and abnormal termination seems undesirable. It is because of this that I have decided to drop the $\langle \pi \rangle \phi$ formulas and replace it by the weaker, but less problematic formulas $[ v:=t ] \top$. Looping behaviour can now no longer be proven to be equivalent to abnormal termination. Furthermore, we avoid problems with formulas that require infinite evaluations, such as $\langle (\?\top)^{*};\?\bot\rangle \phi$.

Because looping behavior and abnormal termination can no longer be proven equal in \dlaf, the relational meaning of \dlaf-instructions now is an instance of the structural operational semantics we defined for our toy language, with the valuations as `states'. Naturally, this is what we want, since it expresses that \dlaf\ is a fully defined system that has the behavior we would expect given our toy language.

This modification also underlines the usefulness of the switch to short-circuit versions of the logical connectives ($\leftor$ and its dual $\leftand$). In QDL, the steering fragment of the program
\begin{equation*}
\text{IF } x:=x+1 \text{ AND } x==2 \text{ THEN } a \text{ ELSE } b
\end{equation*}
can be expressed using $\?(\langle x:=x+1\rangle(x=2))$. In \dlaf\ such an expression now no longer is allowed. However, having $\leftand$ and $\leftor$ in \dlaf\ allows us to provide a perhaps even more natural translation of this program, namely $\?([ x:=x+1 ] \top \leftand x=2)$. The full evaluation versions of these logical connectives ($\land$ and $\lor$) would not do, because the order of the program instructions is important here. As we will see in Chapter \ref{ch:terminology}, we do not \emph{need} $\leftand$ and $\leftor$ in \dlaf, but the fact they provide natural translations of this kind, together with the fact that having logical connectives defined is standard in dynamic logic, is reason enough to keep them.

\chapter{Terminology}
\label{ch:terminology}
In this chapter I will present the terminology I will be using in the remainder of this thesis. In particular, I will present a more fine-grained breakdown of the definitions for formulas, instructions and programs. Furthermore, I will introduce a property of formulas called normal form and use that to prove yet another property of \dlaf\ regarding complex steering fragments. Next, I will introduce a subclass of programs called deterministic programs. Finally, I will introduce a property of deterministic programs called canonical form.

\section{Formulas, instructions and programs}
In this section I will present the more fine-grained breakdown of the definitions for formulas, instructions and programs.

\begin{definition}
\label{def:formulas}
\textbf{Formulas} can either be primitive or compound formulas. \textbf{Primitive formulas} are written as $\varphi$ and defined as follows:
\begin{equation*}
\varphi ::= \top \mid Rt_{1}\ldots t_{n} \mid t_{1}=t_{2} \mid [ v:=t ] \top
\end{equation*}
\textbf{Compound formulas} are written as $\phi$ and defined similarly, but with negation and short-circuit disjunction and conjunction as addition:
\begin{equation*}
\phi ::= \top \mid Rt_{1}\ldots t_{n} \mid t_{1}=t_{2} \mid \lnot \phi \mid \phi_{1} \leftor \phi_{2} \mid \phi_1 \leftand \phi_2 \mid [ v:=t ] \top
\end{equation*}
\end{definition}

\begin{definition}
\label{def:instructions}
\textbf{Instructions} can either be single instructions or basic instructions. \textbf{Single instructions} are written as $\rho$ and defined as follows:
\begin{equation*}
\rho ::= (v:=t) \mid \iq\varphi
\end{equation*}
\textbf{Basic instructions} are written as $\varpi$ and have a little less restrictive definition regarding tests:
\begin{equation*}
\varpi ::= (v:=t) \mid \iq\phi
\end{equation*}
This means that single instructions form a subset of basic instructions:
\begin{equation*}
\rho \subseteq \varpi
\end{equation*}
\end{definition}

\begin{definition}
\textbf{Programs} are written as $\pi$ and consist of one or more basic instructions joined by either concatenation ($;$), union ($\cup$) or repetition ($^{*}$):
\begin{equation*}
\pi ::= \varpi \mid \pi_{1};\pi_{2} \mid \pi_{1} \cup \pi_{2} \mid \pi^{*}
\end{equation*}
\end{definition}

\section{Normal forms of formulas}
\label{sec:normal forms}

In this section I will introduce a property of formulas called normal form and use that to prove a property of \dlaf\ regarding complex steering fragments. I will start with the former.
\begin{definition}
A formula is said to be in its \textbf{normal form} iff all negations (if any) that occur in the formula are on atomic level, that is if the negations only have primitive formulas as their argument (i.e. are of the form $\lnot \varphi$).
\end{definition}
\begin{proposition}
Any formula can be rewritten into its normal form such that its relational meaning is preserved.
\end{proposition}
\begin{proof}
Left-sequential versions of De Morgan's laws are valid for formulas (we come back to this point in Chapter \ref{ch:logicalstructure}): given model $M$ and initial valuation $g$ we prove that
\[M\models_g \neg(\phi_1\leftand \phi_2)\iff M\models_g \neg\phi_1\leftor \neg\phi_2\]
For $\Longrightarrow$, first assume that $M\models_g \phi_1$, thus $M\not\models_h \phi_2$ for $_g\llbracket\Pi_g^M(\phi_1)\rrbracket_h^M$, thus $M\models_h \neg\phi_2$, and thus $M\models_g \neg\phi_1\leftor \neg\phi_2$. If $M\not \models_g \phi_1$, then $M\models_g \neg\phi_1$, and thus also $M\models_g \neg\phi_1\leftor \neg\phi_2$.

In order to show $\Longleftarrow$, first assume that $M\models_g \neg\phi_1$, thus $M\not\models_g \phi_1\leftand\phi_2$, thus $M\models_g \neg(\phi_1\leftand \phi_2)$. If $M\models_g \phi_1$, then $M\models_h \neg\phi_2$ for $_g\llbracket\Pi_g^M(\neg\phi_1)\rrbracket_h^M$, so $M\not\models_g \phi_1\leftand \phi_2$, and thus $M\models_g \neg(\phi_1\leftand \phi_2)$.

The dual statement can also easily be proved.
\end{proof}
The set of side effects caused by the evaluation of a formula does not change under rewritings of this kind. Using normal forms, we can derive an interesting property of \dlaf:
\begin{proposition}
\label{prop:noleftand}
Let $\phi$ be a formula. The program $\iq\phi$ can be rewritten to a form in which only primitive formulas or negations thereof occur in tests, such that its relational meaning is preserved.
\end{proposition}
\begin{proof}
Let $\phi_n$ be a normal form of $\phi$ and assume $\phi_n$ is not a primitive formula or the negation thereof. Then, $\phi_n$ either is of the form $\phi_1 \leftand \phi_2$ or $\phi_1 \leftor \phi_2$. For conjunctions, it is easy to see that the program $\?\phi$ can be rewritten as meant in the proposition:
\begin{equation*}
\?(\phi_1 \leftand \phi_2) = \?\phi_1;\?\phi_2
\end{equation*}
We can assume by induction that $\phi_1$ and $\phi_2$ has been rewritten into a form in which only primitive formulas and negations occur, too. We now need to prove that these programs have the same relational meaning, that is given model $M$ and initial valuation $g$
\begin{equation*}
_g\llbracket \?(\phi_1\leftand \phi_2)\rrbracket_h^M \text{ iff } _g\llbracket \?\phi_1;\?\phi_2 \rrbracket_h^M
\end{equation*}
If $M \not\models_g \phi_1$, then $h$ does not exist in both cases. If, for $_g\llbracket\Pi_g^M(\phi_1)\rrbracket_f^M$, $M \not\models_f \phi_2$, $h$ does not exist in both cases either. Otherwise, on the left hand side, we get $h$ by applying DLA11:
\begin{equation*}
_g\llbracket \Pi_g^M(\phi_1 \leftand\phi_2)\rrbracket_h^M
\end{equation*}
which by definition of the program extraction function, since $M \models_g \phi_1$, equals 
\begin{equation*}
_g\llbracket \Pi_g^M(\phi_1);\Pi_f^M(\phi_2)\rrbracket_h^M
\end{equation*}
On the right hand side, we get $h$ by first applying QDL12, then applying DLA11 twice and finally applying QDL12 again:
\begin{align*}
_g\llbracket \?\phi_1;\?\phi_2\rrbracket_h^M &\text{ iff } \exists f \text{ s.th. } _g\llbracket \?\phi_1\rrbracket_f^M \text{ and } _f\llbracket\?\phi_2\rrbracket_h^M\\
&\text{ iff } \exists f \text{ s.th. } _g\llbracket \Pi_g^M(\phi_1)\rrbracket_f^M \text{ and } _f\llbracket\Pi_f^M(\phi_2)\rrbracket_h^M\\
&\text{ iff }_g\llbracket \Pi_g^M(\phi_1);\Pi_f^M(\phi_2)\rrbracket_h^M
\end{align*}
For disjunctions, the rewritten version is slightly more complex:
\begin{equation*}
\?(\phi_1 \leftor \phi_2) = \?\phi_1 \cup \?\lnot\phi_1;\?\phi_2
\end{equation*}
We can prove that given model $M$ and initial valuation $g$
\begin{equation*}
_g\llbracket\?(\phi_1 \leftor \phi_2)\rrbracket_h^M \text{ iff } _g\llbracket\?\phi_1 \cup \?\lnot\phi_1;\?\phi_2\rrbracket_h^M
\end{equation*}
in a similar fashion as above. If $M \models_g \phi_1$, then in both cases $h$ is obtained by
\begin{equation*}
_g\llbracket \Pi_g^M(\phi_1)\rrbracket_h^M
\end{equation*}
If $M \not\models_g \phi_1$, then if for $_g\llbracket \Pi_g^M(\phi_1)\rrbracket_f^M$, $M \not\models_f \phi_2$, in both cases $h$ does not exist. If $M \models_f \phi_2$, then on the left hand side $h$ is obtained via
\begin{equation*}
_g\llbracket\Pi_g^M(\phi_1 \leftor \phi_2)\rrbracket_h^M = _g\llbracket\Pi_g^M(\phi_1);\Pi_f^M(\phi_2)\rrbracket_h^M
\end{equation*}
And on the right hand side, $h$ is obtained by
\begin{align*}
_g\llbracket \?\lnot\phi_1;\?\phi_2\rrbracket_h^M &\text{ iff } \exists f \text{s.th.} _g\llbracket \?\lnot\phi_1\rrbracket_f^M \text{ and } _f\llbracket\?\phi_2\rrbracket_h^M\\
&\text{ iff } \exists f \text{s.th.} _g\llbracket \Pi_g^M(\phi_1)\rrbracket_f^M \text{ and } _f\llbracket\Pi_f^M(\phi_2)\rrbracket_h^M\\
&\text{ iff }_g\llbracket \Pi_g^M(\phi_1);\Pi_f^M(\phi_2)\rrbracket_h^M
\end{align*}
\end{proof}

On a side note, a similar result can be obtained for QDL. Here the program $\?(\phi_1 \lor \phi_2)$ can be rewritten to
\begin{equation*}
(\?\phi_{1};\?\phi_2) \cup (\?\phi_1;\?\lnot \phi_2) \cup (\?\lnot \phi_1;\?\phi_2)
\end{equation*}
The differences between the \dlaf\ version of the same rule are there because QDL uses full evaluation. Therefore, $\phi_2$ has to be evaluated even when $\phi_1$ is true, although $\phi_2$ does not have to be true anymore.

\section{Deterministic programs and canonical forms}
Defining side effects for entire programs can be complicated. This is because two composition operators, namely union and repetition, can be non-deterministic. We are, however, not interested in (the side effects of) non-deterministic programs, even though they can be expressed in \dlaf.\footnote{In fact, as we already mentioned in Chapter \ref{ch:qdl}, we can ask ourselves if it is reasonable to talk about side effects in non-deterministic programs. We have left this question for future work.} To be exact, we are only interested in \emph{if $\ldots$ then $\ldots$ else} constructions and \emph{while} constructions, which in \dlaf\ are expressed as follows:
\begin{align*}
\text{IF } \phi \text{ THEN } \pi_{1} \text{ ELSE } \pi_{2} &:= (\?\phi;\pi_{1}) \cup (\?\lnot\phi; \pi_{2})\\
\text{WHILE } \phi \text{ DO } \pi &:= (\?\phi;\pi)^{*};\?\lnot\phi
\end{align*}
To formally specify this, we introduce \emph{deterministic programs}, which cf. \cite{Harel, EijckStokhof} are defined as follows:
\begin{definition}
A \textbf{deterministic program} $d\pi$ is a \dlaf-program in one of the following forms:
\begin{equation*}
d\pi ::= \varpi \mid\ d\pi_{1};d\pi_{2} \mid (\iq\phi;d\pi_{1}) \cup (\iq\lnot\phi;d\pi_{2}) \mid ((\iq\phi;d\pi)^{*};\iq\lnot\phi)
\end{equation*}
\end{definition}

There are two interesting properties of deterministic programs. The first is regarding programs of the form $(\?\phi;\pi)^{*};\?\lnot\phi$. In this case there will only ever be exactly one situation in which the program gets evaluated.\footnote{That is unless we are dealing with an infinite loop, but in that case the program has no evaluation and we are not interested in those.} After all, there is exactly one repetition loop for which the test $\?\phi$ succeeds, but will fail the next time it is evaluated. We can formalize this intuition in the following proposition:

\begin{proposition}
\label{prop:n}
Let $d\pi = (\iq\phi;d\pi_{0})^{*};\iq\lnot\phi$ be a deterministic program. Let model $M$ and initial valuation $g$ be given and let $h$ be the valuation such that $_g\llbracket d\pi \rrbracket_h^M$. There is a unique $n \in \mathbb{N}_{0}$ such that
\begin{equation*}
_{g}\llbracket d\pi\rrbracket_{h}^{M} \text{ iff } _{g}\llbracket(\iq\phi;d\pi_{0})^{n};\iq\lnot\phi\rrbracket_{h}^{M}
\end{equation*}
where $(d\pi_{1})^{0};d\pi_{2} = d\pi_{2}$ and $(d\pi_{1})^{n+1};d\pi_{2} = d\pi_{1};(d\pi_{1})^{n};d\pi_{2}$.
\end{proposition}

\begin{proof}
We first prove that there is at least one $n \in \mathbb{N}_{0}$ for which the above equation holds. Assume such an $n$ does not exist. This means that $\?\lnot\phi$ can never be evaluated, which is a contradiction with our requirement that there is a valuation $h$ such that $_g\llbracket d\pi \rrbracket_h^M$.

Next, we have to prove that there is at most one such $n$. Let $g_{i}$ be the valuation such that $_{g}\llbracket (\?\phi;d\pi_{0})^{i}\rrbracket_{g_{i}}^{M}$. By writing this out and then applying DLA11, we know that for $i < n$, we have $M \models_{g_{i}} \?\phi$. Therefore, for valuation $g_{i}$ with $i < n$ we cannot evaluate $\?\lnot\phi$ and thus there is no $i < n$ for which the above equivalence holds.

We know that for $i = n$, we have $M \models_{g_{i}} \?\lnot\phi$. This automatically means that for $i >n$, the above equivalence will not hold either, since we cannot satisfy $\?\phi$. Thus, we have exactly one $n$.
\end{proof}

The second interesting property of a deterministic program is the following:
\begin{definition}
A deterministic program $d\pi$ is said to be in \textbf{canonical form} if only concatenations occur as composition operators.
\end{definition}

This property is going to be very useful, because we can prove that given an initial valuation $g$, any program has a unique canonical form that has the same behavior:
\begin{proposition}
\label{prop:canonical}
Let $d\pi$ be a deterministic program. Let model $M$ and initial valuation $g$ be given and let $h$ be the valuation such that $_g\llbracket d\pi\rrbracket_h^M$. There is a unique deterministic program $d\pi'$ in canonical form such that
\begin{equation*}
_{g}\llbracket d\pi \rrbracket_{h}^{M} \text{ iff } _{g}\llbracket d\pi' \rrbracket_{h}^{M}
\end{equation*}
and $d\pi'$ executes the same basic instructions and the same number of basic instructions as $d\pi$.
\end{proposition}
\begin{proof}
If $d\pi = (\?\phi;d\pi_{1}) \cup (\?\lnot\phi;d\pi_{2})$, then $d\pi'$ depends on the truth of $\phi$:
\begin{align*}
d\pi' =
\begin{cases}
\?\phi;d\pi_{1}' &\text{ if } M \models_{g} \phi\\
\?\lnot\phi;d\pi_{2}' &\text{ o.w.}
\end{cases}
\end{align*}
By induction we can assume that $d\pi_{1}'$ and $d\pi_{2}'$ are the canonical forms of $d\pi_{1}$ and $d\pi_{2}$ (if these are not empty), respectively. The truth of $_{g}\llbracket d\pi \rrbracket_{h}^{M} \text{ iff } _{g}\llbracket d\pi' \rrbracket_{h}^{M}$ follows directly from QDL13 in this case.

If $d\pi = (\?\phi;d\pi_{1})^{*};\?\lnot\phi$, we need to use $n$ as meant in Proposition \ref{prop:n}:
\begin{equation*}
d\pi' = (\?\phi;d\pi_{1}')^{n};\?\lnot\phi
\end{equation*}
Once again we can assume by induction that $d\pi_{1}'$ is the canonical form of $d\pi_{1}$ (once again if $d\pi_1$ is not empty). The truth of $_{g}\llbracket d\pi \rrbracket_{h}^{M} \text{ iff } _{g}\llbracket d\pi' \rrbracket_{h}^{M}$ now follows directly from Proposition \ref{prop:n}.

It is easy to see that in both these cases, $d\pi'$ executes the same basic instructions as $d\pi$. It is also easy to see that $d\pi'$ is unique: we cannot add instructions using union or repetition because then $d\pi'$ will no longer be in canonical form and we cannot add instructions using concatenation because those instructions will be executed, which violates the requirement that $d\pi'$ only executes the same basic instructions as $d\pi$. We cannot alter or remove instructions in $d\pi'$ either because all instructions in $d\pi'$ get executed, so altering or removing one would also violate said requirement.
\end{proof}

\chapter{The logic of formulas in \dlaf}
\label{ch:logicalstructure}

Now that we have \dlaf\ defined and shown how it works, it is time to examine the logic of formulas a little closer. As we have mentioned before, we are making use of short-circuit versions of the $\land$ and $\lor$ connectives, i.e. connectives that prescribe short-circuit evaluation. In \cite{SCL}, different flavours of short-circuit logics (logics that can be defined by short-circuit evaluation) are identified. In this chapter we will give a short overview of these and present the short-circuit logic that underlies the formulas in \dlaf, which turns out to be repetition-proof short-circuit logic (RPSCL).

\section{Proposition algebra}
Short-circuit logic can be defined using \emph{proposition algebra}, an algebra that has short-circuit evaluation as its natural semantics. Proposition algebra is introduced by Bergstra and Ponse in \cite{PA} and makes use of Hoare's ternary connective $x \lef y \rig z$, which is called the \emph{conditional} \cite{Hoare}. A more common expression for this conditional is \emph{if y then x else z}, with $x,y$ and $z$ ranging of propositional statements (including propositional variables). Throughout this thesis, we will use \emph{atom} as a shorthand for propositional variable.

Using a signature which includes this conditional, $\Sigma_{\CP}=\{\top,\bot,\_\lef\_\rig\_\}$, the following set \CP\ of axioms for proposition algebra can be defined:
\begin{align*}
\label{CP1}\tag{CP1} x \lef \top \rig y &= x\\
\label{CP2}\tag{CP2} x \lef \bot \rig y &= y\\
\label{CP3}\tag{CP3} \top \lef x \rig \bot  &= x\\
\label{CP4}\tag{CP4} \qquad x \lef (y \lef z \rig u)\rig v &= (x \lef y \rig v) \lef z \rig (x \lef u \rig v)
\end{align*}
In the earlier mentioned paper \cite{PA}, varieties of so-called \emph{valuation algebras} are defined that serve the interpretation of a logic over $\Sigma_{\CP}$ by means of short-circuit evaluation. The evaluation of the conditional $t_{1} \lef t_{2} \rig t_{3}$ is then as follows: first $t_{2}$ gets evaluated. That yields either $T$, in which case the final evaluation result is determined by the evaluation of $t_{1}$, or $F$, in which case the same goes for $t_{3}$.

All varieties mentioned in \cite{PA} satisfy the above four axioms. The most distinguishing variety is called the variety of \emph{free reactive valuations} and is axiomatized by exactly the four axioms above (further referred to as \emph{conditional propositions} (CP)) and nothing more. The associated valuation congruence is called free valuation congruence and written as $=_{fr}$. Thus, for each pair of closed terms\footnote{Terms that may contain atoms, but not variables.} $t,t'$ over $\Sigma_{\CP}$, we have
\begin{equation*}
\CP\vdash t=t'\iff t=_{fr} t'
\end{equation*}

Using the conditional, we can define negation ($\lnot$), left-sequential conjunction ($\leftand$) and left-sequential disjunction ($\leftor$) as follows:
\begin{align*}
\lnot x &= \bot \lef x \rig \top\\
x \leftand y &= y \lef x \rig \bot\\
x \leftor y &= \top \lef x \rig y
\end{align*}
The above defined connectives are associative and each other's dual. In CP, it is not possible to express the conditional $x \lef y \rig z$ using any set of Boolean connectives, such as $\leftand$ and $\leftor$ \cite{PA}.

By adding axioms to CP, it can be strengthened. The signature and axioms of one such extension are called \emph{memorizing CP}. We write CP$_{mem}$ for this extension that is obtained by adding the axiom CPmem to CP. This axiom expresses that the first evaluation value of $y$ is memorized:
\begin{align*}
\label{CPmem}\tag{CPmem} \qquad x\lef y\rig(z\lef u\rig(v\lef y\rig w))&= x\lef y\rig(z\lef u\rig w)
\end{align*}
With $u=\bot$ and by replacing $y$ by $\lnot y$ we get the contraction law:
\begin{equation*}
\qquad (w\lef y\rig v) \lef y \rig x = w \lef y \rig x
\end{equation*}
A consequence of contraction is the idempotence of $\leftand$. Furthermore, CP$_{mem}$ is the least identifying extension of CP for which the conditional can be expressed using negation, conjunction and disjunction. To be exact, the following holds in CP$_{mem}$:
\begin{equation*}
x \lef y \rig z = (y \leftand x) \leftor (\lnot y \leftand z)
\end{equation*}
We write $=_{mem}$ (memorizing valuation congruence) for the valuation congruence axiomatized by $\CP_{mem}$.

Another extension of CP, the most identifying one distinguised in \cite{PA}, is defined by adding both the contraction law and the axiom below, which expresses how the order of $u$ and $y$ can be swapped, to CP:
\begin{equation*}
\label{CPstat}\tag{CPstat} \qquad (x\lef y\rig z) \lef u \rig v = (x \lef u \rig v) \lef y \rig (z \lef u \rig v)
\end{equation*}
The signature and axioms of this extension, for which we write CP$_{stat}$, are called \emph{static CP}. We write $=_{stat}$ (\emph{static valuation congruence}) for the valuation congruence axiomatized by $\CP_{stat}$. A consequence in $\CP_{stat}$ is $v = v \lef y \rig v$, which can be used to derive the commutativity of $\leftand$: $x \leftand y = y \leftand x$.

CP$_{stat}$ is the most identifying extension of CP because it is `equivalent with' propositional logic, that is, all tautologies in propositional logic can be proved in CP$_{stat}$ using the above translations of its common connectives \cite{SCL}.

\section{Short-Circuit Logics}
In this section we will present the definition of short-circuit logic and its most basic form, free short-circuit logic (FSCL). The definitions are given using \emph{module algebra} \cite{MA}. In module algebra, $S\export X$ is the operation that exports the signature $S$ from module $X$ while declaring other signature elements hidden. Using this operation, short-circuit logics are defined as follows:
\begin{definition}
A \textbf{short-circuit logic} is a logic that implies the consequences of the module expression
\begin{align*}
\SCL=\{\top,\neg,\leftand\}\export(&\CP\\
& + ( \neg x=\bot \lef x \rig \top )\\
& + ( x\leftand y = y \lef x \rig \bot ))
\end{align*}
\end{definition}
Thus, the conditional composition is declared to be an auxiliary operator. In SCL, $\bot$ can be used as a shorthand for $\neg \top$. After all, we have that
\begin{equation*}
\CP + (\neg x = \bot \lef x \rig \top) \vdash \bot = \neg\top
\end{equation*}
With this definition, we can immediately define the most basic short-circuit logic we distinguish:
\begin{definition}
\textbf{$\FreeSCL$ (free short-circuit logic)} is the short-circuit logic that implies no other consequences than those of the module expression \SCL.
\end{definition}
Using these definitions we can provide equations that are derivable from \FreeSCL. The question whether a finite axiomatization of FSCL with only sequential conjunction, negation and $\top$ exists, is open, but the following set \SCLe\ of equations for FSCL is proposed in [5]:\footnote{In [5] it is stated that the authors did not find any equations derivable from FSCL but not from EqFSCL.}
\begin{align}
\label{SCL1}\tag{SCL1} \bot &= \neg \top\\[0mm]
\label{SCL2}\tag{SCL2} x \leftor y &=\neg(\neg x \leftand \neg y)\\[0mm]
\label{SCL3}\tag{SCL3} \neg\neg x &= x\\[0mm]
\label{SCL4}\tag{SCL4} \top \leftand x &= x\\[0mm]
\label{SCL5}\tag{SCL5} x \leftand \top &= x\\[0mm]
\label{SCL6}\tag{SCL6} \bot \leftand x &= \bot\\[0mm]
\label{SCL7}\tag{SCL7} (x \leftand y) \leftand z &= x \leftand (y \leftand z)\\[0mm]
\label{SCL8}\tag{SCL8} \qquad(x \leftor y) \leftand (z \leftand \bot) &= (\neg x \leftor (z \leftand \bot)) \leftand (y \leftand(z \leftand \bot))\\[0mm]
\label{SCL9}\tag{SCL9} (x \leftor y) \leftand (z \leftor \top)&=(x \leftand (z \leftor \top)) \leftor( y \leftand (z \leftor \top))\\[0mm]
\label{SCL10}\tag{SCL10} ((x \leftand \bot) \leftor y) \leftand z &= (x \leftand \bot) \leftor (y \leftand z)
\end{align}
Note that equations SCL2 and SCL3 imply a left-sequential version of De Morgan's laws.

An important equation that is absent is the following:
\begin{equation*}
x \leftand \bot = \bot
\end{equation*}
This is what we would expect, since evaluation of $t \leftand \bot$ (with $t$ a closed term) can generate a side effect that is absent in the evaluation of $\bot$, although we know that evaluation of $t \leftand \bot$ always yields $F$.

We now have the most basic short-circuit logic and some of its equations defined, but of course there also is a ``most liberal'' short-circuit logic below propositional logic. This logic is based on memorizing CP and satisfies idempotence of $\leftand$ (and $\leftor$), but not its commutativity. It is defined as follows:
\begin{definition}
\textbf{$\MSCL$ (memorizing short-circuit logic)} is the short-circuit logic that implies no other consequences than those of the module expression
\begin{align*}
\{\top,\neg,\leftand\}\export(&\CP_{mem}\\
& + ( \neg x= \bot \lef x \rig \top )\\
& + ( x \leftand y = y \lef x \rig \bot ))
\end{align*}
\end{definition}
For the set of axioms EqMSCL, intuitions and an example, and a completeness proof of MSCL we refer the reader to \cite{SCL}. Adding the axiom $x\leftand \bot = \bot$ to MSCL, or equivalently, the axiom $\bot \lef x \rig \bot = \bot$ to CP$_{mem}$,  yields so-called static short-circuit logic (SSCL), which is equivalent with propositional logic (be it in sequential notation and defined by short-circuit evaluation).
\begin{definition}
\textbf{$\SSCL$ (static short-circuit logic)} is the short-circuit logic that implies no other consequences than those of the module expression
\begin{align*}
\{\top,\neg,\leftand\}\export(&\CP_{mem}\\
& + ( \bot \lef x \rig \bot = \bot )\\
& + ( \neg x= \bot \lef x \rig \top )\\
& + ( x \leftand y = y \lef x \rig \bot ))
\end{align*}
\end{definition}

\section{Repetition-Proof Short-Circuit Logic}
With both the most basic as well as the most liberal short-circuit logic we distinguish defined, we can present the variant of short-circuit logic that we are interested in because it underlies the logic of formulas in \dlaf: \emph{repetition-proof short-circuit logic} (RPSCL). This SCL-variant stems from an axiomatization of proposition algebra called repetition-proof CP (CP$_{rp}$) that is in between \CP\ and \CP$_{mem}$ and involves explicit reference to a set $A$ of atoms (propositional variables).

The axiom system CP$_{rp}$ is defined as the extension of CP with the following two axiom schemes (for $a\in A$), which imply that any subsequent evaluation result of an atom $a$ equals the current one:
\begin{align*}
\label{CPrp1}\tag{CPrp1} \qquad (x\lef a\rig y)\lef a\rig z&=(x\lef a\rig x)\lef a\rig z\\
\label{CPrp2}\tag{CPrp2} \qquad x\lef a\rig (y\lef a\rig z)&=x\lef a\rig (z\lef a \rig z)
\end{align*}
We write $\Eq_{rp}(A)$ to denote the set of these axioms schemes in the format of module algebra. In CP$_{rp}$ the conditional cannot be expressed in terms of $\leftand$, $\neg$ and $\top$: in \cite{PA} it is shown that the propositional statement $a \lef b \rig c$ (for atoms $a,b,c \in A$) cannot be expressed modulo repetition-proof valuation congruence, that is, the valuation congruence axiomatized by CP$_{rp}$. The definition of RPSCL then becomes:
\begin{definition}
\textbf{$\RPSCL$ (repetition-proof short-circuit logic)} is the short-circuit logic that implies no other consequences than those of the module expression
\begin{align*}
\{\top,\neg,\leftand,a \mid a \in A\} \export (&\CP +\Eq_{rp}(A)\\
& + ( \neg x = \bot \lef x \rig \top )\\
& + ( x \leftand y = y \lef x \rig \bot ))
\end{align*}
\end{definition}

The equations defined by \RPSCL\ include those that are defined by \SCLe\ as well as for $a \in A$:
\begin{align*}
\label{rp1}\tag{RP1} a\leftand(a\leftor x)&=a\leftand(a\leftor y)\\
\label{rp2}\tag{RP2} a\leftor(a\leftand x)&=a\leftor(a\leftand y)\\
\label{rp3}\tag{RP3} (a\leftor\neg a)\leftand x&=(\neg a\leftand a)\leftor x\\
\label{rp4}\tag{RP4} (\neg a\leftor a)\leftand x&=(a\leftand \neg a)\leftor x\\
\label{rp5}\tag{RP5} (a\leftand\neg a)\leftand x &=a\leftand\neg a\\
\label{rp6}\tag{RP6} (\neg a\leftand a)\leftand x &=\neg a\leftand a\\
\label{rp7}\tag{RP7} (x\leftor y) \leftand (a\leftand \neg a) &= (\neg x \leftor (a\leftand\neg a)) \leftand (y\leftand(a\leftand \neg a))\\
\label{rp8}\tag{RP8} (x\leftor y) \leftand (\neg a\leftand a) &= (\neg x \leftor (\neg a\leftand a)) \leftand (y\leftand(\neg a\leftand a))\\
\label{rp9}\tag{RP9} (x\leftor y)\leftand(a\leftor\neg a)&=(x\leftand (a\leftor\neg a))\leftor(y\leftand (a\leftor\neg a))\\
\label{rp10}\tag{RP10} (x\leftor y)\leftand(\neg a\leftor a)&=(x\leftand (\neg a\leftor a))\leftor(y\leftand (\neg a\leftor a))\\
\label{rp11}\tag{RP11} ((a\leftand\neg a)\leftor y)\leftand z&=(a\leftand \neg a)\leftor (y\leftand z)\\
\label{rp12}\tag{RP12} ((\neg a\leftand a)\leftor y)\leftand z&=(\neg a\leftand a)\leftor (y\leftand z)
\end{align*}
It is an open question whether the equations SCL1-SCL10 and the equation schemes RP1-RP12 axiomatize RPSCL, but it will be shown below that RPSCL is the logic that models equivalence of formulas in \dlaf, where
\begin{equation*}
A = \{Rt_{1}\ldots t_{n}, t_{1}=t_{2}, [ v:=t]\top \}
\end{equation*}
For this reason, we add the conditional $\phi_{1} \lef \phi_{2} \rig \phi_{3}$ and the constant $\bot$ to \dlaf\ (thus making $\leftor$ and $\neg$ definable). In order to decide whether different \dlaf\ formulas are equivalent, just translate these to CP$_{rp}$ and decide their equivalence (either by axiomatic reasoning or by checking their repetition-proof valuation congruence). So, we extend the formulas in \dlaf\ in order to characterize the logic that models their equivalence. In this extension of \dlaf, which we baptize \dlcaf\ (for Dynamic Logic with the Conditional and Assignments as Formulas), truth in $M$ relative an initial valuation $g$ for the conditional is defined as follows:
\begin{align*}
\label{DLCA}\tag{DLCA}
M \models_{g} (\phi_{2} \lef \phi_{1} \rig \phi_{3}) \text{ iff for } _{g}\llbracket \Pi_{g}^{M}(\phi_{1})\rrbracket_{h}^{M},
\begin{cases}
M \models_{h} \phi_{2} &\text{ if } M \models_{g} \phi_{1}\\
M \models_{h} \phi_{3} &\text{ o.w.}
\end{cases}
\end{align*}
This means that we need an extra equation for the program extraction function $\Pi$ too which handles the conditional. For model $M$, initial valuation $g$ and $_{g}\llbracket \phi_{1} \rrbracket_{h}^{M}$
\begin{align*}
\Pi_{g}^{M}(\phi_{2} \lef \phi_{1} \rig \phi_{3}) = 
\begin{cases}
\Pi_{g}^{M}(\phi_{1});\Pi_{h}^{M}(\phi_{2}) &\text{ if } M \models_{g} \phi_{1}\\
\Pi_{g}^{M}(\phi_{1});\Pi_{h}^{M}(\phi_{3}) &\text{ o.w. }
\end{cases}
\end{align*}
In the remainder of this section we consider formulas over this signature, thus formulas over $A$ composed with $\_\lef\_\rig\_$. Below we will prove for all mentioned axioms that they are valid in \dlcaf.
\begin{proposition}
Let $M$ be a model for \dlcaf. The axiom CP1, that is
\begin{equation*}
\tag{CP1}x \lef \top \rig y = x
\end{equation*}
is valid in $M$.
\end{proposition}
\begin{proof}
Let $t_{1},t_{2}$ be arbitrary formulas and let $g$ be an initial valuation. Regardless of $g$, we have $M \models_{g} \top$ (by QDL3), so by DLCA, we get $M \models_{g} (t_{1} \lef \top \rig t_{2})$ iff for $_{g}\llbracket \?\top \rrbracket_{h}^{M}$, $M \models_{h} t_{1}$. Since $g=h$, we indeed have that $M \models_{g} (t_{1} \lef \top \rig t_{2})$ iff $M \models_{g} t_{1}$.
\end{proof}

\begin{proposition}
Let $M$ be a model for \dlcaf. The axiom CP2, that is
\begin{equation*}
\tag{CP2}
x \lef \bot \rig y = y
\end{equation*}
is valid in $M$.
\end{proposition}
\begin{proof}
Let $t_{1},t_{2}$ be arbitrary formulas and let $g$ be an initial valuation. $\bot$ is a shorthand for $\lnot \top$, so we first need QDL6, which states that $M \models_{g} \lnot \top$ iff not $M \models_{g} \top$, which is never the case. So for any initial valuation $g$, $M \models_{g} \bot$ is false. Thus by DLCA, we get $M \models_{g} (t_{1} \lef \bot \rig t_{2})$ iff for $_{g}\llbracket \?\top \rrbracket_{h}^{M}$, $M \models_{h} t_{1}$. Since $g=h$, we indeed have that $M \models_{g} (t_{1} \lef \bot \rig t_{2})$ iff $M \models_{g} t_{2}$.
\end{proof}

\begin{proposition}
Let $M$ be a model for \dlcaf. The axiom CP3, that is
\begin{equation*}
\tag{CP3}
\top \lef x \rig \bot = x
\end{equation*}
is valid in $M$.
\end{proposition}
\begin{proof}
Let $t$ be an arbitrary formula and let $g$ be an initial valuation. If $M \models_{g} t$ then by DLCA we get for $_{g}\llbracket \Pi_{g}^{M} (t)\rrbracket_{h}^{M}$, $M \models_{h} \top$, which also is true. If $M \not\models_{g} t$ then by DLCA we obtain $M \models_{h} \bot$ (note that also in this case, $h$ is defined), which also is false. Thus $M \models_{g} t$ iff $M \models_{g} \top \lef t \rig \bot$ and hence the axiom CP3 is valid.
\end{proof}

\begin{proposition}
Let $M$ be a model for \dlcaf. The axiom CP4, that is
\begin{equation*}
\tag{CP4}
x \lef (y \lef z \rig v) \rig u = (x \lef y \rig u) \lef z \rig (x \lef v \rig u)
\end{equation*}
is valid in $M$.
\end{proposition}
\begin{proof}
Let $t_{1},t_{2},t_{3},t_{4},t_{5}$ be arbitrary formulas and let $g$ be an initial valuation. We are going to have to show that
\begin{equation*}
M \models_g t_1 \lef (t_2 \lef t_3 \rig t_4) \rig t_5 \text{ iff } M \models_g (t_1 \lef t_2 \rig t_5) \lef t_3 \rig (t_1 \lef t_4 \rig t_5)
\end{equation*}
We have to apply DLCA multiple times here. By applying it to the left hand side we get for $_{g}\llbracket \Pi_{g}^{M}( t_{2} \lef t_{3} \rig t_{4}) \rrbracket_{f}^{M}$
\begin{align*}
M \models_g t_{1} \lef (t_{2} \lef t_{3} \rig t_{4}) \rig t_{5} \text{ iff }
\begin{cases}
M \models_{f} t_{1} &\text{ if } M \models_{g} (t_{2} \lef t_{3} \rig t_{4})\\
M \models_{f} t_{5} &\text{ o.w. }
\end{cases}
\end{align*}
By applying DLCA again to $M \models_{g} (t_{2} \lef t_{3} \rig t_{4})$ we get for $_{g}\llbracket \Pi_{g}^{M}(t_{3}) \rrbracket_{f'}^{M}$
\begin{align*}
M \models_{g} (t_{2} \lef t_{3} \rig t_{4}) \text{ iff }
\begin{cases}
M \models_{f'} t_{2} &\text{ if } M \models_{g} t_{3}\\
M \models_{f'} t_{4} &\text{ o.w.}
\end{cases}
\end{align*}
So if $M \models_{g} t_{3}$ and $M \models_{f'} t_{2}$, we get $M \models_{f} t_{1}$. If on the other hand $M \not\models_{g} t_{3}$ but $M \models_{f'} t_{4}$, we also get $M \models_{f} t_{1}$. In all other situations we get $M \models_{f} t_{5}$.

Let us now consider the right hand side of the equation. Here we get for $_{g}\llbracket \Pi_{g}^{M}(t_{3})\rrbracket_{h'}^{M}$:
\begin{equation*}
M \models_{g} (t_{1} \lef t_{2} \rig t_{5}) \lef t_{3} \rig (t_{1} \lef t_{4} \rig t_{5}) \text{ iff }
\begin{cases}
M \models_{h'} (t_{1} \lef t_{2} \rig t_{5}) \text{ if } M \models_{g} t_{3}\\
M \models_{h'} (t_{1} \lef t_{4} \rig t_{5}) \text{ o.w.}
\end{cases}
\end{equation*}
Let us first turn our attention to the situation where $M \models_{g} t_{3}$. We need to apply DLCA again and get for $_{h'}\llbracket \Pi_{h'}^{M}(t_{2})\rrbracket_{h}^{M}$
\begin{align*}
M \models_{h'} (t_{1} \lef t_{2} \rig t_{5}) \text{ iff }
\begin{cases}
M \models_{h} t_{1} &\text{ if } M \models_{h'} t_{2}\\
M \models_{h} t_{5} &\text{ o.w.}
\end{cases}
\end{align*}
In the situation where $M \not\models_{g} t_{3}$, we get for $_{h'}\llbracket \Pi_{h'}^{M}(t_{4})\rrbracket_{h''}^{M}$
\begin{align*}
M \models_{h'} (t_{1} \lef t_{4} \rig t_{5}) \text{ iff }
\begin{cases}
M \models_{h''} t_{1} &\text{ if } M \models_{h'} t_{4}\\
M \models_{h''} t_{5} &\text{ o.w.}
\end{cases}
\end{align*}
So on the right hand side, if $M \models_{g} t_{3}$ and $M \models_{h'} t_{2}$, we get $M \models_{h} t_{1}$. If $M \not\models_{g} t_{3}$ but $M \models_{h'} t_{4}$, we also get $M \models_{h''} t_{1}$. In the other situations we get either $M \models_{h} t_{5}$ or $M \models_{h''} t_{5}$.

To prove that is the same result as on the left-hand side, we need to prove that $f' = h'$, $f = h$ if $M \not\models_{g} t_{3}$, and $f = h''$ if $M \models_{g} t_{3}$. The last two statements seem contradictory, but as we will see $f$ can actually take two different valuations depending on the truth of $t_{3}$. The mentioned variations are all determined using the program extraction function. To recap, we have the following:
\begin{gather*}
_{g}\llbracket \Pi_{g}^{M}( t_{2} \lef t_{3} \rig t_{4}) \rrbracket_{f}^{M}\\
_{g}\llbracket \Pi_{g}^{M}(t_{3}) \rrbracket_{f'}^{M}\\
_{g}\llbracket \Pi_{g}^{M}(t_{3})\rrbracket_{h'}^{M}\\
_{h'}\llbracket \Pi_{h'}^{M}(t_{2})\rrbracket_{h}^{M}\\
_{h'}\llbracket \Pi_{h'}^{M}(t_{4})\rrbracket_{h''}^{M}
\end{gather*}
We can immediately see that $f' = h'$. Using the updated definition for the program extraction function we get that
\begin{align*}
_{g}\llbracket \Pi_{g}^{M}( t_{2} \lef t_{3} \rig t_{4}) \rrbracket_{f}^{M} \text{ iff }
\begin{cases}
_{g}\llbracket \Pi_{g}^{M}( t_{3} );\Pi_{h'}^{M}( t_{2} )  \rrbracket_{f}^{M} &\text{ if } M \models_g t_3\\
_{g}\llbracket \Pi_{g}^{M}( t_{3} );\Pi_{h'}^{M}( t_{4} )  \rrbracket_{f}^{M} &\text{ o.w.}
\end{cases}
\end{align*}
Using the new rule for the conditional, we get that:
\begin{align*}
_{g}\llbracket \Pi_{g}^{M}(t_{3});\Pi_{h'}^{M}(t_{2}) \rrbracket_{f}^{M} &\text{ if } M \models_{g} t_{3}\\
_{g}\llbracket \Pi_{g}^{M}(t_{3});\Pi_{h'}^{M}(t_{4}) \rrbracket_{f}^{M} &\text{ if } M \not\models_{g} t_{3}
\end{align*}
To determine if $f = h$, we need to have $M \models_{g} t_{3}$ and we need to evaluate:
\begin{equation*}
_{g}\llbracket \Pi_{g}^{M}(t_{3})\rrbracket_{h'}^{M} \text{ and } _{h'}\llbracket \Pi_{h'}^{M}(t_{2})\rrbracket_{h}^{M}
\end{equation*}
By QDL12, we know that is equivalent to
\begin{equation*}
_{g}\llbracket \Pi_{g}^{M}(t_{3});\Pi_{h'}^{M}(t_{2})\rrbracket_{h}^{M}
\end{equation*}
So indeed we have that if $M \models_{g} t_{3}$, then $f = h$. Using the same argument, we get that if $M \not\models_{g} t_{3}$, then
\begin{equation*}
_{g}\llbracket \Pi_{g}^{M}(t_{3});\Pi_{h'}^{M}(t_{4})\rrbracket_{h''}^{M}
\end{equation*}
Therefore, if $M \not\models_{g} t_{3}$ then $f = h''$.
\end{proof}

With those four axioms proven, we already know for a fact that the logic of formulas in \dlaf\ indeed is a short-circuit logic. To prove that it is a repetition-free short-circuit logic, we need to prove the axiom schemes CPrp1 and CPrp2, too. Those axiom schemes make use of atoms $a \in A$.
\begin{proposition}
Let $M$ be a model for \dlcaf. The axiom CPrp1, that is
\begin{equation*}
\tag{CPrp1}
(x \lef a \rig y) \lef a \rig z = (x \lef a \rig x) \lef a \rig z
\end{equation*}
is valid in $M$.
\end{proposition}
\begin{proof}
Let $t_{1},t_{2},t_{3}$ be arbitrary formulas and $g$ an initial valuation. $M \models_{g} a$ can either be true or false. If it is false, both the left hand side and the right hand side, by DLCA, are determined for $_{g}\llbracket \Pi_{g}^{M}(a)\rrbracket_{h}^{M}$ by $M \models_{h} t_{3}$. If it is true, the question if $M \models_{h} a$ is asked. We have to prove that for every atom $a \in A$, the reply to this will be the same as the reply to $M \models_{g} a$ (namely, true), that is:
\begin{equation*}
M \models_{h} a \text{ iff } M \models_{g} a
\end{equation*}
Recall that $a$ can be of the forms $\{Rt'_{1}\ldots t'_{n}, t'_{1}=t'_{2}, [ v:=t']\top\}$. For the first two atoms we can immediately see our claim is true, since $\Pi_{g}^{M}(a) = \?\top$ and therefore $g = h$. For $[ v:=t']\top$ the claim immediately follows from DLA9: it is, regardless of the valuation, always true.
\end{proof}

\begin{proposition}
Let $M$ be a model for \dlcaf. The axiom CPrp2, that is
\begin{equation*}
\tag{CPrp2}
x \lef a \rig (y \lef a \rig z) = x \lef a \rig (z \lef a \rig z)
\end{equation*}
is valid in $M$.
\end{proposition}
\begin{proof}
This is the symmetric variant of CPrp1 and proven similarly.
\end{proof}

By proving the validity of these axiom schemes in \dlcaf\ we have proven that the equations SCL1-SCL10 together with RP1-RP12 are axioms for formulas in \dlcaf. CP$_{rp}$ indeed is the most identifying extension of CP which is valid for formulas. After all, the first more identifying extension we distinguish is CP$_{con}$ (\emph{contractive CP}) \cite{SCL}, from which amongst others the following weak contraction rule can be derived: for $a \in A$
\begin{equation*}
a \leftand a = a
\end{equation*}
Clearly this is not valid for \dlaf-formulas such as $[ x:=x+1]\top$.

\chapter{A treatment of side effects}
\label{ch:treatment}

\section{Introduction}
\label{sec:introduction}
Now that we have defined a system to model program instructions and program states, we can return to our original problem: that of formally defining side effects. Like I said in Section \ref{sec:introsideeffects}, the basic idea is that a side effect has occurred in the execution of a program if there is a difference between the actual evaluation and the expected evaluation of a program given an initial valuation. 

We can immediately see however, that we cannot build a definition of side effects based on the actual and expected evaluation of an entire program. Such a definition will get into trouble when there are multiple side effects, especially if those cancel each other out or reinforce each other. Consider for example the following program:
\begin{equation*}
\pi = \?([ x:=x+1]\top);\?([ x:=x+1]\top)
\end{equation*}
If we are only going to look at the entire program, we will detect one side effect here, that has incremented the value of $x$ by two. However, it appears to be more acceptable to say that \emph{two} side effects have occurred, that happen to affect the same value.

It gets even more interesting if there is a formula in between the two clauses above and the clauses themself cancel each other out:
\begin{align*}
\pi = \?([ x:=x+1]\top \leftand \phi \leftand [ x:=x\monus1]\top)
\end{align*}
If we again only look at the entire program, we will detect no side effects (unless side effects occur in $\phi$). However, because $\phi$ might use or modify $x$ as well, it seems we will have to pay attention to the side effect of the first clause, even though it will be cancelled out on by the last clause. 

So instead of building a definition of side effects by looking only at the actual and expected evaluation of an entire program, we are going to build it up starting at the instruction level.

\section{Side effects in single instructions}
\label{sec:singleinstructions}
As said, we are going to use a bottom-up approach to define side effects, so we will first define side effects for single instructions, then move up to basic instructions and end with a full definition of side effects for programs.

The idea is that the side effect of a single instruction is the difference between the actual and expected evaluation of a single instruction. This difference is essentially the difference between the resulting valuations after, respectively, the actual and expected evaluation of the single instruction. The difference between two valuations is defined as follows:
\begin{definition}
Given a model $M$, the \textbf{difference} between valuations $g$ and $h$ is defined as those variables that have a different assignment in $g$ and $h$:
\begin{equation*}
(x \mapsto k') \in \delta^{M}(g,h) \text{ iff } g(x) = k, h(x)=k' \text{ and } M \not\models k = k'
\end{equation*}
\end{definition}
This notion of difference is not symmetric.

We already know what the actual evaluation of a single instruction is: for this we can use \dlaf. This leaves us to define the expected evaluation. For this we need to know for each single instruction how we expect it to evaluate, that is, what changes we expect it to make to the initial valuation. We have the following expectations of each single instruction:
\begin{itemize}
\item \textbf{Assignments} change the initial valuation by updating the variable assignment of the variable under consideration to the (interpretation of the) new variable assignment.
\item \textbf{Tests} do not change the initial valuation: they only yield $T$ or $F$ and steer the rest of the program accordingly.
\end{itemize}

We need the following equations for determining the expected evaluation $\mathcal{E}$ of a single instruction:
\begin{align*}
\label{EV1}\tag{EV1}M \models_g^{\mathcal{E}} \top & \text{ always}\\
\label{EV2}\tag{EV2}M \models_g^{\mathcal{E}} Rt_{1}\ldots t_{n} &\text{ iff } (\llbracket t_{1}\rrbracket_{g}^{M},\ldots,\llbracket t_{n}\rrbracket_{g}^{M})\in R^{M}\\
\label{EV3}\tag{EV3}M \models_g^{\mathcal{E}} t_{1}=t_{2} &\text{ iff } \llbracket t_{1}\rrbracket_{g}^{M} \text{ is the same as }\llbracket t_{2}\rrbracket_{g}^{M}\\
\label{EV4}\tag{EV4}M \models_{g}^{\mathcal{E}} [ v:=t ] \top & \text{ always}
\end{align*}
\begin{align*}
\label{EV5}\tag{EV5}_{g}\llbracket v:=t \rrbracket_{h}^{M,\mathcal{E}} \text{ iff } & h = g[v \mapsto \llbracket t \rrbracket_{g}^{M}]\\
\label{EV6}\tag{EV6}_{g}\llbracket \?\varphi \rrbracket_{h}^{M,\mathcal{E}} \text{ iff } & g=h \text{ and }M \models_{g}^{\mathcal{E}} \varphi
\end{align*}

Now that we have the actual and the expected evaluation of a single instruction, we can define its side effects. As said, this is going to be the difference between the two resulting valuations.
\begin{definition}
Let $\rho$ be a single instruction. Let model $M$ be given and let $g$ be an initial valuation. Furthermore, let $h$ be a valuation such that $_g\llbracket \rho \rrbracket_h$ and let $h'$ be a valuation such that $_g\llbracket \rho \rrbracket_{h'}^{\mathcal{E}}$. The set of side effects of single instruction $\rho$ given model $M$ and initial valuation $g$ is defined as
\begin{equation*}
\mathcal{S}_{g}^{M}(\rho) = \delta^{M}(h',h)
\end{equation*}
\end{definition}
It is important to note that the valuations $h$ and $h'$ as meant in the above definition may not exist. We are not interested in those situations, however. If $h$ and $h'$ do exist, they are unique. Also note that $\delta^{M}(h',h)$ returns the variable assignment of valuation $h$ if there is a difference with the variable assignment of valuation $h'$. Thus, the set of side effects is defined as a set containing those variables that have a different assignment after the actual and expected valuation, with as assignments the ones the variables actually get (that is, the assignments they will have after evaluating the single instruction with the actual evaluation).

We will illustrate this with two examples. First, consider the single instruction $\rho = (x:=1)$, evaluated under model $M$ in initial valuation $g$ with $g(x)=0$. We want to know if this causes a side effect, so we need to know the actual evaluation and the expected evaluation. To calculate the actual evaluation, we need to know if $_{g}\llbracket x:=1\rrbracket_{h}^{M}$ and if yes, for which valuation $h$. The equations for \dlaf\ immediately give us the answer, in this case via QDL10: $h = g[x \mapsto \llbracket 1 \rrbracket_{g}^{M}]$. So we get $h(x) = 1$.

Getting the expected evaluation works in a similar fashion, but instead of \dlaf\ we now use the equations above to evaluate $\rho$. Since the equation for evaluating an assignment (EV5) is the same as QDL10, we now get the exact same expected evaluation as the actual evaluation. Thus we get $h' = g[x \mapsto \llbracket 1 \rrbracket_{g}^{M}]$ and therefore $h'(x) = 1$. We can immediately see that this results in the set of side effects being empty:
\begin{equation*}
\mathcal{S}_{g}^{M}(x:=1) = \delta^M(h',h) = \emptyset
\end{equation*}
This is of course what we would expect: an assignment should not have a side effect if it does not occur in a steering fragment. Let us now consider an example where we do expect a side effect: namely if an assignment does occur in a steering fragment: $\rho = \?([ x:=1 ]\top)$. We use the same initial valuation $g$. First we try to find the actual evaluation again, which we do by evaluating $_{g}\llbracket\?([ x:=1 ]\top)\rrbracket_{h}^{M}$. We now need DLA11, which tells us that (in this case) $_{g}\llbracket\?([ x:=1 ]\top)\rrbracket_{h}^{M}$ iff $M \models_{g}([ x:=1]\top)$ and $_{g}\llbracket\Pi_{g}^{M}([ x:=1]\top)\rrbracket_{h}^{M} =\ _{g}\llbracket x:=1 \rrbracket_{h}^{M}$. Both evaluate to true, the latter with $h = g[x \mapsto 1]$. 

The expected update once again takes us to the equations above; we need to determine $h'$ such that $_{g}\llbracket \?([x:=1]\top)\rrbracket_{h'}^{M,\mathcal{E}}$. For tests, the demands are fairly simple: $g=h'$ and $M \models_{g}^{\mathcal{E}} [ x:=1]\top$ (see EV6). The latter is by EV4 defined to be always true. As a result, we get $h'(x) = g(x) = 0$. Thus we get the following set of side effects:
\begin{align*}
\mathcal{S}_{g}^{M}(\?[ x:=1 ]\top) &= \delta^{M}(h',h)\\
&= \{x \mapsto 1\}
\end{align*}
Again, this is exactly what we want: since we expect formulas to only yield true or false, the change this formula makes to the program state upon evaluation is a side effect.

\section{Side effects in basic instructions}
\label{sec:basicinstructions}
With side effects for single instructions defined, we can move up a step to side effects in basic instructions. The difference between single and basic instructions is that in basic instructions, complex steering fragments are allowed. This means that we are going to have to define how side effects are handled in tests that contain a disjunction ($\leftor$), conjunction ($\leftand$) or negation ($\lnot$). The idea is that the set of side effects of the whole formula is the union of the sets of side effects of its primitive parts. However, we also have to pay attention to the short-circuit character of $\leftor$. Only the primitive formulas that get evaluated can contribute to the set of side effects.

With this in mind, we can give the definition for side effects in (possibly) complex steering fragments. Like before, we are only interested in the side effects if the test actually succeeds. We need to define this for disjunctions, conjuctions and negations:
\begin{definition}
Let $\phi = \phi_{1} \leftor \phi_{2}$ be a disjunction. Let model $M$ and initial valuation $g$ be given, with $M \models_{g} \phi$ and where $\phi$ is in its normal form. Furthermore, let $f$ be the valuation after evaluation of formula $\phi_{1}$, that is, $_{g}\llbracket \iq\phi_{1}\rrbracket_{f}^{M}$. The set of side effects $\mathcal{S}_{g}^{M}(\iq\phi)$ is defined as:
\begin{align*}
\mathcal{S}_{g}^{M}(\iq\phi) = 
\begin{cases}
\mathcal{S}_{g}^{M}(\iq\phi_{1}) &\text{ if } M \models_{g} \phi_{1}\\
\mathcal{S}_{g}^{M}(\iq\phi_{1}) \cup \mathcal{S}_{f}^{M}(\iq\phi_{2}) &\text{ o.w.}
\end{cases}
\end{align*}
\end{definition}
The case distinction is in place because of the short-circuit character of $\leftor$. For the definition of its dual $\leftand$ we do not need this case distinction, because since we are again only interested in the side effects if the (entire) formula succeeds, all the formulas in the conjunction have to yield true. Therefore, the definition for conjunction is a bit easier:
\begin{definition}
Let $\phi = \phi_{1} \leftand \phi_{2}$ be a conjunction. Let model $M$ and initial valuation $g$ be given, with $M \models_{g} \phi$ and where $\phi$ is in its normal form. Furthermore, let $f$ be the valuation after evaluation of primitive formula $\phi_{1}$, that is, $_{g}\llbracket \iq\phi_{1} \rrbracket_{f}^{M}$. The set of side effects $\mathcal{S}_{g}^{M}(\iq\phi)$ is defined as:
\begin{equation*}
\mathcal{S}_{g}^{M}(\iq\phi) = \mathcal{S}_{g}^{M}(\iq\phi_{1}) \cup \mathcal{S}_{f}^{M}(\iq\phi_{2})
\end{equation*}
\end{definition}
The recursive definitions for disjunction and conjunction work because eventually, a primitive formula will be encountered, for which the side effects are already defined. Unfortunately, we cannot use a similar construction for negation. This is because the side effects in a primitive formula are only defined if that formula yields true upon evaluation, so we cannot simply treat negation as a transparent operator (that is, it is typically not true that $\mathcal{S}_g^M(\lnot \phi) = \mathcal{S}_g^M(\phi)$). So we will have to define negation the hard way instead. Because we are using formulas in normal form in the other definitions, we only have to define negation for primitive formulas:
\begin{definition}
Let $\lnot\varphi$ be a negation. Let model $M$ be given and let $g$ be an initial valuation. Furthermore, let $h$ be a valuation such that $_g\llbracket \iq\lnot\varphi \rrbracket_h$ and let $h'$ be a valuation such that $_g\llbracket \iq\lnot\varphi \rrbracket_{h'}^{\mathcal{E}}$. The set of side effects of basic instruction $\iq\lnot\varphi$ given model $M$ and initial valuation $g$ is defined as
\begin{equation*}
\mathcal{S}_{g}^{M}(\iq\lnot\varphi) = \delta^{M}(h',h)
\end{equation*}
\end{definition}

Now that we have a definition for side effects in (complex) steering fragments, the extension of our definition of side effects in single instructions to side effects in basic instructions is trivial:
\begin{definition}
Let $\varpi$ be a basic instruction. Let model $M$ and initial valuation $g$ be given and let $h$ be a valuation such that $_{g}\llbracket\varpi \rrbracket_{h}^{M}$. The set of side effects $\mathcal{S}_{g}^{M}(\varpi)$ is defined as:
\begin{align*}
\mathcal{S}_{g}^{M}(\varpi) = 
\begin{cases}
\mathcal{S}_{g}^{M}(\rho) &\text{ if } \varpi = \rho\\
\mathcal{S}_{g}^{M}(\iq\phi) &\text{ if } \varpi = \iq\phi' \text{ and }\phi \text{ is the normal form of } \phi'
\end{cases}
\end{align*}
\end{definition}

We can illustrate this with a simple, yet interesting example. Consider the following basic instruction: $\varpi = ?([ x:=x+1]\top \leftand [ x:=x\monus1]\top)$ with initial valuation $g$ such that $g(x)=1$. In this situation we have two side effects that happen to cancel each other out. The resulting valuation after the actual evaluation of this basic instruction will be the same as the initial valuation $g$.

First we observe that the formula in this basic instruction is in its normal form, a trivial observation since no negations occur in it. There are two primitive formulas in this conjunction, so the set of side effects is:
\begin{align*}
\mathcal{S}_{g}^{M}(?([ x:=x+1]\top \leftand [ x:=x\monus1]\top)) =\ &\mathcal{S}_{g}^{M}(?([ x:=x+1]\top))\ \cup\\ &\mathcal{S}_{g_{1}}^{M}(?([ x:=x\monus1]\top))
\end{align*}
Here $g_{1}$ is determined by $_{g}\llbracket ?([ x:=x+1]\top)\rrbracket_{g_{1}}^{M}$, so we get $g_{1}(x)=2$. We have already seen in the previous section how the parts of the union above evaluate, so we get:
\begin{align*}
\mathcal{S}_{g}^{M}(?([ x:=x+1]\top \leftand [ x:=x\monus1]\top)) &= \{x \mapsto 2\} \cup \{x \mapsto 1\}\\
&= \{x \mapsto 2, x \mapsto 1\}
\end{align*}
So with this definition we have avoided the trap of not detecting any side effects when there are two side effects that cancel each other out. Instead we have two side effects here, the last of which happens to restore the valuation of $x$ to its original one.

\section{Side effects in programs}
\label{sec:programs}
If we are going to extend our definition to that of side effects in programs, we are going to have to define how concatenation, union and repetition are handled.

Defining side effects for entire programs is more complicated than defining side effects for single and basic instructions. This is because two composition operators, namely union and repetition, can be non-deterministic. As we have mentioned before, however, we are only interested in (the side effects of) deterministic programs. This leaves us to define how side effects are calculated for the composition operators of deterministic programs. For concatenation, this is trivial. We once again require that the entire program can be evaluated with the given initial valuation. The set of side effects of a program then is the union of the side effects in its basic instructions that are executed given some initial valuation:
\begin{definition}
Let $d\pi = d\pi_{1}; d\pi_{2}$ be a deterministic program. Let model $M$ and initial valuation $g$ be given and let $h$ be the valuation such that $_g\llbracket d\pi \rrbracket_h^M$. Furthermore, let $f$ be the valuation such that $_{g}\llbracket d\pi_{1} \rrbracket_{f}^{M}$. The set of side effects $\mathcal{S}_{g}^{M}(d\pi)$ is defined by:
\begin{equation*}
\mathcal{S}_{g}^{M}(d\pi) = \mathcal{S}_{g}^{M}(d\pi_{1}) \cup \mathcal{S}_{f}^{M}(d\pi_{2})
\end{equation*}
\end{definition}
This works in a similar fashion as the definition of side effects in complex steering fragments. We can return now to our example given in the Introduction of this chapter: $d\pi = \?([ x:=x+1]\top);\?([ x:=x+1]\top)$. The above definition indeed avoids the trap presented there, namely that this program only yields a single side effect. To see this, consider initial valuation $g$ such that $g(x)=0$. We will then get $_{g}\llbracket \?([ x:=x+1]\top)\rrbracket_{f}^{M}$ and therefore $f(x) = 1$, so the set of side effects becomes:
\begin{align*}
\mathcal{S}_{g}^{M}(d\pi) &= \mathcal{S}_{g}^{M}(\?([ x:=x+1]\top)) \cup \mathcal{S}_{f}^{M}(\?([ x:=x+1]\top))\\
&= \{x \mapsto 1\} \cup \{x \mapsto 2\}\\
&= \{x \mapsto 1, x \mapsto 2\}
\end{align*}
Similarly, side effects that cancel each other out, such as in $d\pi = \?([ x:=x+1]\top);\?([ x:=x\monus 1]\top)$ will now perfectly be detected, resulting for the same initial valuation $g$ in a set of side effects $\mathcal{S}_{g}^{M}(d\pi) = \{x \mapsto 1, x \mapsto 0\}$.

Another interesting observation is that the transformation as defined in Proposition \ref{prop:noleftand}, which eliminates occurences of $\leftand$ and $\leftor$ in steering fragments, not only preserves the relational meaning, but also the side effects of such a steering fragment. The programs $?([x:=x+1]\top \leftand [x:=x\monus 1]\top)$ and its transformed version $?([x:=x+1]\top); ?([x:=x\monus 1]\top)$ are an illustration of this: we can easily see that both have the same set of side effects.

With concatenation defined, we can move on to the next composition operators: union and repetition. For this we can use the property that given an initial valuation, every (terminating) deterministic program has a unique canonical form that executes the same basic instructions (see Proposition \ref{prop:canonical} in Chapter \ref{ch:terminology}). This makes the definition of side effects for programs containing a union or repetition straight-forward:

\begin{definition}
Let $d\pi$ be a deterministic program. Let model $M$ and initial valuation $g$ be given and let $h$ be the valuation such that $_g\llbracket d\pi \rrbracket_h^M$. Furthermore, let $d\pi'$ be the deterministic program in canonical form as meant in Proposition \ref{prop:canonical}. The set of side effects $\mathcal{S}_{g}^{M}(d\pi)$ is defined by:
\begin{equation*}
\mathcal{S}_{g}^{M}(d\pi) = \mathcal{S}_{g}^{M}(d\pi')
\end{equation*}
\end{definition}

We can illustrate how this works by returning to our running example, discussed in detail in Section \ref{sec:example}:
\begin{flalign*}
&x := 1;&\\
&\text{IF }(x:=x+1 \leftand x = 2)&\\
&\text{THEN }y:=1&\\
&\text{ELSE } y:=2&
\end{flalign*}
In \dlaf, this translates to the following deterministic program $d\pi$:
\begin{flalign*}
&x:=1;&\\
&(\?([ x:=x+1]\top \leftand x=2);y:=1)&\\
&\cup&\\
&(\?\lnot([ x:=x+1]\top \leftand x=2);y:=2)&
\end{flalign*}
We have already seen that for $g(x) = g(y)=0$, there is a valuation $h$ such that $_g\llbracket d\pi \rrbracket_h^M$ (namely $h = g[x \mapsto 2,y \mapsto 1]$). We can break this program down as follows:
\begin{align*}
d\pi &::= \rho_{1};d\pi{1}\\
\rho_{1} &::= (x:=1)\\
d\pi_{1} &::= (\?\phi_{0}; \rho_{2}) \cup (\?\lnot \phi_{0}; \rho_{3})\\
\rho_{2} &::= (y:=1)\\
\rho_{3} &::= (y:=2)\\
\phi_{0} &::= \varphi_{1} \leftand \varphi_{2}\\
\varphi_{1} &::= [ x:=x+1]\top\\
\varphi_{2} &::= (x=2)
\end{align*}
We want to know the set of side effects in this program. This is determined as follows:
\begin{align*}
\mathcal{S}_{g}^{M}(d\pi) &= \mathcal{S}_{g}^{M}(\rho_{1};d\pi{1})\\
&= \mathcal{S}_{g}^{M}(\rho_{1}) \cup \mathcal{S}_{f}^{M}(d\pi{1})
\end{align*}
where we get $f$ by evaluating $_{g}\llbracket x:=1\rrbracket_{f}^{M}$. Thus, $f = g[x \mapsto 1]$. We can easily see that the first set of side effects $\mathcal{S}_{g}^{M}(\rho_{1}) = \emptyset$. The interesting part is the second set of side effects, since we now have a deterministic program of the form $d\pi_1 = (\?\phi;d\pi_{2}) \cup (\?\lnot\phi; d\pi_{3})$. Here $\phi = \phi_{0}, d\pi_{2} = \rho_{2}$ and $d\pi_{3} = \rho_{3}$.

We now have to ask ourselves what the canonical form of $d\pi_1$ given valuation $f$ is. This is determined by the outcome of the test
\begin{equation*}
?([ x:=x+1]\top \leftand x=2)
\end{equation*}
It is easy to see that this yields true. Thus, the canonical form $d\pi'$ of $d\pi_1$ is
\begin{equation*}
d\pi' = \?\phi_{0};\rho_{2}
\end{equation*}
Therefore according to our definition, for $_f\llbracket \?\phi_{0}\rrbracket_h^{M}$:
\begin{align*}
\mathcal{S}_{f}^{M}(d\pi_{1}) &= \mathcal{S}_{f}^{M}(d\pi')\\
&= \mathcal{S}_{f}^{M}(\?\phi_{0};\rho_{2})\\
&= \mathcal{S}_{f}^{M}(\?\phi_{0}) \cup \mathcal{S}_{h}^{M}(\rho_{2})
\end{align*}
We can once again immediately see that the second set of side effects $\mathcal{S}_{h}^{M}(\rho_{2}) = \emptyset$. The first set of side effects is determined in a similar fashion as in the example in the previous section. In the end, it gives us:
\begin{align*}
\mathcal{S}_{f}^{M}(\?\phi_{1}) &= \mathcal{S}_{f}^{M}(\?([ x:=x+1]\top \leftand (x=2)))\\
&= \mathcal{S}_{f}^{M}(\?([ x:=x+1]\top) \cup \mathcal{S}_{f'}^{M}(\?(x=2)))
\end{align*}
So we again get a union of two sets of side effects, where we get $f'$ by evaluating $_{f}\llbracket [ x:=x+1]\top\rrbracket_{f'}^{M}$. Thus, $f' = f[x \mapsto 2]$. It should be clear by now that the first set of side effects contains one side effect, namely $\{x \mapsto 2\}$, whereas the latter does not contain any side effects. This gives us as final set of side effects:
\begin{align*}
\mathcal{S}_{g}^{M}(d\pi) &= \mathcal{S}_{g}^{M}(\rho_{1}) \cup ( (  \mathcal{S}_{f}^{M}(\?([ x:=x+1]\top) \cup \mathcal{S}_{f'}^{M}(\?(x=2)))   ) \cup \mathcal{S}_{h}^{M}(\rho_{2})   )\\
&= \emptyset \cup ( (\{x \mapsto 2\}  \cup \emptyset  ) \cup \emptyset)\\
&= \{x \mapsto 2\}
\end{align*}
This is exactly the side effect we have come to expect from our running example.

We can now move on to an example of side effects in programs containing a repetition. Recall that repetition is defined as follows:
\begin{equation*}
_{g}\llbracket \pi^{*} \rrbracket_{h}^{M} \text{ iff } g = h \text{ or } _{g}\llbracket \pi;\pi^{*} \rrbracket_{h}^{M}\tag{QDL14}
\end{equation*}
So, $\pi$ either gets executed not at all or at least once. The form of programs we are interested in is
\begin{equation*}
d\pi = (\?\phi;\pi)^{*};\?\lnot\phi
\end{equation*}
In this case there will only ever be exactly one situation in which the program gets evaluated (see Proposition \ref{prop:n} in Chapter \ref{ch:terminology}). Our definition of canonical forms tells us that given an initial valuation $g$ and $n$ as meant in Proposition \ref{prop:n}, the canonical form $d\pi'$ of $d\pi$ is
\begin{equation*}
d\pi' = (\pi_{r})^{n};\?\lnot\phi
\end{equation*}
Using this we get the following set of side effects of a deterministic program of the above form:
\begin{equation*}
\mathcal{S}_{g}^{M}(d\pi) = \mathcal{S}_{g}^{M}((\pi_{r})^{n};\?\lnot\phi)
\end{equation*}
As an example of this, we can return to a slightly modified version of the example we gave in Section \ref{sec:whileinqdla}.
\begin{flalign*}
&x := 0;&\\
&y := 0;&\\
&\text{WHILE }(x:=x+1 \leftand x \leq 3)&\\
&\text{DO }y:= y+ 1&
\end{flalign*}
In \dlaf, this translates to the following deterministic program $d\pi$ given model $M$ and initial valuation $g$ such that $g(x)= g(y)=0$:
\begin{equation*}
d\pi = (\?([ x:=x+1]\top \leftand (x\leq 3));y:=y+1)^{*};
\?\lnot([ x:=x+1]\top \leftand (x\leq 3))
\end{equation*}
Clearly this is a deterministic program in the form we are interested in and there is a valuation $h$ such that $_g\llbracket d\pi\rrbracket_h^M$. In this case we have $\pi_{r} = \?\phi;y:=y+1$ with $\phi = [ x:=x+1]\top \leftand (x\leq 3)$. To get the canonical form $d\pi'$ of $d\pi$, we need to find the iteration $n$ for which $\?\phi$ will succeed, but for which the test will not succeed another time. This will be for $n=3$. After all, after three iterations we will have valuation $g_{3} = g[x \mapsto 3, y \mapsto 3]$. With this valuation, the test $\?([ x:=x+1]\top \leftand (x\leq 3))$ will fail, or to put it formally: $M \not\models_{g_{3}} [ x:=x+1]\top \leftand (x\leq 3)$. This means that we will get the following set of side effects:
\begin{align*}
\mathcal{S}_{g}^{M}(d\pi) &= \mathcal{S}_{g}^{M}(d\pi')\\
&= \mathcal{S}_{g}^{M}((\pi_{r})^{3};\?\lnot\phi)\\
&= \mathcal{S}_{g}^{M}((\pi_{r})^{3}) \cup \mathcal{S}_{g_{3}}^{M}(\?\lnot\phi)\\
&= \mathcal{S}_{g}^{M}(\pi_{r};\pi_{r};\pi_{r}) \cup \mathcal{S}_{g_{3}}^{M}(\?\lnot\phi)\\
&= \{x \mapsto 1, x \mapsto 2, x \mapsto 3\} \cup \{x \mapsto 4\}\\
&= \{x \mapsto 1, x \mapsto 2, x \mapsto 3, x \mapsto 4\}
\end{align*}
Is this the result we would expect? The answer is yes. It is clear that for each time the test is evaluated, a side effect occurs. The test is performed four times: three times it succeeds (after which the program executes the body of its loop) and the fourth time it fails, but not after updating the valuation of $x$. The program evaluates with as final valuation $h= g[x \mapsto 4, y \mapsto 3]$.

\section{Side effects outside steering fragments}
\label{sec:outside}

The keen observer will have noticed by now that under our current definition, side effects can only occur in steering fragments. I have been going through quite some trouble, however, to make my definitions of side effects as general as possible. Even though in this thesis I am only interested in side effects in steering fragments, I am fully aware that views can differ on what the main effect and what the side effect of an instruction is. That may either be a matter of opinion or a matter of necessity, as in different systems, the same instruction may have a side effect in one system and not in the other.

The way my definitions of side effects\footnote{As well as the definitions of classes of side effects presented in Chapter \ref{ch:classification}.} are built up, one need only change the \emph{expected evaluation} of an instruction in order to change if it is viewed as a side effect in a certain context. Consider, for example, the sometimes accepted view that an assignment causes a side effect, no matter where it occurs in a program. This view is for example expressed by Norrish in \cite{Norrish}. The only change we would need to make to our system to incorporate that view is a change to the expected evaluation of the assignment, which would then become:
\begin{equation*}
_{g}\llbracket v:=t \rrbracket_{h}^{M,\mathcal{E}} \text{ iff } g=h
\end{equation*}
The consequence of this in our current setting would be that the expected evaluation of every program always has a resulting valuation $h$ that is equal to the initial valuation $g$, since only assignments can make changes to a valuation currently and by the above definition we do not expect any assignment to do so, wherever it occurs in the program. As a consequence, any change to the valuation (caused by the actual evaluation) will automatically be a side effect.

It is almost as simple to add new instructions to our setting. I definitely do not want to claim that the instructions I have defined in \dlaf\ are exhaustive, so this need may arise. If we were, for instance, to re-introduce the random assignment $v:=\?$, all we would have to do was to define the actual and expected evaluation of this. The actual evaluation is already given by Harel in \cite{Harel} and Van Eijck in \cite{EijckStokhof}:
\begin{equation*}
_{g}\llbracket v:=\?\rrbracket_{h}^{M} \text{ iff } g \sim_{v} h
\end{equation*}
If we also would want to allow random assignments in tests, we would have to add a rule for that as well, similar to the one already in place for normal assignments:
\begin{equation*}
M \models_{g} [ v:=\?]\top \text{ iff } _{g}\llbracket v:=\?\rrbracket_{h}^{M}
\end{equation*}
The definition of the expected evaluation is dictated by what we really expect the random assignment to do. This can be the same as what it actually does, in which case we have to define the expected evaluation to be the same as the actual evaluation above:
\begin{align*}
_{g}\llbracket v:=\?\rrbracket_{h}^{M,\mathcal{E}} &\text{ iff } _{g}\llbracket v:=\?\rrbracket_{h}^{M}\\
M \models_{g}^{\mathcal{E}} [ v:=\?]\top &\text{ iff } M \models_{g} [ v:=\?]\top
\end{align*}
If we expect random assignments to do something different, all we have to do is define the expected evaluation accordingly. This expected evaluation can literally be anything: from simply not updating the valuation at all to always setting a completely unrelated variable to 42:
\begin{equation*}
_{g}\llbracket v:=\?\rrbracket_{h}^{M,\mathcal{E}} \text{ iff } h = g[\text{\emph{the answer to life, the universe and everything}} \mapsto 42] 
\end{equation*}
On a side note, this example poses some interesting questions about `negative' side effects. Under our current definition, setting the above mentioned variable to 42 registers as a side effect, but in a somewhat strange fashion. After all $v:=\?$ is a single instruction and for $_g\llbracket \rho\rrbracket_h^M$ and $_g\llbracket \rho\rrbracket_{h'}^{M,\mathcal{E}}$, $\mathcal{S}_{g}^{M}(\rho) = \delta(h',h)$. There will actually be two differences between valuations $h'$ and $h$ here: the actual evaluation updates variable $v$, whereas the expected evaluation leaves $v$ alone but does update the variable \emph{the answer to life, the universe and everything}. Both variables will show up in the set of side effects, both with the assignment the actual evaluation has assigned to them.

This fails to capture what has actually happened here: after all, not only did an unexpected change to the initial valuation happen (a `regular' side effect), but an expected change also did \emph{not} happen (a `negative' side effect). At least part of the information what should have happened is lost, namely the value the variable \emph{the answer to life, the universe and everything} was supposed to get.\footnote{Which is quite a shame, considering the trouble it cost to get it.} It is an open question if we should even allow these somewhat odd situations where the actual evaluation does something completely different than we expect, thereby generating a negative side effect. We leave this question, as well as the question how we should handle these situations if we do choose to allow them, for future work.

\chapter{A classification of side effects}
\label{ch:classification}

\section{Introduction}
In this chapter we will take a closer look at side effects in steering fragments. In particular, we will give a classification of side effects. This classification gives us a measure of the impact of a side effect.

As we have already mentioned in our introduction in Chapter \ref{ch:introduction}, Bergstra has given an informal classification of side effects in \cite{SF}. Bergstra makes a distinction between steering instructions and working instructions. This distinction is based on a setting called Program Algebra (PGA). In PGA, there is no distinction between formulas and single instructions other than formulas, which is why the proposed distinction by Bergstra is meaningful in that setting. Every basic instruction $a$ in PGA yields a Boolean reply upon execution and can therefore be made into a positive or negative test instruction $+a$ or $-a$.  Naturally, this cannot be done in our setting of \dlaf, so instead of giving an overview of Bergstra's paper, I will just present the major classes of side effects Bergstra distinguishes and what they come down to in our setting.

Bergstra's first class of side effects is what he calls `trivial side effects'. By this he means side effects that can only be found in e.g. consequences for the length of the program or its running time. We are usually not interested in those kinds of side effects, which is exactly why Bergstra calls them trivial and why we would say that no side effects occur at all. An instruction that only returns a meaningful Boolean reply (that is, a Boolean reply that may differ depending on the valuation the instruction is evaluated in) is an instruction that only has trivial side effects. Examples of such instructions are the comparision instructions such as ($x=2$) or ($x \leq 2$). These instructions can be turned into meaningful test instructions by prefixing them with a $+$ or $-$ symbol. We will return to this in our explanation of PGA in Chapter \ref{ch:pga}. In our terms, these kinds of instructions can only be formulas, occuring in steering fragments such as $\?(x=2)$ or $\?(x \leq 2)$. To be precise, they can only be formulas that have the same actual and expected evaluation, and thus no side effects.

The above described situation, where only trivial side effects occur, is one extreme. The other extreme is when an instruction always yields the same Boolean reply, regardless of when it is executed. Bergstra says that in that case, only `trivial Boolean results' occur and that the instruction should be classified as a working instruction (that is, a single instruction not being a formula). In our setting this is also true with one notable exception: that of assignments. As we know, assignments always return true, so their Boolean result is trivial. Still, we allow them in formulas, too. If an instruction with trivial Boolean results occurs outside a formula, its only relevance would be its effect other than the Boolean reply, in which case you can hardly call that effect a side effect. If it occurs in a formula, however, the Boolean result --- albeit trivial --- does have relevance, so the effect other than the Boolean reply can indeed be called a side effect. This is exactly what happens in our setting.

What the classification between steering instructions and working instructions gives us in the end, is a recommendation on how to use a particular kind of instruction. Instructions such as comparision ($x \leq 2$), that only give a Boolean reply, have no meaning as a working instruction and therefore ideally should only occur in steering fragments. Other instructions such as assignment ($x := 2$) can be both steering instructions as well as working instructions and can thus occur both inside as well as outside steering fragments. Finally, instructions such as writing to the screen (\verb$write x$) do not return a meaningful Boolean reply and should therefore ideally not occur in steering fragments.

\section{Marginal side effects}
\subsection{Introduction}
Having seen the base class of side effects, we can move on to the next level, that of \emph{marginal side effects}. The intuition behind a marginal side effect is fairly simple: the side effect of a single instruction is marginal if the remainder of the execution of the program is unaffected by the occurrence of the side effect. The following program is a typical example of one where a marginal side effect occurs:
\begin{equation*}
d\pi = d\pi_{1};\?([ x:=x+1]\top);y:=1
\end{equation*}
Here $d\pi_{1}$ can be any (deterministic) program. The side effect occurs in the test. However, since the variable $x$ is no longer used in the remainder of the program (which only consists of the single instruction $y:=1$), the remainder of the program is unaffected by the occurrence of the side effect. Therefore, this side effect is marginal.

So what if $x$ does occur in the remainder of the program, for example in this program:
\begin{equation*}
d\pi = d\pi_{1};\?([ x:=x+1]\top);x:=x+1
\end{equation*}
This is a typical example of a program in which the occuring side effect is not marginal. The reason is that the assignment in the remainder of the program ($x:=x+1$) has a different effect on the variable $x$ than when it would have had if the side effect had not occurred. For instance, for initial valuation $g$ such that $g(x) = 1$ (and assuming $x$ does not occur in $\pi_{1}$), the assignment maps $x$ to $3$. If the side effect had not occurred, it would have had a different effect on $x$ (namely, it would have mapped it to $2$).

Another typical example of a program in which an occuring side effect is not marginal is our running example:
\begin{equation*}
d\pi = d\pi_{1};\?([ x:=x+1]\top \leftand (x=2));y:=1
\end{equation*}
Here $d\pi_{1}$ can again be any deterministic program and the side effect occurs in the same place as in our first example. However, the test is now a complex test and in the second part of the test, $x$ is used. Suppose the valuation after evaluation of $d\pi_{1}$ is $f$ such that $f(x) = 1, f(y) = 2$. The second part of the test ($x=2$) will now give a different reply if a side effect does not occur in the first part (or if that side effect would have affected a different variable). As a result, the remainder of the program is affected by the side effect: it will be executed differently if a side effect occurs.

Perhaps the answer to the question if the side effect is marginal is less clear when the initial valuation in the previous example would not have been $g$ with $g(x) =1$, but for example with $g(x)= 42$. It is still the case that the variable $x$, that is affected by a side effect, is used again in the remainder of the program, but now it does not change the outcome of the (complex) test. Is that side effect still not marginal then? The same question can be posed about the following example:
\begin{equation*}
d\pi = d\pi_{1};\?([ x:=x+1]\top);x:=42
\end{equation*}
Regardless of initial valuation $g$, at the end of this program (assuming $d\pi_{1}$ terminates), $x$ will always be mapped to 42. So is the side effect in the test marginal or not? The answer can be found by checking if the remainder of the program is executed in the same way, or more formally: if the actual update of the remainder of the program is the same regardless of whether a side effect has occurred. In both our last examples, the answer to that last question is yes. After all, in the first example the test $x=2$ will fail whether $x$ has been incremented first or not, and in the second example $x$ will always be mapped to $42$, again regardless of the side effect that incremented $x$ earlier. Therefore, the side effects in the discussed instructions are marginal.

\subsection{Marginal side effects in single instructions}
Although the intuition of marginal side effects should be clear enough by now, formally defining it is tricky because we have to define precisely what the remainder of a (deterministic) program $d\pi$ given a single instruction $\rho$ and an initial valuation $g$ is. Before we can define that, we also need to know the \emph{history} of that same program given single instruction $\rho$, which is loosely described as those (single or basic) instructions that have already been evaluated when $\rho$ is about to get evaluated.

In what follows we are going to assume that in a certain deterministic program $d\pi$ a single instruction $\rho$ occurs that is causing a side effect. Furthermore, we are going to use that given initial valuation $g$, any deterministic program has a unique canonical form that has the same behavior (see Proposition \ref{prop:canonical} in Chapter \ref{ch:terminology}). Defining the history and remainder of a deterministic program is straight-forward if that program is in canonical form. Also, we can actually immediately give a more general definition than what we need here, namely the history and remainder of a deterministic program given a basic instruction. This extra generality will come in handy later on.
\begin{definition}
Let $d\pi$ be a deterministic program in canonical form. Let model $M$ and initial valuation $g$ be given and let $h$ be the valuation such that $_g\llbracket d\pi\rrbracket_h^M$. Let $\varpi$ be a basic instruction occuring in $d\pi$, that is, $d\pi$ is of the form $d\pi_{1};\varpi;d\pi_{2}$, with $d\pi_{1}$ and $d\pi_{2}$ being possibly empty deterministic programs in canonical form. The \textbf{history} of program $d\pi$ given basic instruction $\varpi$ is defined as:
\begin{align*}
\mathcal{H}_{g}^{M}(d\pi,\varpi) =
\begin{cases}
\iq\top &\text{ if } d\pi_{1} \text{ is empty}\\
d\pi_{1} &\text{ o.w.}
\end{cases}
\end{align*}
The \textbf{remainder} of program $d\pi$ given basic instruction $\varpi$ is defined as:
\begin{align*}
\mathcal{R}_{g}^{M}(d\pi,\varpi) =
\begin{cases}
\iq\top &\text{ if } d\pi_{2} \text{ is empty}\\
d\pi_{2} &\text{ o.w.}
\end{cases}
\end{align*}
\end{definition}

Using Proposition \ref{prop:canonical} the extension of the definitions of history and remainder of a program to all deterministic programs (not just the ones in canonical form) is trivial:
\begin{definition}
\label{def:historybasic}
Let $d\pi$ be a deterministic program. Let model $M$ and initial valuation $g$ be given and let $h$ be the valuation such that $_g\llbracket d\pi\rrbracket_h^M$. Furthermore, let $d\pi'$ be the deterministic program in canonical form as meant in Proposition \ref{prop:canonical}. The \textbf{history} of program $d\pi$ given basic instruction $\varpi$ is defined as:
\begin{equation*}
\mathcal{H}_{g}^{M}(d\pi,\varpi) = \mathcal{H}_{g}^{M}(d\pi',\varpi)
\end{equation*}
The \textbf{remainder} of program $d\pi$ given basic instruction $\varpi$ is defined as:
\begin{equation*}
\mathcal{R}_{g}^{M}(d\pi,\varpi) = \mathcal{R}_{g}^{M}(d\pi',\varpi)
\end{equation*}
\end{definition}

With definitions for the history and the remainder of a program in hand, we can define marginal side effects. According to our intuition, a side effect should be marginal if the evaluation of the remainder of the program is the same regardless of whether the side effect occurred. We can tell if that is the case by evaluating the remainder of the program with two different valuations: one in which the single instruction in which the side effect occurs has been evaluated using the actual evaluation, and one in which is has been evaluated using the expected evaluation.\footnote{We now need to restrict ourselves again to single instructions because the expected evaluation is (currently) undefined for complex steering fragments.} If the only difference between those two valuations is exactly the side effect that occurred in the single instruction, or if there is no difference between those two valuations at all, then we can say that the evaluation of the remainder of the program has been the same. This is formally defined as follows:

\begin{definition}
Let $d\pi$ be a deterministic program. Let model $M$ and initial valuation $g$ be given and let $h_{A}$ be the valuation such that $_g\llbracket d\pi\rrbracket_{h_A}^M$. Let $\rho$ be a single instruction in program $d\pi$ causing a side effect, that is, for $_{g}\llbracket \mathcal{H}_{g}^{M}(d\pi,\rho)\rrbracket_{f}^{M}$, $\mathcal{S}_{f}^{M}(\rho) \neq \emptyset$. Let $f_{A}$ be the valuation such that $_{f}\llbracket\rho\rrbracket_{f_{A}}^{M}$ and let $f_{E}$ be the valuation such that $_{f}\llbracket\rho\rrbracket_{f_{E}}^{M,\mathcal{E}}$. The side effect in $\rho$ is marginal iff for $_{f_{A}}\llbracket \mathcal{R}_{g}^{M}(d\pi,\rho)\rrbracket_{h_{A}}^{M}$
\begin{equation*}
\exists h_{E} \text{ s.th. } _{f_{E}}\llbracket \mathcal{R}_{g}^{M}(d\pi,\rho)\rrbracket_{h_{E}}^{M,\mathcal{E}} \text{ and } \delta^{M}(h_{E},h_{A}) = (\mathcal{S}_{f}^M(\rho) \text{ or } \emptyset)
\end{equation*}
\end{definition}

So what happens here exactly? To show this, we return to the examples we have given earlier in this section. First, consider the program $d\pi = x:=1;\?([ x:=x+1]\top);y:=1$, with initial valuation $g$ such that $g(x) = g(y)=0$. We can observe that $d\pi$ is in canonical form. In this program, a side effect occurs in the single instruction $\rho = \?([ x:=x+1]\top)$. So is this side effect marginal or not? Here we have the following:
\begin{align*}
\mathcal{H}_{g}^{M}(d\pi,\rho) &= (x:=1)\\
\mathcal{R}_{g}^{M}(d\pi,\rho) &= (y:=1)\\
f &= g[x \mapsto 1,y \mapsto 0]\\
f_{A} &= f[x \mapsto 2, y \mapsto 0]\\
f_{E} &= f[x \mapsto 1, y \mapsto 0]\\
h_{A} &= f_{A}[x \mapsto 2, y \mapsto 1]\\
h_{E} &= f_{E}[x \mapsto 1, y \mapsto 1]
\end{align*}
As we can see, the valuations $f$ and $f_{E}$ are the same. Using our current definition of the expected evaluation, this will always be the case, so we could just use valuation $f$ here. However, as I have said in Section \ref{sec:outside} of Chapter \ref{ch:treatment}, I want to keep generality in the definitions of side effects. We might want to change the definition of the expected evaluation in the future or add new instructions or connectives that do modify the initial valuation. Therefore, we use valuation $f_{E}$, the resulting valuation after evaluating the single instruction $\rho$ with the expected evaluation.

To determine if the side effects are marginal, we have to ask ourselves if
\begin{equation*}
\delta^{M}(h_{E},h_{A}) = \mathcal{S}_{f}^M(\rho) \text{ or } \emptyset
\end{equation*}
We know how to calculate the set of side effects; it is $\{x \mapsto 2\}$. In this case, $\delta^{M}(h_{E},h_{A})$ is $\{x \mapsto 2\}$ too, so the side effect occurring in $\rho$ is marginal, which is what we want. We can also clearly see in this case that it is no coincidence that we are testing $\delta^{M}(h_{E},h_{A})$ and not $\delta^{M}(h_{A},h_{E})$: we need the valuation that is the result of evaluating the single instruction using the actual evaluation in order to properly compare this with the set of side effects.

We can now take a look at an example in which the side effect should not be marginal. Consider the program $d\pi = x:=1;\?([ x:=x+1]\top);x:=x+1$, with initial valuation $g$ such that $g(x) = 0$. This program is in canonical form too and the side effect occurs in the same single instruction $\rho$. This time we get the following:
\begin{align*}
\mathcal{H}_{g}^{M}(d\pi,\rho) &= (x:=1)\\
\mathcal{R}_{g}^{M}(d\pi,\rho) &= (x:=x+1)\\
f &= g[x \mapsto 1]\\
f_{A} &= f[x \mapsto 2]\\
f_{E} &= f[x \mapsto 1]\\
h_{A} &= f_{A}[x \mapsto 3]\\
h_{E} &= f_{E}[x \mapsto 1]
\end{align*}
We have the same set of side effects: $\{x \mapsto 2\}$. However, $\delta^{M}(h_{E},h_{A})$ now is $\{x \mapsto 3\}$. Therefore the side effect is not marginal, which is again what we would expect.

We have given a third example which closely resembles the ones we have discussed above, namely $d\pi = x:=1;\?([ x:=x+1]\top);x:=42$. If we take the same initial valuation $g$ as above, everything except the remainder of the program given $\rho$ will be the same:
\begin{align*}
\mathcal{H}_{g}^{M}(d\pi,\rho) &= (x:=1)\\
\mathcal{R}_{g}^{M}(d\pi,\rho) &= (x:=42)\\
f &= g[x \mapsto 1]\\
f_{A} &= f[x \mapsto 2]\\
f_{E} &= f[x \mapsto 1]\\
h_{A} &= f_{A}[x \mapsto 42]\\
h_{E} &= f_{E}[x \mapsto 42]
\end{align*}
With this example we can see why our definition of marginal side effects allows the difference between $h_{A}$ and $h_{E}$ to be $\emptyset$, too. We have seen before that in situations like these, the side effects should be marginal, and by allowing the difference to be $\emptyset$, that indeed is the case.

\subsection{Marginal side effects caused by primitive formulas}
As we have seen, our current definition of marginal side effects is capable of determining whether a side effect occurring in a single instruction is marginal or not. We still have to define marginal side effects for basic instructions. In particular, we need to have a definition for the situation in which a primitive formula in a complex test causes a side effect\footnote{We say that a primitive formula causes a side effect here because a side effect cannot occur \emph{in} a primitive formula. It can, however, occur in a single or basic instruction which tests that formula.} and in that same test, the variable affected by that side effect is used again, such as in the following program: $d\pi = d\pi_{1};\?([ x:=x+1]\top \leftand (x=2));y:=1$. In order to define how to determine if a side effect is marginal or not in these situations, we need to extend our definitions of the history and remainder of a program such that it not only works given a single instruction, but also given a primitive formula. Before we can give that definition, we first need to define the history and remainder of a compound formula given a primitive formula. We are once again only interested in those two concepts if the primitive formula $\varphi$ gets evaluated. 

To get an idea of what the history and the remainder of a compound formula given a primitive formula should be, consider the following example:
\begin{align*}
\varphi &= [x:=6]\top\\
\phi &= \neg\varphi \leftor (x \leq 10)\\
&=\neg([x:=6]\top) \leftor (x \leq 10)
\end{align*}
In this example, the history of $\phi$ given $\varphi$ and given model $M$ and initial valuation $g$ is empty. The remainder, however, is not:
\begin{equation*}
\mathcal{R}_g(\phi,\varphi) = x \leq 10
\end{equation*}
Notice that this remainder should be empty if $\lnot\varphi$ would have been true.

The history of a formula of course is not always empty. To illustrate that, we will first introduce a notational convention.
\begin{notation}
We will write $\phi(\underline{\varphi})$ to refer to the primitive formula $\varphi$ occurring in formula $\phi$ at a specific position.
\end{notation}
As an example of this, compare the formulas $\phi_{1}(\underline{\varphi}) = \underline{\varphi} \leftand \varphi$ and $\phi_{2}(\underline{\varphi}) = \varphi \leftand \underline{\varphi}$. The difference between the formulas $\phi_1(\underline{\varphi})$ and $\phi_2(\underline{\varphi})$ is in the instance of primitive formula $\varphi$ we are referring to.

Let $\varphi = [x:=6]\top$ and $\phi(\underline{\varphi})$ as in the example above. Now consider the following example:
\begin{equation*}
\psi(\underline{\varphi}) = (x=2\leftand \phi(\underline{\varphi}))
\end{equation*}
Here the history of $\psi$ given $\underline{\varphi}$ and given model M and initial valuation $g$ such that $g(x) = 2$ is not empty:
\begin{equation*}
\mathcal{H}_g^M(\psi,\underline{\varphi}) = (x=2)
\end{equation*}
Now that we have given an intuition what the history and remainder of a formula given a primitive formula and an initial valuation are going to be, we can move on to giving the actual definitions. In what follows we will assume that the $\phi$ in $\mathcal{H}_f(\phi,\underline{\varphi})$ is in normal form and that the specific primitive formula $\underline{\varphi}$ actually appears exactly once in formula $\phi(\underline{\varphi})$ (although other instances of $\varphi$ may occur in the formula). $\phi(\underline{\varphi})$ can take the following forms:
\begin{equation*}
\varphi(\underline{\varphi}),\quad
\neg\varphi(\underline{\varphi}),\quad
\phi_1(\underline{\varphi})\leftor\phi_2,\quad
\phi_1\leftor \phi_2(\underline{\varphi}),\quad
\phi_1(\underline{\varphi})\leftand\phi_2,\quad
\phi_1\leftand \phi_2(\underline{\varphi})
\end{equation*}
Here $\varphi(\underline{\varphi})$ is the same as $\varphi$. For each of these forms, we will have to define how the history and the remainder is calculated.

\begin{definition}
\label{def:historyremainderformula}
Let $\phi$ be a formula of one of the above forms. Let model $M$ and initial valuation $g$ be given. Let $\underline{\varphi}$ be a primitive formula occurring in $\phi$ such that $\underline{\varphi}$ gets evaluated during the evaluation of $\phi$ given initial valuation $g$. The \textbf{history} of formula $\phi$ given primitive formula $\underline{\varphi}$ is defined as:
\begin{align*}
\mathcal{H}_{g}^{M}(\varphi(\underline{\varphi}),\underline{\varphi})&=\top\\
\mathcal{H}_{g}^{M}(\neg\varphi(\underline{\varphi}),\underline{\varphi})&=\top\\
\mathcal{H}_{g}^{M}(\phi_1(\underline{\varphi})\leftor\phi_2,\underline{\varphi})&=\mathcal{H}_{g}^{M}(\phi_1(\underline{\varphi}),\underline{\varphi})\\
\mathcal{H}_{g}^{M}(\phi_1\leftor \phi_2(\underline{\varphi}),\underline{\varphi})&=\phi_{1} \leftor \mathcal{H}_{g}^{M}(\phi_2(\underline{\varphi}),\underline{\varphi})\\
\mathcal{H}_{g}^{M}(\phi_1(\underline{\varphi})\leftand\phi_2,\underline{\varphi})&=\mathcal{H}_{g}^{M}(\phi_1(\underline{\varphi}),\underline{\varphi})\\
\mathcal{H}_{g}^{M}(\phi_1\leftand \phi_2(\underline{\varphi}),\underline{\varphi})&=\phi_{1} \leftand \mathcal{H}_{g}^{M}(\phi_2(\underline{\varphi}),\underline{\varphi})
\end{align*}
The \textbf{remainder} of formula $\phi$ given primitive formula $\underline{\varphi}$ is defined as:
\begin{align*}
\mathcal{R}_{g}^{M}(\varphi(\underline{\varphi}),\underline{\varphi})&=\top\\
\mathcal{R}_{g}^{M}(\neg\varphi(\underline{\varphi}),\underline{\varphi})&=\top\\
\mathcal{R}_{g}^{M}(\phi_1(\underline{\varphi})\leftor\phi_2,\underline{\varphi})&=\mathcal{R}_{g}^{M}(\phi_1(\underline{\varphi}),\underline{\varphi}) \leftor \phi_2\\
\mathcal{R}_{g}^{M}(\phi_1\leftor \phi_2(\underline{\varphi}),\underline{\varphi})&=\mathcal{R}_{g}^{M}(\phi_2(\underline{\varphi}),\underline{\varphi})\\
\mathcal{R}_{g}^{M}(\phi_1(\underline{\varphi})\leftand\phi_2,\underline{\varphi})&=\mathcal{R}_{g}^{M}(\phi_1(\underline{\varphi}),\underline{\varphi}) \leftand \phi_2\\
\mathcal{R}_{g}^{M}(\phi_1\leftand \phi_2(\underline{\varphi}),\underline{\varphi})&=\mathcal{R}_{g}^{M}(\phi_2(\underline{\varphi}),\underline{\varphi})
\end{align*}
\end{definition}
The reason we are only interested in the history and remainder of a primitive formula if that formula is actually evaluated, is straight-forward: we use these definitions to calculate the side effects caused by that primitive formula and those side effects only exist if the primitive formula is evaluated. As straight-forward as this is, the restriction is an important one. Because we know that $\underline{\varphi}$ gets evaluated (not be be confused with `yielding true'), we do not have to take potentially troublesome formulas into account such as $\bot \leftand \underline{\varphi}$.

The above definitions make the history and remainder of a formula given a primitive formula, partial functions. To see in which situations the history and remainder are defined and for which they are not, consider the following formula:
\begin{equation*}
\phi = (x=5 \leftand [x:=x+1]\top) \leftor [x:=x+2]\top
\end{equation*}
Now assume we want to know the history of $\phi$ given $\varphi = [x:=x+1]\top$. This history $\mathcal{H}_g^M(\phi(\underline{\varphi}),\underline{\varphi})$ is only defined if $[x:=x+1]\top$ gets evaluated, which in turn only is the case if we have a initial valuation $g$ such that $g(x) = 5$. For all initial valuations $g'$ such that $g(x)\neq 5$, the history of $\phi$ given $\varphi$ is undefined. If we would be interested in the history of $\phi$ given $\varphi' = [x:=x+2]\top$, the situation would be reversed: in that case the history $\mathcal{H}_g^M(\phi(\underline{\varphi'}),\underline{\varphi'})$ is only undefined with initial valuation $g$ such that $g(x) = 5$.

That the history (and the remainder) is undefined in these cases is not problematic because as said, we are going to use these definitions to check if the side effects caused by $\underline{\varphi}$ are marginal and $\underline{\varphi}$ can only cause side effects if it gets evaluated.

Using these definitions, we can move on to define the history and remainder of a program given a primitive formula:

\begin{definition}
\label{def:historyformula}
Let $d\pi$ be a deterministic program in canonical form. Let model $M$ and initial valuation $g$ be given and let $h$ be the valuation such that $_g\llbracket d\pi\rrbracket_{h}^M$. Let $\iq\phi$ be a test occurring in program $d\pi$, where $\phi$ is a formula in normal form. Finally, let $\underline{\varphi}$ be a primitive formula occuring in $\phi$ such that $\underline{\varphi}$ gets evaluated during the evaluation of $\phi$ given initial valuation $g$. The \textbf{history} of program $d\pi$ given primitive formula $\underline{\varphi}$ is, for $_{g}\llbracket \iq\mathcal{H}_{g}^{M}(d\pi,\iq\phi) \rrbracket_{f}^{M}$, defined as:
\begin{equation*}
\mathcal{H}_{g}^{M}(d\pi,\underline{\varphi}) = \mathcal{H}_{g}^{M}(d\pi,\iq\phi);\iq\mathcal{H}_{f}^{M}(\phi(\underline{\varphi}),\underline{\varphi})
\end{equation*}
The \textbf{remainder} of program $d\pi$ given primitive formula $\underline{\varphi}$ is defined as:
\begin{equation*}
\mathcal{R}_{g}^{M}(d\pi,\underline{\varphi}) = \iq\mathcal{R}_{f}^{M}(\phi(\underline{\varphi}),\underline{\varphi}); \mathcal{R}_{g}^{M}(d\pi,\iq\phi)
\end{equation*}
\end{definition}

The final step is to give a definition to determine if a side effect occurring in a primitive formula is marginal. Given the above, this definition should not be surprising:

\begin{definition}
Let $d\pi$ be a deterministic program. Let model $M$ and initial valuation $g$ be given and let $h_A$ be the valuation such that $_g\llbracket d\pi\rrbracket_{h_A}^M$. Let $\underline{\varphi}$ be a primitive formula in program $d\pi$ causing one of the side effects of $d\pi$. Let $f$ be the valuation such that $_{g}\llbracket \mathcal{H}_{g}^{M}(d\pi,\underline{\varphi})\rrbracket_{f}^{M}$. Let $f_{A}$ be the valuation such that $_{f}\llbracket\iq\varphi\rrbracket_{f_{A}}^{M}$ or $_{f}\llbracket\iq\lnot\varphi\rrbracket_{f_{A}}^{M}$ and let $f_{E}$ be the valuation such that $_{f}\llbracket\iq\varphi\rrbracket_{f_{E}}^{M,\mathcal{E}}$ or $_{f}\llbracket\iq\lnot\varphi\rrbracket_{f_{E}}^{M,\mathcal{E}}$.\footnote{This distinction is necessary because we can only evaluate a test if its argument yields true. $M \models_f \varphi$ might actually yield false if $\varphi$ is part of a larger formula $\phi$ that despite that yields true, such as $\phi = \varphi \leftor \phi_1$ such that $M \models_{f_A} \phi_1$. Thus, we need either $\varphi$ or $\lnot\varphi$.} The side effect caused by $\underline{\varphi}$ is marginal iff for $_{f_{A}}\llbracket \mathcal{R}_{g}^{M}(d\pi,\underline{\varphi}) \rrbracket_{h_{A}}^{M}$
\begin{equation*}
\exists h_{E} \text{ s.th. } _{f_{E}}\llbracket \iq\mathcal{R}_{g}^{M}(d\pi,\underline{\varphi})\rrbracket_{h_{E}}^{M,\mathcal{E}} \text{ and }\delta^{M}(h_{E},h) = (\mathcal{S}_f^M(\iq\underline{\varphi}) \text{ or } \emptyset)
\end{equation*}
\end{definition}

To show how this works, we return to the example given in the beginning of this section: $d\pi = x:=1;\?([ x:=x+1]\top \leftand (x=2));y:=1$, with initial valuation $g$ such that $g(x)= g(y)=0$. Here the primitive formula $\varphi = [ x:=x+1]\top$ causes a side effect. We can now use our definition to find out if that side effect is marginal. For that, we first need the history of $d\pi$ given primitive formula $\underline{\varphi}$. To calculate $\mathcal{H}_{g}^{M}(d\pi,\underline{\varphi})$, we first observe that $\phi$ is in normal form. This gives us a go to use Definition \ref{def:historyformula}. This definition tells us to first calculate valuation $f$, which we get by evaluating $_{g}\llbracket \mathcal{H}_{g}^{M}(d\pi,\?\phi)\rrbracket_{f}^{M}$. Here $\?\phi$ is a basic instruction, so we can use Definition \ref{def:historybasic} to calculate it. We have seen before how that evaluates:
\begin{equation*}
\mathcal{H}_{g}^{M}(d\pi,\?\phi) = (x:=1)
\end{equation*}
Thus we get $_{g}\llbracket x:=1\rrbracket_{f}^{M}$, so $f = g[x \mapsto 1, y \mapsto 0]$.

All we need to do now to get the history we are looking for, is the history of formula $\phi$ given primitive formula $\underline{\varphi}$: $\mathcal{H}_{f}^{M}(\phi(\underline{\varphi}),\underline{\varphi})$. We can use Definition \ref{def:historyremainderformula} here and are in the situation where $\phi(\underline{\varphi}) = \phi_{1}(\underline{\varphi}) \leftand \phi_{2}$. Here $\phi_{1} = \varphi$ and $\phi_{2} = (x=2)$, so as history we get:
\begin{align*}
\mathcal{H}_{f}^{M}(\phi(\underline{\varphi}),\underline{\varphi}) &= \mathcal{H}_{f}^{M}(\phi_{1}(\underline{\varphi}) \leftand \phi_{2},\underline{\varphi})\\
&= \mathcal{H}_{f}^{M}(\phi_1(\underline{\varphi}),\underline{\varphi})\\
&= \mathcal{H}_{f}^{M}(\varphi(\underline{\varphi})),\underline{\varphi})\\
&= \top
\end{align*}
Thus, the history of program $d\pi$ given primitive formula $\underline{\varphi}$ is:
\begin{align*}
\mathcal{H}_{g}^{M}(d\pi,\underline{\varphi}) &= \mathcal{H}_{g}^{M}(d\pi,\?\phi);\?\mathcal{H}_{f}^{M}(\phi(\underline{\varphi}),\underline{\varphi})\\
&= (x:=1);\?\top
\end{align*}
With the information above we can also immediately calculate the remainder of formula $\phi$ given primitive formula $\underline{\varphi}$:
\begin{align*}
\mathcal{R}_{f}^{M}(\phi(\underline{\varphi}),\underline{\varphi}) &= \mathcal{R}_{f}^{M}(\phi_{1}(\underline{\varphi}) \leftand \phi_{2},\underline{\varphi})\\
&= \mathcal{R}_{f}^{M}(\phi_1(\underline{\varphi}),\underline{\varphi}) \leftand \phi_2\\
&= \mathcal{R}_{f}^{M}(\varphi(\underline{\varphi}),\underline{\varphi}) \leftand \phi_2\\
&= \top \leftand (x=2)
\end{align*}
Then all we need to determine the remainder of program $d\pi$ given primitive formula $\underline{\varphi}$ is the remainder of program $d\pi$ given basic instruction $\?\phi$. To see how this evaluates, see the previous section. We can use Definition \ref{def:historybasic} for this again and get:
\begin{align*}
f_{A} &= f[x \mapsto 2,y \mapsto 0]\\
\mathcal{R}_{f_{A}}^{M}(d\pi,\underline{\varphi}) &= (y:=1)
\end{align*}
So the remainder of program $d\pi$ given primitive formula $\underline{\varphi}$ is:
\begin{align*}
\mathcal{R}_{g}^{M}(d\pi,\underline{\varphi}) &= \?\mathcal{R}_{f}^{M}(\phi(\underline{\varphi}),\underline{\varphi}); \mathcal{R}_{f_{A}}^{M}(d\pi,?\phi)\\
&= \?(\top \leftand (x=2));(y:=1)
\end{align*}
Now that we have the history and the remainder of $d\pi$ given $\underline{\varphi}$, we can finally determine if the side effect occurring in $\underline{\varphi}$ is marginal. To quickly recap, we have:
\begin{align*}
\mathcal{H}_{g}^{M}(d\pi,\underline{\varphi}) &= (x:=1);?\top\\ 
\mathcal{R}_{g}^{M}(d\pi,\underline{\varphi}) &= \?(\top \leftand (x=2));(y:=1)\\
f &= g[x \mapsto 1, y \mapsto 0]\\
f_{A} &= f[x \mapsto 2, y \mapsto 0]\\
f_{E} &= f[x \mapsto 1, y \mapsto 0]\\
h_{A} &= f_{A}[x \mapsto 2, y \mapsto 1]\\
h_{E} & \text{ does not exist}
\end{align*}
Here we have an example where we do not even have to determine if $\delta^{M}(h_{E},h_{A})$ is the same as $\mathcal{S}_f^{M}(\?\underline{\varphi})$, because there is no valuation $h_{E}$ such that
\begin{equation*}
_{f_{E}}\llbracket \mathcal{R}_{g}^{M}(d\pi,\underline{\varphi})\rrbracket_{h_{E}}^{M,\mathcal{E}}
\end{equation*}
This is because for valuation $f_{E}$ the test $\?(\top \leftand (x=2))$ will fail. Therefore, the side effect in $\underline{\varphi}$ is `automatically' not marginal, which is indeed what we wanted.

\section{Other classes of side effects}
There are two more classes of side effects that I want to discuss. The first is the class \emph{detectible side effects}. According to Bergstra, a side effect in an instruction is detectible if the fact that that side effect has occured can be measured by means of a steering fragment containing that instruction \cite{SF}. This is the most general class of side effects: in my terms, any difference between the actual and the expected evaluation of a single instruction is a detectible side effect.

The presence of detectible side effects suggests there are non-detectible side effects as well. This can indeed be the case. A side effect is undetectible if the evaluation of a (single) instruction causing a side effect would normally change the program state, but because of the specific initial valuation, it does not. As a simple example, consider the single instruction $\?([ v:=1 ]\top)$. Under any initial valuation $g$ this would change the program state and cause a side effect, with one exception: namely if $g(v)=1$. We can formally define this as follows:
\begin{definition}
Let $\rho$ be a single instruction in model $M$ under initial valuation $g$, updating the valuation of a variable $v$.\footnote{In \dlaf, this would mean that $\rho$ either is $v:=t$ or $\?[ v:=t ]\top$.} Furthermore, let $\mathcal{S}_{g}^{M}(\rho) = \emptyset$. $\rho$ contains an \textbf{undetectible side effect} iff for $h$ such that $h(v) \neq g(v)$:
\begin{equation*}
\mathcal{S}_{h}^{M}(\rho) \neq \emptyset
\end{equation*}
\end{definition}
It remains to be seen whether these non-detectible side effects are worth our attention. After all, not being able to detect side effects suggests that the presence of the side effects does not make much difference, in any case not to the further execution of the program. Possible exceptions to this are the execution speed or the efficiency of the program, especially if there are a lot of undetectible side effects.

In contrast to non-detectible side effects, marginal side effects can potentially be very useful because they can occur far more often. Like non-detectible side effects, they are a measure of the impact of a side effect. If a side effect is marginal, that means that the rest of the program is unaffected by it and therefore, the side effect is essentially pretty harmless. One could at this point imagine a claim that a program in which only marginal side effects occur can be considered a well-written program, whereas a program in which non-marginal side effects occur is one that should probably be rewritten to avoid unexpected behavior. We will leave further investigation of this claim for future work, however.

\chapter{A case study: Program Algebra}
\label{ch:pga}

In Chapter \ref{ch:treatment}, I presented the system I will be using for the treatment of side effects. In this chapter I will provide a case study to see my system in action. For this, we will use \emph{Program Algebra} (PGA) \cite{PGA}. Since PGA is a basic framework for sequential programming, it provides an ideal case study for our treatment of side effects. By showing how side effects are determined in the very general setting of PGA, we are essentially showing how they are dealt with on a host of different, more specific programming languages.

I will first summarize PGA and explain how we can use it. Next, some extensions necessary for our purpose will be presented. Finally, I will present some examples to see in full how my system deals with side effects.

\section{Program Algebra}
\label{sec:pga}

\subsection{Basics of PGA}
PGA is built from a set $A$ of basic instructions (not to be confused with the \dlaf-notion by the same name), which are regarded as indivisible units. Basic instructions always provide a Boolean reply, which may be used for program control (i.e.\ in steering fragments). There are two composition constructs: concatenation and repetition. If $X$ and $Y$ are programs, then so is their concatenation $X; Y$ and its repetition $X^{\omega}$. PGA has the following primitive instructions:
\begin{itemize}
\item \textbf{Basic instruction} Basic instructions are typically notated as \verb$a,b,$\ldots. As said they generate a Boolean value. Especially important for our purpose is that their associated behavior may modify a (program) state.
\item \textbf{Termination instruction} This instruction, notated as $!$, terminates the program.
\item \textbf{Test instruction} Test instructions come in two flavours: the positive test instruction, notated as $+a$ (where $a$ is a basic instruction), and its negative counterpart, $-a$. For the positive test instruction, $a$ is evaluated and if it yields \verb$true$, all remaining instructions are executed. If it yields \verb$false$, the next instruction is skipped and evaluation continues with the instruction after that. For the negative test instruction, this is the other way around.
\item \textbf{Forward jump instruction} A jump instruction, notated as $\#k$ where $k$ can be any natural number. This instruction prescribes a jump to $k$ instructions from the current one. If $k=0$, the program jumps to the same instruction and inaction occurs. If $k=1$, the program jumps to the next instruction (so this is essentially useless). If $k=2$, the next instruction is skipped and the program proceeds with the one after that, and so on.
\end{itemize}

If two programs execute identical sequences of instructions, \emph{instruction sequence congruence} holds between them. This can be axiomatized by the following four axioms:
\begin{align*}
(X;Y);Z &= X;(Y;Z)\tag{PGA1}\\
(X^{n})^{\omega} &= X^{\omega}\tag{PGA2}\\
X^{\omega};Y &= X^{\omega}\tag{PGA3}\\
(X;Y)^{\omega} &= X;(Y;X)^{\omega}\tag{PGA4}
\end{align*}
The \emph{first canonical form} of a PGA program is then defined to be a PGA program which is in one of the following two forms:
\begin{enumerate}
\item $X$ not containing a repetition
\item $X;Y^{\omega}$, with both $X$ and $Y$ not containing a repetition
\end{enumerate}
Any PGA program can be rewritten into a first canonical form using the above four equations. The next four axiom schemes for PGA deal with the simplification of chained jumps:
\begin{align*}
\#n+1;u_{1};\ldots;u_{n};\#0 &= \#0;u_{1};\ldots;u_{n};\#0\tag{PGA5}\\
\#n+1;u_{1};\ldots;u_{n};\#m &= \#n+m+1;u_{1};\ldots;u_{n};\#m\tag{PGA6}\\
(\#n+k+1;u_{1};\ldots;u_{n})^{\omega} &= (\#k;u_{1};\ldots;u_{n})^{\omega}\tag{PGA7}\\
X = u_{1};\ldots;u_{n};(v_{1};\ldots;v_{m+1})^{\omega} \rightarrow\\ \#n+m+k+2;X &= \#n+k+1;X\tag{PGA8}
\end{align*}
Programs are considered to be \emph{structurally congruent} if they can be proven equal using the axioms PGA1-8.

The \emph{second canonical form} of a PGA program is defined to be a PGA program in first canonical form for which additionally the following holds:
\begin{enumerate}
\item There are no chained jumps
\item Counters used for a jump into the repeating part of the expression are as short as possible
\end{enumerate}
Each PGA expression can be rewritten into a shortest structurally equivalent second canonical form using the above eight equations \cite{PGA}.

\subsection{Behavior extraction}
The previous section describes the forms a PGA program can take. In this section I will explain the behavioral semantics defined in \cite{PGA}. The process of determining the behavior of a PGA program given its instructions is called \emph{behavior extraction}. The behavioral semantics itself is based on thread algebra, TA in short.

Like PGA, TA has a set $A$ of basic instructions, which in this setting are referred to as actions. Furthermore, it has the following two constants and two composition mechanisms:
\begin{itemize}
\item \textbf{Termination} This is notated as \verb$S$ (for Stop) and terminates the behavior.
\item \textbf{Divergent behavior} This is notated as \verb$D$ (for Divergence). Divergence (or inaction) means there no longer is active behavior. For instance, infinite jump loops cause divergent behavior since the program only makes jumps and does not perform any actions.
\item \textbf{Postconditional composition} This is notated as $P \unlhd a \unrhd Q$ and means that first $a$ is executed; if its reply is \verb$true$ then the behavior proceeds with $P$, otherwise it proceeds with $Q$.
\item \textbf{Action prefix} This is notated as $a \circ P$ and is a shorthand for $P \unlhd a \unrhd P$: regardless of the reply of $a$, the behavior will proceed with $P$.
\end{itemize}

As said, behavior extraction determines the behavior of a PGA program given its instructions. For that, the behavior extraction operator, notated as $|\_|$, is defined. If a program ends without an explicit termination instruction, it is defined to end in inaction by the following equation:
\begin{equation}
|X| = |X;(\#0)^{\omega}|
\end{equation}
A termination instruction followed by other instructions ends in termination and nothing else, which is defined by the following equation:
\begin{equation}
|\e;X| = \text{S}
\end{equation}
Behavior extraction is further defined by the following equations dealing with the composition mechanisms:
\begin{gather}
|a;X| = a \circ |X|\\
|\text{+}a;u;X| = |u;X| \unlhd a \unrhd |X|\\
|\text{$-$}a;u;X| = |X| \unlhd a \unrhd |u;X|
\end{gather}
The jump instruction requires a set of equations as well. The first equation defines that a jump instruction which is jumping to itself leads to inaction. The second and third define how a jump instruction can skip subsequent instructions.
\begin{gather}
|\#0;X| = \text{D}\\
|\#1;X| = |X|\\
|\#k+2;u;X| = |\#k+1;X|
\end{gather}

\subsection{Extensions of PGA}
PGA is a most basic framework \cite{PGAu}. However, there are many extensions that introduce more `advanced' programming features such as goto's and backward jump instructions. Via projections, each of these extensions can be projected to PGA in such a way that the resulting PGA-program is behaviorally equivalent to the original program. Examples of such extensions are \emph{PGLB}, in which PGA is extended with a backward jump instruction (\verb$\$$\#k$) and \emph{PGLB$_{g}$}, in which PGLB is further extended with a label catch instruction ($L\sigma$) and an absolute goto instruction ($\#\#L\sigma$).

Of particular interest for our purpose is the extension of PGA with the \emph{unit instruction operator} (PGA$_{u}$), introduced in \cite{PGAu}. The idea of the unit instruction operator, notated as $\un(\_)$, is to wrap a sequence of instructions into a single unit of length 1. That way, a more flexible style of PGA-programming is possible. In particular, programs of the form
\begin{verbatim}
if a then {
    b, c, d
} else {
    f, g, h
}\end{verbatim}
now have a more intuitive translation: $+a; \un(b;c;d;\#4;);f;g;h$.\footnote{The jump is necessary to prevent the instructions $f$, $g$ and $h$ from being executed when $a$ yields true.} Because, thanks to the unit instruction operator, the instructions $b$, $c$, $d$ and $\#4$ are viewed as a single instruction, the execution of those is skipped when $a$ yields \verb$false$.

\section{Logical connectives in PGA}

\subsection{Introduction}
\label{sec:lcintroduction}
As mentioned in Section \ref{sec:pga}, in PGA a lot of basic notations for assembly-like programming languages are defined, especially with its extension with unit instruction operators (PGA$_{u}$) \cite{PGAu}. However, one important basic notation is missing: that of complex tests, of the form \verb$if(a and b) then c$. As we have seen, currently there are positive and negative test instructions in PGA, which can only test the Boolean reply of a single instruction. More complex constructions such as the one in the working example of Section \ref{sec:example} are however very common in programming practice and also appear in research papers such as \cite{SF}, where they are referred to as complex steering fragments. This means that for our purpose, PGA will have to be extended to accommodate for complex steering fragments. I will do so below.

Atomic steering fragments (that is, steering fragments containing only one instruction) are already present in PGA in the form of the positive and negative test instruction ($+a$ and $-a$ respectively). If we were to extend this with complex steering fragments, an obvious notation would be $+\phi$ and $-\phi$. The question now is what forms $\phi$ can take and what it means to have such a complex test.

Since the instructions in the steering fragment need to produce a Boolean reply, the answer to the question above in my opinion should be that a complex test can only be meaningful if all the instructions in the complex test may be used to determine the reply. It is not necessary that all instructions are always used to determine the reply: for instance when using short-circuit evaluation, in some situations not all components of a complex test have to be (and therefore are not) used. However, my claim here is that if a certain instruction is \emph{never} necessary to determine the Boolean reply of the whole steering fragment, then is should not be in the steering fragment.

Currently, PGA has two composition constructs (composition and repetition). Neither of those define anything, however, about the Boolean value of multiple instructions. That is, the Boolean value of $\phi;\ldots;\psi$ and of $\phi^{\omega}$ is undefined. The intuitive way to determine the Boolean reply of a sequence of instructions is via logical connections such as And ($\land$) and Or ($\lor$). However, these are not present yet in PGA. This means that I will have to introduce them in an extension of PGA$_{u}$, which we baptize \pgaul.

Before I do so, however, I need to say something more about the type of And and Or I will be using. There are multiple flavours available:
\begin{itemize}
\item \textbf{Logical And / Or} These versions are notated as $\land$ and $\lor$, respectively. They use full evaluation and the order of evaluation is undefined.
\item \textbf{Short-circuit Left And / Or} These versions are the ones we use in \dlaf\ (see Chapter \ref{ch:treatment}). They are notated as $\leftand$ and $\leftor$. From here on I will refer to them as SCLAnd and SCLOr. They use short-circuit evaluation and are therefore not commutative. The left conjunct or disjunct is evaluated first. There naturally are right-hand versions as well, but I will not be using them.
\item \textbf{Logical Left And / Or} These versions are a combination of the other two: they use full evaluation, but the left conjunct or disjunct is evaluated first. I will notate this as $\&$ and $|$, respectively and refer to them as LLAnd and LLOr. I will not discuss right-hand versions.
\end{itemize}
The latter two are interesting for our purpose, because they are very suitable to demonstrate side effects. However, since we currently only have SCLAnd and SCLOr at our disposal in \dlaf, I will concentrate on those connectives. Although LLAnd and LLOr can be added to both PGA and \dlaf, this would raise more questions than it answers, for instance with regard to the logic which would then be behind the system, which is why we leave it for future work.

The above connectives will almost always be used in combination with either a positive or a negative test. This will be written as $+(a \leftand b)$ (and similar for the negative test and the $\leftor$ connective).

\subsection{Implementation of SCLAnd and SCLOr}
\label{sec:leftand}

If I am to introduce the mentioned logical connectives in \pgaul, I will have to be able to project this extention into PGA. Since the projection of PGA$_u$ to PGA is already given in \cite{PGAu}, it is sufficient to project \pgaul\ to PGA$_{u}$ to show that the former can be projected to PGA. Below is a proposal of a projection of the SCLAnd ($\leftandpf$) connective from \pgaul\ to PGA$_{u}$, for $a,b \in A$:
\begin{equation}\label{nosideeffects}
\proj(+(a \leftand b)) = \un(+a;\un(+b;\#2);\#2)
\end{equation}
To see why this projection works, consider the following example: suppose we have the sequence $+\phi;c;d$ with $\phi = a \leftand b$. This means that if $a$ and $b$ are true, $c$ and $d$ will be executed. Otherwise, only $d$ will be executed. In \pgaul\ this sequence would be $+(a \leftand b);c;d$. The projection to PGA$_{u}$ would then be $\un(+a;$\un$(+b;\#2);\#2);c;d$. If $a$ is false, the execution skips the unit and executes the jump instruction, ending up executing $d$. If $a$ is true, the unit is entered, starting with the test $b$. If $b$ is false, the execution again arrives at the same jump as before, skipping $c$ and executing $d$. If $b$ is true, a different jump is executed which makes the program jump to $c$ first and only then moves on to $d$, which is exactly the desired behaviour.

The entire projection is wrapped in a unit because, as we will see later, the SCLAnd and other operators we define here also are to be considered units. Therefore, a program sequence prior to (or after) the operators discussed here cannot jump into the execution of that operator. By wrapping the projection into a unit I ensure that cannot happen after the projection either.

For the SCLOr connective, the projection is a little easier. It looks like this, again for $a,b \in A$:
\begin{equation}
\proj(+(a \leftor b)) = \un(-a;+b)
\end{equation}
To see why this projection works, consider the same example as above: $+\phi;c;d$, but now with $\phi = a \leftor b$. So, if $a$ and / or $b$ are true, $c$ and $d$ should be executed. If they are both false, only $d$ should be executed. In \pgaul\ this looks like this: $+(a \leftor b);c;d$. The projection to PGA$_{u}$ then is $\un(-a;+b);c;d$. So, if $a$ is true, execution skips testing $b$ and moves on directly to $c$. If $a$ is false, $b$ is tested first. If $b$ is also false, execution skips $c$ and $d$ is executed. If $b$ is true, $c$ gets executed first: exactly the desired behaviour.

So far, we have only been considering programs of the form $+\phi;c;d$, that is, with a positive test. Of course, we also have the negative test instruction. For a negative test, the projection of SCLAnd resembles that of SCLOr. This comes as no surprise since SCLAnd and SCLOr are each other's dual. It looks like this, again for $a,b \in A$:
\begin{equation}
\proj(-(a \leftand b)) = \un(+a;-b)
\end{equation}

The projection of $\leftor$ for a negative test resembles the projection of $\leftand$ for a positive test:
\begin{equation}
\proj(-(a \leftor b)) = \un(-a;\un(-b;\#2);\#2)
\end{equation}

\subsection{Complex Steering Fragments}
\label{subsec:complex}

The implementations in the previous section work for steering fragments containing a single logical connective (that is, with disjuncts or conjuncts $a,b \in A$). However, we also need to define what happens for larger complex steering fragments (for instance $a \leftand (b \leftor c$)). In order to accommodate this, we need one more property for the $\leftand$ and $\leftor$ operators in PGA: they have to be treated as units. If we do this, we can give a recursive definition for the projection, with as base cases the ones given in the previous sections.

In what follows, the formulas $\phi_1$ and $\phi_2$ can take the following form:
\begin{equation}
\phi ::= \top \text{ \large{$\mid$} } a \in A \text{ \large{$\mid$} } \neg \phi \text{ \large{$\mid$} } \phi \leftand \psi \text{ \large{$\mid$} } \phi \leftor \psi
\end{equation}
As we can see, this includes negation. For more on negation, see the next section. We get the following projections:
\begin{align*}
\proj(+(\phi_1 \leftand \phi_2)) &= \un(\proj(+\phi_1);\un(\proj(+\phi_2);\#2);\#2)\\
\proj(+(\phi_1 \leftor \phi_2)) &= \un(\proj(-\phi_1);\proj(+\phi_2))\\
\proj(-(\phi_1 \leftand \phi_2)) &= \un(\proj(+\phi_1);\proj(-\phi_2))\\
\proj(-(\phi_1 \leftor \phi_2)) &= \un(\proj(-\phi_1);\un(\proj(-\phi_2);\#2);\#2)
\end{align*}

This works as follows. Consider the example $+\phi;d;\e$,  with $\phi = a \leftand (b \leftand c)$. In \pgaul\ this would be written as:
\begin{equation}
+(a \leftand (b \leftand c));d;\e
\end{equation}
We can use our new recursive definition of $\leftand$ and get:
\begin{align*}
\proj(+(a \leftand (b \leftand c));d;\e) = \un\big(&\proj(+a);\\
&\un(\proj(+(b \leftand c));\#2);\\
&\#2\big);d;\e
\end{align*}
The projections left now are base cases of $+a$ and $+(b \leftand c)$, respectively. Thus, we get
\begin{align*}
\proj(+(a \leftand (b \leftand c));d;\e) = \un\big(&\proj(+a);\\
&\un(\proj(+(b \leftand c));\#2);\\
&\#2\big);d;\e\\
= \un\big(&+a;\\
&\un(\un(+b;\un(+c;\#2);\#2);\#2);\\
&\#2\big);d;\e
\end{align*}
An interesting question is whether these projections make $\leftand$ an associative operator. To find out, we compare the above with the example $+\phi;d;\e$ where this time $\phi = (a \leftand b) \leftand c$. We get:
\begin{align*}
\proj(+((a \leftand b) \leftand c);d;\e) = \un\big(&\proj(+(a \leftand b));\\
&\un(\proj(+c);\#2);\\
&\#2\big);d;\e\\
= \un\big(&\un(+a;\un(+b;\#2);\#2);\\
&\un(+c;\#2);\\
&\#2\big);d;\e
\end{align*}
We can use behavior extraction to check if these programs are behavioral equivalent. It turns out that both programs indeed have the same behavior:
\begin{equation*}
((d \circ S\unlhd c \unrhd S) \unlhd b \unrhd S) \unlhd a \unrhd S
\end{equation*}
Thus, we can conclude that $\leftand$ is associative in \pgaul, as we would expect given SCL7. We can analyze $\leftor$ in a similar manner.

\subsection{Negation}
\label{subsec:negation}

Now that we have the projections for positive and negative tests defined, we can turn our attention to one more operator that is common both in programming practice and in logic: negation. In PGA, negation is absent, so we need to define it here. Not all instructions or sequences of instructions can be negated: after all, there is no intuition for the meaning of the negation of a certain behavior. We can, however, negate basic instructions: by this we mean its Boolean reply changes value. Sequences of instructions consisting of the operators I have defined above can be negated as well, which I will write as $\neg \phi$. First, I define the following standard projection rules:
\begin{gather}
\label{first}
+(\neg \phi) = -\phi\\
-(\neg \phi) = +\phi\\
\label{last}
\neg \neg \phi = \phi
\end{gather}
Now that we have this, we need to take a look at how negation interacts with the $\leftand$ and $\leftor$ connectives. In particular, we are interested in what happens if one or both of the instructions in such a connective are negated. For this, the De Morgan's laws will come in handy:
\begin{gather}
\neg(\phi_1 \leftand \phi_2) = \neg \phi_1 \leftor \neg \phi_2\\
\neg(\phi_1 \leftor \phi_2) = \neg \phi_1 \leftand \neg \phi_2
\end{gather}
With the above equations in combination with the equations \ref{first}-\ref{last}, we already have the projections for two possible cases (namely when no instructions are negated and when both instructions are negated). That leaves us two other cases for both $\leftand$ and $\leftor$: one in which the first instruction is negated, and one in which the other is. Below are the projections of these cases:
\begin{gather}
\begin{split}
\proj(+(\neg \phi_1 \leftand \phi_2)) & = \proj(-(\phi_1 \leftor \neg \phi_2))\\
& = \un(\proj(+\phi_1);\#3;\proj(+\phi_2))
\end{split}\\
\begin{split}
\proj(+(\phi_1 \leftand \neg \phi_2)) & = \proj(-(\neg \phi_1 \leftor \phi_2))\\
& = \un(\proj(-\phi_1);\#3;\proj(-\phi_2))
\end{split}\\
\begin{split}
\proj(+(\neg \phi_1 \leftor \phi_2)) & = \proj(-(\phi_1 \leftand \neg \phi_2))\\ 
& = \un(\proj(-\phi_1);\#2;\proj(+\phi_2))
\end{split}\\
\begin{split}
\proj(+(\phi_1 \leftor \neg \phi_2)) & = \proj(-(\neg \phi_1 \leftand \phi_2))\\
& = \un(\proj(+\phi_1);\#2;\proj(-\phi_2))
\end{split}
\end{gather}

For more on the $\leftand$ and $\leftor$ connectives and the rules that apply to them, see the paper by Bergstra and Ponse on short-circuit logic \cite{SCL} as well as Chapter \ref{ch:logicalstructure}.

\subsection{Other instructions}
In the previous subsections we have seen what the projections of the new logical connectives in \pgaul\ to \pgau\ look like. To complete the list of projections, we have to define the projections for the `regular' instructions, as well as how concatenation and repetition are projected. This is trivial, since these `regular' instructions are the same in \pgaul\ and \pgau. We get for $a \in A$ and \pgaul-programs $X,Y$
\begin{align*}
\proj(a) &= a\\
\proj(+a) &= +a\\
\proj(-a) &= -a\\
\proj(\e) &= \e\\
\proj(\#k) &= \#k\\
\proj(X;Y) &= \proj(X);\proj(Y)\\
\proj(X^\omega) &= (\proj(X))^\omega\\
\proj(\un(X)) &= \un(\proj(X))
\end{align*}

\section{Detecting side effects in PGA}
\label{sec:sideeffectspga}
In this section I will show how to detect side effects in a \pgaul\ program using our treatment of side effects. In essence, all we have to do is translate the \pgaul\ program to an equivalent \dlaf-program, which can then be used to determine the side effects that occur.

To recap, we have the following operators in \pgaul\ that have to be translated:
\begin{itemize}
\item Concatenation ($X;Y$)
\item Repetition ($X^{\omega}$)
\item Unit instruction operator ($\un(\_)$)
\item Termination ($!$)
\item Positive and negative tests ($+\phi,-\phi$)
\item Only in tests: conjunction, disjunction and negation ($\phi_1 \leftand \phi_2$,$\phi_1 \leftor \phi_2,\lnot \phi$)
\end{itemize}
There are two notable differences between \pgaul\ and \dlaf. The first is that in \pgaul\, a program unsuccessfully terminates unless explicitly instructed otherwise by the termination instruction, whereas in \dlaf\ the default is a successful termination. This is an issue that has to be addressed to properly translate \pgaul\ to \dlaf\ and the best way to do this, is to add the termination instruction to \dlaf. This illustrates the point I made in Section \ref{sec:outside} in Chapter \ref{ch:treatment}: the instructions I defined so far in \dlaf\ are by no means exhaustive and new instructions may have to be added to them. This can usually be done by simply defining the actual and expected evaluation of the new instruction.

The nature of the termination instruction requires us to do a little more than just that. After all, the termination instruction has a control element to it: just like for instance the test instruction it has an influence on which instructions are to be evaluated next. To be exact, \emph{no} instructions are to be evaluated next when a termination instruction is encountered during evaluation of a program. Because of this, we have to slightly modify the concatenation operator in \dlaf\ too when we introduce the termination instruction. We baptize the extension of \dlaf\ with the termination instruction \dltaf\ (for Dynamic Logic with Termination and Assignment in Formulas).

The equation for the relational meaning of $!$ in a given model $M$ and initial valuation $g$ is straight-forward. Execution simply finishes with the same resulting valuation as the initial valuation:
\begin{equation*}
_{g}\llbracket \e \rrbracket_{h}^{M} \text{ iff } g=h \tag{DLTA15}
\end{equation*}
The updated rule for concatenation has to express that when a termination instruction is encountered, nothing should be evaluated afterwards. We use a case distinction for this on the first instruction of a concatenation:
\begin{align*}
\tag{DLTA12} _{g}\llbracket \varpi ; d\pi\rrbracket_{h}^{M} &\text{ iff }
\begin{cases}
g=h &\text{ if } \varpi = \e\\
\exists f \text{ s.th. } _{g}\llbracket \varpi \rrbracket_{f} \text{ and } _{f}\llbracket d\pi \rrbracket_{h}^{M} &\text{ o.w.}
\end{cases}
\end{align*}
We only define the termination instruction in the setting of deterministic programs here. This is sufficient because this is the only setting we are currently interested in. DLTA12 replaces QDL12, but keeps the associative character of concatenation intact:
\begin{equation*}
_{g}\llbracket (d\pi_{0};d\pi_{1});d\pi_{2}\rrbracket_{h}^{M} =\ _{g}\llbracket d\pi_{0};(d\pi_{1};d\pi_{2})\rrbracket_{h}^{M}
\end{equation*}
The addition of the termination instruction allows us to easily express \pgaul-programs such as $+a;\e;b$ in \dltaf. They would otherwise have caused a problem because there would have been no easy way to stop the evaluation of the program from continuing to evaluating $b$, which it of course is not supposed to do if $a$ yields true.

The other notable difference between \pgaul\ and \dlaf\ is that in the former, anything can be used as a basic instruction. That includes what we refer to in \dlaf\ as primitive formulas such as $x \leq 2$ or $t_{1} = t_{2}$. In PGA the execution of an instruction always succeeds, even if the Boolean reply that it generates, is false. To model this in \dltaf, we have to add the primitive formulas $\varphi$ to the set of instructions, as follows:
\begin{equation*}
\pi ::= \varphi \mid \e \mid v:=t \mid \?\phi \mid \pi_{1};\pi_{2} \mid \pi_{1} \cup \pi_{2} \mid \pi^{*}
\end{equation*}
The relational meaning in $M$ given initial valuation $g$ for these new instructions is simply that they always succeed without modifying $g$:
\begin{align*}
_g\llbracket \varphi \rrbracket_h^M &\text{ iff } g=h
\end{align*}

With the termination instruction and the formulas-as-instructions defined, we can take a first look at the mapping from \pgaul\ to \dltaf. For this we define a translation function $f_{t}:$ \pgaul $ \rightarrow $ \dltaf. We define this translation function for PGA programs in first or second canonical form only; this is sufficient because as we have seen, every PGA program can be rewritten to first and second canonical form.

First, we define the set $A$ of basic instructions in PGA to be equal to the set of primitive formulas and single instructions, not being tests, in \dlaf:
\begin{equation*}
A ::= \varphi \mid \rho^{-}
\end{equation*}
where $\rho^{-}$ denotes the set of single instructions not being tests. In \dlaf, this set only consists of the assignment instruction $v:=t$.

For finite sequences of instructions with length $n=1$, $a,b \in A$ and $k \in \mathbb{N}_{0}$, and $\phi$ a formula as meant in section \ref{subsec:complex}, $f_t$ is defined as follows:
\begin{align*}
f_t(a) &= a;\?\bot\\
f_t(+\phi) &= \?\phi;\?\bot\\
f_t(-\phi) &= \?\lnot \phi;\?\bot\\
f_t(\#k) &= \?\bot\\
f_t(\e) &= \e\\
f_t(\un(a_1; \ldots ;a_k)) &= f_t(a_1;\ldots;a_k)
\end{align*}
Here we can clearly see what effect it has that \pgaul\ has unsuccessful termination as its default. We have to explicitly introduce unsuccessful termination in \dltaf\ by adding $\?\bot$ (a test that always fails) at the end of every instruction. Furthermore, notice the unit instruction operator that here has length $n=1$, but is transparent when it has to be translated and thus becomes a sequence of instructions with length $k$ that is potentially larger than $1$. Finally, notice that there is no need to translate possibly compound formulas $\phi$. This is because formulas have the exact same syntax in \pgaul\ and \dltaf.

Next, we can show the definition of $f_t$ for finite sequences of instructions with length $n = m+1$. For $a,b_{1},\ldots,b_{m} \in A$, $k \in \mathbb{N}_{0}$ and $\phi$ a formula as meant in section \ref{subsec:complex}, we have
\begin{align*}
f_t(a;b_{1};\ldots;b_{m}) &= a;f_t(b_{1};\ldots;b_{m})\\
f_t(+\phi;b_{1};\ldots;b_{m}) &= 
\begin{cases}
(\?\phi;f_t(b_{1})) \cup (\?\lnot \phi;\?\bot) &\text{if } m\text{=1}\\
(\?\phi;f_t(b_{1};\ldots;b_{m}))\ \cup \\\text{ }\text{ }\text{ }(\?\lnot \phi;f_t(b_{2};\ldots;b_{m})) &\text{o.w.}
\end{cases}\\
f_t(-\phi;b_{1};\ldots;b_{m}) &= 
\begin{cases}
(\?\phi;\?\bot) \cup (\?\lnot\phi;f_t(b_1)) &\text{if } m\text{=1}\\
(\?\phi;f_t(b_{2};\ldots;b_{m}))\ \cup \\\text{ }\text{ }\text{ }(\?\lnot\phi;f_t(b_{1};\ldots;b_{m})) &\text{o.w.}
\end{cases}\\
f_t(\#0;b_{1};\ldots;b_{m}) &= \?\bot\\
f_t(\#1;b_{1};\ldots;b_{m}) &= f_t(b_{1};\ldots;b_{m})\\
f_t(\#(\text{2+}k);b_{1};\ldots;b_{m}) &=
\begin{cases}
f_t(b_{k+2};\ldots;b_{m}) &\text{ if } k+2 < m\\
\?\bot &\text{ o.w.}
\end{cases}\\
f_t(\e;b_{1};\ldots;b_{m}) &= \e\\
f_t(\un(a_1;\ldots;a_k);b_{1};\ldots;b_{m}) &= f_t(a_1;\ldots;a_k;b_{1};\ldots;b_{m})
\end{align*}

With the above translation rules, we can now translate finite \pgaul-programs to their \dltaf-versions. A complete translation would require a translation of repetition as well. This, however, is quite a complex task. The reason for that becomes clear when considering examples like these:
\begin{gather*}
(a;b;+c)^\omega\\
(+a;+b;+c)^\omega\\
(a;+b;\#5;c;+d;\!)^\omega
\end{gather*}
Because of the behavior of $+c$, we get into trouble here if we attempt to use the regular translation. The problem is that $+c$ can possibly skip the first instruction of the next repetition loop, which is behavior that is hard to translate without explicitly introducing this variant of repetition ($^\omega$) in \dlaf. The same problem arises with the jump instruction. At first glance, the best solution there is to introduce the jump instruction to \dlaf\ as well. In that case the second canonical form of PGA-programs comes in handy, as it is designed to manipulate expressions with repetition such that no infinite jumps occur.

Since this case study is meant as a relatively clear example of how to use \dlaf\ to model side effects in other systems such as PGA, it is beyond our interest here to present these rather complex translations of repetition. Instead, we restrict ourselves to finite \pgaul-programs and leave the relational semantics for \dlaf\, which models side effects, as the basis for future work on PGA involving repetition.

\section{A working example}
In this section I will present a working example of the translation from finite \pgaul-programs, which we write as \pgaulf, to \dltaf. In addition, I will show that we get sufficiently similar results if we first translate \pgaulf\ to \dltaf\ compared to first projecting \pgaulf\ to \pgauf\ and then translating that to \dltaf. To be exact, we are going to show that the following diagram defines a program transformation $E$ on finite deterministic programs in \dltaf:
\begin{displaymath}
    \xymatrix{
        \text{\pgaulf} \ar[r]^{f_t} \ar[d]_{\text{pgaul2pgau}} & \text{\dltaf} \ar@{.>}[d]^{E} \\
        \text{\pgauf} \ar[r]_{f_t}       & \text{\dltaf} }
\end{displaymath}
Here $E$ is a reduction function on \dltaf\ that yields deterministic \dltaf-programs where occurrences of $\leftand$ and $\leftor$ have been eliminated.

For the working example, we return to a variant of our running example. Consider the \pgaulf-program
\begin{equation*}
X=+([x:=x+1]\top\leftand x=2);\un(w[x=2];!);w[x\ne 2];!
\end{equation*}
where $w[...]$ suggests a write command. This is a program of the form
\begin{equation*}
+(b \leftand c);\un(d;\e);e;\e
\end{equation*}
with $b = [x:=x+1]\top, c=(x=2), d = w[x=2]$ and $e = w[x\ne 2]$. Thus, we get the following translation, where we for clarity have underlined the instruction that we are going to translate next:
\begin{flalign*}
&f_t(\underline{+(b \leftand c)};\un(d;\e);e;\e)=&\\
&(\?(b \leftand c);f_t(\underline{\un(d;\e)};e;\e)) \cup (\?\lnot(b \leftand c);f_t(e;\e))=&\\
&(\?(b \leftand c);f_t(\underline{d};\e;e;\e)) \cup (\?\lnot(b \leftand c);f_t(e;\e))=&\\
&(\?(b \leftand c);d;f_t(\underline{\e};e;\e)) \cup (\?\lnot(b \leftand c);f_t(e;\e))=&\\
&(\?(b \leftand c);d;\e) \cup (\?\lnot(b \leftand c);f_t(\underline{e};\e))=&\\
&(\?(b \leftand c);d;\e) \cup (\?\lnot(b \leftand c);e;f_t(\underline{\e}))=&\\
&(\?(b \leftand c);d;\e) \cup (\?\lnot(b \leftand c);e;\e)&
\end{flalign*}
So there we have it: if we replace the shorthands with their original instructions or formulas again, we get the following \dltaf-program, which we baptize $d\pi_{ul}$:
\begin{align*}
d\pi_{ul} =\ &(\?([x:=x+1]\top \leftand (x=2));w[x=2];\e)&\\
&\cup&\\
&(\?\lnot([x:=x+1]\top \leftand (x=2));w[x\ne 2];\e)&
\end{align*}
Clearly, given model $M$, $_g\llbracket f_t(X)\rrbracket_h^M$ implies that $h=g[x\mapsto g(x)+1]$. So, if $g(x)=1$, the instruction $w[x=2]$ is executed, after which the program terminates, while for $g(x)\ne 1$, the instruction $w[x\ne 2]$ is executed after which the program terminates.

Now let $Y=\proj(X)$, so
\begin{equation*}
Y =\un\big(+([x:=x+1]\top);\un(+(x=2);\#2);
\#2\big);\un(w[x=2];!);w[x\ne 2];!
\end{equation*}
We compute
\begin{align*}
f_t(Y)&=f_t(+([x:=x+1]\top);\un(+(x=2);\#2);
\#2;\un(w[x=2];!);w[x\ne 2];!)\\
&=\begin{array}[t]{l}
(? ([x:=x+1]\top);f_t(+(x=2);\#2;\#2;\un(w[x=2];!);w[x\ne 2];!))\\
\cup\\
(?\neg ([x:=x+1]\top);f_t(\#2;\un(w[x=2];!);w[x\ne 2];!))
\end{array}\\[2mm]
&=\begin{array}[t]{l}
(? ([x:=x+1]\top);
  \begin{array}[t]{l}
  (\\
  (?(x=2);f_t(\#2;\#2;\un(w[x=2];!);w[x\ne 2];!))\\
  \cup\\
  (?\neg(x=2);f_t(\#2;\un(w[x=2];!);w[x\ne 2];!))\\
  )
  \end{array}
\\
)
\\
\cup\\
(?\neg ([x:=x+1]\top);w[x\ne 2];!)
\end{array}
\\
&=\begin{array}[t]{l}
(? ([x:=x+1]\top);
  \begin{array}[t]{l}
  (\\
  (?(x=2);w[x=2];!)\\
  \cup\\
  (?\neg(x=2);w[x\ne 2];!)\\
  )
  \end{array}
\\
)
\\
\cup\\
(?\neg ([x:=x+1]\top);w[x\ne 2];!)
\end{array}
\end{align*}
Note that for each model $M$ and initial valuation $g$, $M\not\models_g \neg ([x:=x+1]\top)$, so
\begin{equation*}
_g\llbracket f_t(Y)\rrbracket_h^M \text{ iff }
_g\llbracket \?([x:=x+1]\top);
  \begin{array}[t]{l}
  (\\
  (?(x=2);w[x=2];!)\\
  \cup\\
  (?\neg(x=2);w[x\ne 2];!)\\
  )\rrbracket_h^M
  \end{array}
\end{equation*}
Thus, writing $d\pi_u$ for the rightmost deterministic \dltaf-program,
we find 
\begin{equation*}
_g\llbracket f_t(Y)\rrbracket_h^M \text{ iff } _g\llbracket d\pi_u \rrbracket_h^M
\end{equation*}
We now need to ask ourselves if $d\pi_u$ is `sufficiently similar' to the earlier derived $d\pi_{ul}$. Intuitively, we would say that in this working example, this indeed is the case. After all, $[x:=x+1]\top$ always yields true, so the truth of $[x:=x+1]\top \leftand (x=2)$ depends solely on the Boolean reply that $x=2$ yields. It therefore does not matter if we lift $\?[x:=x+1]\top$ out of the union, which is essentially what we have done in the case of $d\pi_u$.

We can call two programs `sufficiently similar' if they evaluate the same single instructions, not being tests, or primitive formulas in the same order. We can formalize that notion with the following proposition:

\begin{proposition}
\label{prop:sufficient}
Let $X$ be a program in \pgaulf, let $d\pi_{ul}=f_t(X)$ and let $d\pi_u = f_t(\iproj(X))$. Let model $M$ be given and let $g$ be an initial valuation such that there exists a valuation $h$ such that $_g\llbracket d\pi_{ul}\rrbracket_h^M$. Then
\begin{equation*}
_g\llbracket d\pi_{ul}\rrbracket_h^M \text{ iff } _g\llbracket d\pi_{u}\rrbracket_h^M
\end{equation*}
and the same single instructions, not being tests, and primitive formulas are evaluated in the same order during evaluation of $d\pi_{ul}$ and $d\pi_{u}$ given $g$.
\end{proposition}

As said, we do not consider repetition as program constructor in our case study. Furthermore, our model of side effects is limited to terminating programs, as opposed to programs that can either end in termination or in divergence. A proof of this proposition might be found, but is for these reasons perhaps not very much to the point. In Chapter \ref{ch:conclusions} (Conclusions) we return to this issue.

It is, however, worthwhile to check the proposition for our working example. Recall that we have the following $d\pi_{ul}$ and $d\pi_u$:
\begin{align*}
d\pi_{ul} =\ &(\?([x:=x+1]\top \leftand (x=2));w[x=2];\e)\\
&\cup\\
&(\?\lnot([x:=x+1]\top \leftand (x=2));w[x\ne 2];\e)\\
d\pi_u =\ &\?([x:=x+1]\top);\\
&(?(x=2);w[x=2];!)\\
&\cup\\
&(?\neg(x=2);w[x\ne 2];!)
\end{align*}
It is not hard to check in this case that for any model $M$ and initial valuation $g$ such that $d\pi_{ul}$ can be evaluated, $_g\llbracket d\pi_{ul}\rrbracket_h^M \text{ iff } _g\llbracket d\pi_{u}\rrbracket_h^M$. It is also easy to see that the same single instructions, not being tests, and primitive formulas are evaluated (in the same order). After all, $d\pi_{ul}$, first evaluates the primitive formulas $[x:=x+1]\top$ and $x=2$ and uses those to determine the reply of $[x:=x+1]\top \leftand (x=2)$. Depending on the reply, it then either evaluates the single instructions $w[x=2]$ and $\e$, or $w[x \ne 2]$ and $\e$.

Almost the same goes for $d\pi_u$. It first evaluates the primitive formula $[x:=x+1]\top$ and depending on the reply (which happens to be always true), either stops evaluation (which therefore is never the case) or continues with the evaluation of primitive formula $x=2$. Depending on the reply, it like $d\pi_{ul}$ then either evaluates the single instructions $w[x=2]$ and $\e$, or $w[x \ne 2]$ and $\e$. So at least in our working example, Proposition \ref{prop:sufficient} holds.

In a similar way, we can analyze the \pgaulf-program
\begin{equation*}
+(\neg[x:=x+1]\top\leftor x=2);\un(w[x=2];\e);w[x\ne2];\e
\end{equation*}
We can compute $d\pi_{ul} = f_t(X)$:
\begin{align*}
f_t(X) &= f_t(+(\neg[x:=x+1]\top\leftor x=2);\un(w[x=2];\e);w[x\ne2];\e)\\
&= \begin{array}[t]{l}
(\?(\neg[x:=x+1]\top\leftor x=2);f_t(\un(w[x=2];\e);w[x\ne2];\e)\\
\cup\\
(?\neg(\neg[x:=x+1]\top \leftor x=2);f_t(w[x\ne 2];\e))
\end{array}\\
&= \begin{array}[t]{l}
(\?(\neg[x:=x+1]\top\leftor x=2);f_t(w[x=2];\e;w[x\ne2];\e)\\
\cup\\
(?\neg(\neg[x:=x+1]\top \leftor x=2);w[x\ne 2];f_t(\e))
\end{array}\\
&= \begin{array}[t]{l}
(\?(\neg[x:=x+1]\top\leftor x=2);w[x=2];f_t(\e;w[x\ne2];\e)\\
\cup\\
(?\neg(\neg[x:=x+1]\top \leftor x=2);w[x\ne 2];\e)
\end{array}\\
&= \begin{array}[t]{l}
(\?(\neg[x:=x+1]\top\leftor x=2);w[x=2];\e)\\
\cup\\
(?\neg(\neg[x:=x+1]\top \leftor x=2);w[x\ne 2];\e)
\end{array}
\end{align*}
We once again define $Y=\proj(X)$, so
\begin{equation*}
Y =\un\big(-([x:=x+1]\top);\#2;+(x=2)\big);\un(w[x=2];!);w[x\ne 2];\e
\end{equation*}
We compute
\begin{align*}
f_t(Y) &= f_t(-([x:=x+1]\top);\#2;+(x=2);\un(w[x=2];\e);w[x\ne 2];\e)\\
&= \begin{array}[t]{l}
(\?(\neg([x:=x+1]\top));f_t(\#2;+(x=2);\un(w[x=2];\e);w[x\ne2];\e)\\
\cup\\
(?\neg(\neg([x:=x+1]\top));f_t(+(x=2);\un(w[x=2];\e);w[x\ne2];\e)
\end{array}\\
&= \begin{array}[t]{l}
(\?(\neg([x:=x+1]\top));f_t(\un(w[x=2];\e);w[x\ne2];\e)\\
\cup\\
(?\neg(\neg[x:=x+1]\top);
\begin{array}[t]{l}
  (\\
  (?(x=2);f_t(w[x=2];\e;w[x\ne2];\e))\\
  \cup\\
  (?\neg(x=2);f_t(w[x\ne 2];!))\\
  )
  \end{array}
\end{array}\\
&= \begin{array}[t]{l}
(\?(\neg([x:=x+1]\top));w[x=2];\e)\\
\cup\\
(?\neg(\neg([x:=x+1]\top));
\begin{array}[t]{l}
  (\\
  (?(x=2);w[x=2];\e)\\
  \cup\\
  (?\neg(x=2);w[x\ne 2];\e)\\
  )
  \end{array}
\end{array}
\end{align*}
We can directly eliminate a situation: $\neg([x:=x+1]\top)$ is false for any initial valuation $g$. Thus, writing $d\pi_u$ for the second part of the topmost union:
\begin{equation*}
d\pi_u =\ \?\neg(\neg([x:=x+1]\top));
\begin{array}[t]{l}
  (\\
  (?(x=2);w[x=2];\e)\\
  \cup\\
  (?\neg(x=2);w[x\ne 2];\e)\\
  )
  \end{array}
\end{equation*}
we get given model $M$ for any initial valuation $g$
\begin{equation*}
_g\llbracket f(Y) \rrbracket_h^M \text{ iff } _g\llbracket d\pi_u\rrbracket_h^M
\end{equation*}
We can check in similar fashion as before that Proposition \ref{prop:sufficient} holds (for any initial valuation $g$). We can conclude that at least for these working examples, the mentioned proposition is valid. As said, we leave the proof for future work.


This case study started from the abstract approach to attempt decomposition of complex steering fragments in instruction sequences in \pgaulf\ as advocated in \cite{SCL}. We show that we can apply this approach to a rather concrete instance in imperative programming (namely the set $A$ of basic instructions given in this chapter) and we obtain some interesting results. In the first place, it inspired our definition of \dltaf\ and the analysis and classification of side effects as discussed in this thesis. Secondly, by the preservation property formulated in Proposition \ref{prop:sufficient}, it justifies our proposal for the projection function $\proj$. It is an interesting result that we are able to show that the projection $\proj$, which does not have to anything to do with valuations, preserves the relational semantics (and therefore side effects) of a program via the diagram at the beginning of this section, which is based on a very natural translation.

\chapter{Conclusions and future work}
\label{ch:conclusions}
In this thesis I have given a formal definition of side effects. I have done so by modifying a system for modelling program instructions and program states, Quantified Dynamic Logic, to a system called \dlaf\ (Dynamic Logic with Assignments as Formulas), which in contrast to QDL allows assignments in formulas and makes use of short-circuit evaluation. I have shown the underlying logic in those formulas to be a variant of short-circuit logic called repetition-proof short-circuit logic.

Using \dlaf\ I have defined the actual and the expected evaluation of a single instruction. The side effects are then defined to be the difference between the two. I have given rules for composing those side effects in single instructions, thus scaling up our definition of side effects to a definition of side effects in deterministic \dlaf-programs. Using this definition I have given a classification of side effects, introducing as most important class that of marginal side effects. Finally, I have shown how to use our system for calculating the side effects in a real system such as PGA.

Our definition gives us an intuitive way to calculate the side effects in a program. Because of the definition in terms of actual and expected evaluation, one can easily adapt the system to ones own needs without having to change the definition of side effects. All one has to do is update the expected evaluation of a single instruction, or if an entirely new single instruction is added to the system, define the actual and expected evaluation for it.

In Chapter \ref{ch:logicalstructure} we have seen how a sound axiomatization of the formulas in \dlaf\ can be given using the signature $\{\top,\bot,\_\lef\_\rig\_\}$. I have not used this signature in the first place because I wanted to stick to the conventions in dynamic logic. It is noteworthy, however, that this alternative and possibly more elegant signature exists, especially because an axiomatization can be given for it.

The definition of side effects given here can point the way to a lot more research. I can see future work being done in the following areas:
\begin{itemize}
\item I do not want to claim that the instructions I have defined in \dlaf\ are exhaustive. Finding out what possible other instructions might have to be added to \dlaf\ can be an interesting project.
\item Another possible subject for future work is the issue of `negative' side effects I briefly touched upon in Section \ref{sec:outside}. It is an open question whether or not we should allow situations in which `negative' side effects occur and if so, how we should handle them.
\item In this thesis, we have mostly been looking at imperative programs. It should be interesting to see if our definition can be extended to, for example, functional programs. Perhaps the work done by Van Eijck in \cite{EijckFunctional}, in which he defines functional programs making use of program states, can be used for this.
\item Another interesting question, which has been raised before in Chapters \ref{ch:qdl} and \ref{ch:treatment}, is that of side effects in non-deterministic programs. It warrants further research if it is reasonable to talk about side effects there. One can imagine that if the set of side effects in all possibilities of a non-deterministic program are the same, the side effects of the whole can be defined as exactly that set. What needs to be done if that's not the case however, or if we should even want to define side effects of such programs, are open questions.
\item In Chapter \ref{ch:classification}, the concept of marginal side effects was introduced and the suggestion was made that this notion can be linked to claims about how well-written a program is. I have not pursued such claims, but can imagine further research being done in that area.
\item To develop a direct modelling of side effects for the variant of PGA discussed in Chapter \ref{ch:pga}, one can introduce valuation functions as program states and define a relational meaning that separates termination from deadlock/inaction, say
\begin{equation*}
_g\llbracket\![ X \rrbracket\!]_h
\end{equation*}
The idea of this would be to evaluate $X$ as far as possible, which is a reasonable requirement if $X$ is in second canonical form. In addition, we could define a termination predicate, e.g. Term$(X,g)$, which states that $X$ terminates for initial valuation $g$. Using this we could define a ``behavioral equivalence'' on programs $X$ and $Y$ as follows:
\begin{equation*}
\forall g, _g\llbracket\![ X \rrbracket\!]_h \text{ iff } _g\llbracket\![ Y \rrbracket\!]_h \text{ AND } \text{Term}(X,g) \text{ iff Term}(Y,g)
\end{equation*}
Using this, Proposition \ref{prop:sufficient} can probably be proven, especially considering the in Chapter \ref{ch:terminology} proven property of \dlaf\ that any program can be rewritten into a form in which its steering fragments only contain primitive formulas and their negations.
\item Also mentioned in Chapter \ref{ch:pga} is the possibility to introduce extra logical operators, namely Logical Left And (LLAnd) and its dual Logical Left Or (LLOr). Introducing these in \dlaf\ is fairly straight-forward: one only needs to define its truth in $M$:
\begin{align*}
M \models_{g} \phi_{1} \llor \phi_{2} &\text{ iff } M \models_{g} \phi_{1} \leftor \phi_{2} \tag{DLA7c} \\
M \models_{g} \phi_{1} \lland \phi_{2} &\text{ iff } M \models_{g} \phi_{1} \leftand \phi_{2} \tag{DLA7d}
\end{align*}
as well as update the program extraction function:
\begin{equation*}
\Pi^M_g( \phi_1 \Box \phi_2) = \Pi^M_g(\phi_1);\Pi^M_h(\phi_2) \text{ if } _g\llbracket\Pi_g^M(\phi_1)\rrbracket_h^M \text{ and } \Box\in\{|,\&\}
\end{equation*}
To introduce the same operator in \pgaul, projection functions in the same style as the ones given in Chapter \ref{ch:pga} for SCLAnd and SCLOr need to be defined.
\item Another possible matter for further study is whether side effects can be used in natural language. In the Introduction, we have already seen that they can occur in the pregnant wife example, where your wife told you to do the grocery shopping if she did not call you, which she later did, but to tell you that she was pregnant. Possibly there is a role for side effects when explaining misunderstandings. There is no doubt that side effects can be the cause of misunderstandings. The pregnant wife example illustrates that: you could decide to do grocery shopping to be on the safe side after her call, claiming her call indicated you might have to shop, only to run into your wife at the store also shopping (who, of course, didn't want to convey the message that you should shop at all).

When we take the Dynamic Epistemic Logic system mentioned in \cite{EijckVisser}, the knowledge of two communicating agents is captured by an epistemic state, one for each agent. The agents also have an epistemic state for what they think is the (relevant) knowledge of the other agent with whom they are in conversation. A misunderstanding has occurred when an agent updates his own epistemic state in a different way than the other agents expects him to. There are a lot of ways in which this can happen, but relevant for us is that one of those ways is, when a side effect from an utterance occurs of which one of the agents is not aware.

If one of the agents is aware of the side effect and also of the fact the other agent might not be aware of it, it may be recommended to point out this side effect to the other agent. In our example of the pregnant wife calling, this would mean that you would have to ask your wife on the phone that the fact she called leaves you in doubt about the grocery shopping. Naturally, though, we recommend a more enthusiastic response to the news she is pregnant first.
\end{itemize}

\end{document}